\documentclass[letters,a4paper,fleqn,usenatbib]{mnras}

\usepackage{newtxtext,newtxmath}

\usepackage[T1]{fontenc}
\usepackage{ae,aecompl}

\usepackage{graphicx}	
\usepackage{amsmath}	
\usepackage{hyperref}
\usepackage{pdflscape}
\usepackage{wasysym}
\usepackage{siunitx}
\usepackage{mathtools}
\usepackage{color}
\usepackage{gensymb}

\title[Asteroids of the Taurid Complex]{A dynamical analysis of the Taurid Complex: evidence for past orbital convergences}

\author[A. Egal et al.]{
A. Egal,$^{1,2,3}$\thanks{E-mail: aegal@uwo.ca (AE)}
P. Wiegert,$^{1,2}$
P. G. Brown,$^{1,2}$
P. Spurn\'y,$^{4}$
J. Borovi\v{c}ka$^{4}$
and G. B. Valsecchi$^{5,6}$
\\
$^{1}$Department of Physics and Astronomy, The University of Western Ontario, London, Ontario N6A 3K7, Canada\\
$^{2}$ Institute for Earth and Space Exploration (IESX), The University of Western Ontario, London, Ontario N6A 3K7, Canada\\
$^{3}$IMCCE, Observatoire de Paris, PSL Research University, CNRS, Sorbonne Universit\'{e}s, UPMC Univ. Paris 06, Univ. Lille, France\\
$^{4}$Astronomical Institute of the Czech Academy of Sciences, Fri\v{c}ova 298, CZ-25165 Ond\v{r}ejov, Czech Republic\\
$^{5}$INAF, Istituto di Astrofisica e Planetologia Spaziali Via del Fosso del Cavaliere 100, 00133 Roma, RM, Italy\\
$^{6}$CNR, Istituto di Fisica Applicata Nello Carrara Via Madonna del Piano, 10 50019 Sesto Fiorentino (FI), Italy
}

\date{Accepted XYZ; Received XYZ; in original form XYZ}

\pubyear{2021}

\begin{document}
\label{firstpage}
\pagerange{\pageref{firstpage}--\pageref{lastpage}}
\maketitle

\begin{abstract}

The goal of this work is to determine if the dynamics of individual Taurid Complex (TC) objects are consistent with the formation of the complex via fragmentation of a larger body, or if the current orbital affinities between the TC members result from other dynamical processes. To this end, the orbital similarity through time of comet 2P/Encke, fifty-one Near-Earth Asteroids (NEAs) and sixteen Taurid fireballs was explored. Clones of each body were numerically simulated backwards in time, and epochs when significant fractions of the clones of any two bodies approached each other with both a low Minimum Orbit Intersection Distance and small relative velocity were identified. Only twelve pairs of bodies in our sample show such an association in the past 20 000 years, primarily circa 3200 BCE. These include 2P/Encke and NEAs 2004 TG10, 2005 TF50, 2005 UR, 2015 TX24, and several Southern Taurid fireballs. We find this orbital convergence to be compatible with the fragmentation of a large parent body five to six thousand years ago,  resulting in the separation of 2P/Encke and several NEAs associated with the TC, as well as some larger meteoroids now recorded in the Taurid stream. However, the influence of purely dynamical processes may offer an alternative explanation for this orbital rapprochement without requiring a common origin between these objects. In order to discriminate between these two hypotheses, future spectral surveys of the TC asteroids are required.

\end{abstract}




\section{Introduction}

The Taurid meteor shower is arguably the most enigmatic of all the major showers currently observable at Earth. It was linked to the dynamically peculiar comet 2P/Encke by \cite{Whipple1940} on the basis of high precision photographic trajectories of a handful of Taurid meteors. The relationship between 2P/Encke and its large complex of associated streams has been examined in various studies in the interim \citep{Whipple1952, Jones1986, Steel1991, Stohl1992,Asher1993b}. 

Among the unusual aspects of the Taurid shower are the long duration of the traditional N and S Taurid branches, reaching a peak in the fall months \citep{Brown2010} and the presence of meter-sized meteoroids in the stream \citep{Spurny2017}, a property unique to the Taurids. Periodic enhancements of bright meteors associated with the Northern (N) and Southern (S) Taurids also occur in October-November of some years \citep{Beech2004a, Dubietis2007}.

The multiple daytime and night-time showers linked to the Taurid Complex (hereafter TC) \citep{Steel1991, Babadzhanov1992} may reflect a gradation between a younger, TC stream-component and an even older, broader dust component populating the Helion and Anti-helion sporadic sources \citep{Stohl1983, Stohl1992}. The latter has been historically interpreted to suggest that 2P/Encke and by extension the TC may be the dominant contributor to the zodiacal dust complex \citep{Whipple1952, Whipple1967, Wiegert2009}.

Through a synthesis of the observational properties of the TC, \cite{Clube1984} expanded on the hypothesis of \cite{Whipple1940} and \cite{Whipple1952} that the Taurid stream was formed as a result of collisional disruption of a large progenitor. \cite{Clube1984} suggested that the Taurid Complex consisted of both the known meteoroid streams at small sizes as well as a number of km-sized asteroids and 2P/Encke which collectively are remnants of the hierarchical fragmentation of a giant comet. 

This hypothesis became known as coherent catastrophism \citep{Asher1994}, reflecting the fact that large ($\sim$100m) TC members may collide with the Earth at shorter time intervals than would be expected from a random background population. In particular, the proximity of TC orbits to the 7:2 mean motion resonance (MMR) with Jupiter may act to concentrate TC members. This sub-component of the TC, termed the Taurid Swarm Complex (TSC), includes objects protected from dynamical diffusion by trapping in the 7:2 MMR. This model of the TSC has been successfully used to explain the observed periodicity in the enhancement of Taurid fireball activity when Earth approaches the centre of the resonant swarm, i.e., within a mean anomaly difference $\Delta M$ of 30-40$\degree$ of the swarm centre \citep{Asher1993b,Asher1998,McBeath1999,Dubietis2007,Spurny2017}. 

To explain the dispersion of orbits in the Taurid complex, the large progenitor would need to have been injected into the inner solar system some 20,000 years ago \citep{Steel1991, Asher1993}. Refinement of this model suggested that the large progenitor comet was active until roughly 10 ka and then experienced a series of fragmentation/splitting events to produce both sub-streams of the TC and sibling fragments now present as large Near Earth Asteroids (NEAs) within the complex \citep{Steel1991}. Much additional work has now been published arguing for a genetic linkage of TC material spanning a large range of sizes, between $10^{-6}$ and $10^3$ m \cite[e.g., in][] {Clube1984,Asher1993,Asher1993b,Asher1994,Steel1996,Steel1996b,Napier2010,Napier2015,Napier2019}.

\cite{Asher1993, Asher1994, Steel1996} identified potential NEAs whose orbits suggested association with the TC consistent with the giant comet breakup hypothesis. Subsequent authors greatly expanded the list of potential NEAs associated with the TC \citep[e.g.,][]{Babadzhanov2001, Porubcan2006, Babadzhanov2008, Jopek2011, Dumitru2017} based on present day orbital similarity. A genetic relationship among TC NEAs has been called into question, however, by the spectral reflectance measurements of many of these NEAs \citep{Popescu2014,Tubiana2015} which do not show expected primitive NEA spectra (C, D or P classes) nor significant common spectral type. \cite{Valsecchi1995} has noted that the Taurid dynamical region is a natural pathway for both main belt asteroids and Jupiter Family Comets moving to Earth-crossing orbits. Moreover, the often cited clustering of perihelion longitudes of TC candidates around 2P/Encke and Hephaistos as evidence for a common association \citep{Asher1993} may be simply due to an observational bias, and not a consequence of the parent body's fragmentation \citep{Valsecchi1999}. 

In this work we focus on exploring dynamical linkages among TC objects. If the TC has been produced by ongoing fragmentation of larger objects, it would be natural to expect some NEAs and large meteoroids in the TC to show closer orbital associations in the recent past signalling a potential common epoch of breakup. Our goal is to identify potential dynamical associations of groups of TC objects in specific epochs. This  might be consistent with the last epoch of major fragmentation events . 

If no TC objects show such an association, it would indicate that current TC objects are likely not the result of a recent fragmentation of a common progenitor. In contrast, if some subset of the TC show a common epoch of orbital similarity, this may represent either a dynamical process which concentrates orbits, or evidence for a past fragmentation episode. In the case of the latter, identifying potential TC NEAs with near-term dynamical association will provide a compelling list of targets for future spectral measurements as a more stringent test of the giant comet fragmentation hypothesis for the TC.  

\section{Taurid Complex candidates}

\subsection{The literature on TC asteroid candidates} \label{sec:litterature}

In previous studies, empirical metrics based on the similarity of present day orbital elements were used to identify potential members of the TC. \cite{Steel1991} selected high-quality orbits of Taurid meteoroids available in the IAU Meteor Data Center, and computed a mean perihelion distance, eccentricity and inclination of $\Bar{q}=0.375$ AU, $\Bar{e}=0.82$ and $\Bar{i}=4\degree$ for the stream. Meteoroids belonging to the complex were identified with a reduced form of the D-criterion of \cite{Southworth1963}: 
\begin{equation}
  D^2=\left(\Bar{q}-q\right)^2+(\Bar{e}-e)^2+\left(2\sin\left(\frac{\Bar{i}-i}{2}\right)\right)^2\text{,  D }\leq0.15
\end{equation}
Following this approach, \cite{Asher1993,Asher1994} used a similar criterion to identify potential TC asteroids, involving the asteroids' semi-major axis instead: 
\begin{equation}
D^2=\left(\frac{\Bar{a}-a}{3}\right)^2+(\Bar{e}-e)^2+\left(2\sin\left(\frac{\Bar{i}-i}{2}\right)\right)^2
\end{equation}
with $\Bar{a}=2.1$ AU, $\Bar{e}=0.82$ and $\Bar{i}=4\degree$. Asteroids with small D values and a longitude of perihelion $\varpi$ aligned with the TC were retained. The range of $\varpi$ corresponding to the TC was taken to be 140$\pm$40$\degree$ \citep{Steel1991} and 25 asteroids with $D\leq0.3$ were identified, 12 of them having a longitude of perihelion aligned with comet 2P/Encke \citep{Asher1993}. The authors also found a cluster of perihelion longitudes centred around the $\varpi$ of (2212) Hephaistos. 

Since those studies, several new associations between NEAs and the Taurid complex have been proposed. The number of NEA candidates for the TC published in the literature is now well in excess of 100 \citep{Tubiana2015}. We will briefly review these later studies below, but for the reader's reference, the TC NEA candidates that also have a longitude of perihelion aligned with 2P/Encke are listed in Table \ref{tab:TC_asteroids}. 

\cite{Babadzhanov2001} and \cite{Babadzhanov2008} adopted \cite{Asher1993}'s modified D-criterion (coupled with further restrictions on the asteroids' aphelion distance and period) to identify additional Taurid NEAs in the NeoDys catalog\footnote{\url{http://newton.dm.unipi.it/neodys/neodys.cat}}. 

 \begin{table*}
 \centering
  \resizebox{\textwidth}{!}{
  \begin{tabular}{llllll}
  \hline
  Designation & Number & Name & H magnitude & Geometric albedo & References \\
  \hline
  $^*$2P/Encke  &         &     &     & 0.046 $\pm$ 0.023 & e.g., \cite{Whipple1940,Whipple1952,Asher1993,Babadzhanov2001}    \\
                &         &     &     &    & \mbox{ }\hspace{0.5cm} \cite{Porubcan2006,Jopek2011,Olech2016}    \\
  \hline
  $^*$2004 TG10 &         &       & 19.4  & 0.018 $\pm$ 0.037 &  \cite{Porubcan2006,Babadzhanov2008,Jopek2011,Dumitru2017,Spurny2017} \\
  $^*$2005 TF50 &         &       & 20.3  &    &  \cite{Porubcan2006,Jopek2011,Olech2016,Spurny2017} \\
  $^*$2005 UR   &         &       & 21.6  &    &  \cite{Jopek2011,Olech2016,Spurny2017} \\
  $^*$2015 TX24 &         &       & 21.5  & 0.070 $\pm$ 0.018 & \cite{Spurny2017} \\
  $^*$1947 XC   & 2201    & Oljato & 15.3 & 0.433 $\pm$ 0.030 & \cite{Asher1993,Babadzhanov2001,Jopek2011,Popescu2014} \\
  $^*$1959 LM   & 4183    & Cuno   &  14.1 & 0.097 $\pm$ 0.025 & \cite{Asher1993,Babadzhanov2001,Popescu2014} \\	
  $^*$1984 KB   & 6063    & Jason  & 15.9 & 0.21 & \cite{Asher1993,Babadzhanov2001,Tubiana2015} \\
  $^*$1987 SB   & 4486    & Mithra & 15.5 & 0.297 $\pm$ 0.056 & \cite{Asher1993,Popescu2014} \\	
  $^*$1991 VL   & 5143    & Heracles & 13.9 & 0.227 $\pm$ 0.054 & \cite{Asher1993,Babadzhanov2001,Popescu2014} \\
  $^*$1996 RG3  & 269690  &     &  18.4   & 0.098 $\pm$ 0.107 & \cite{Babadzhanov2001,Popescu2014} \\	
  $^*$1998 QS52 & 16960   &     &  14.3   &    & \cite{Babadzhanov2008,Tubiana2015} \\
  $^*$1998 SS49 & 85713   &     &  15.7   & 0.076 $\pm$ 0.039 & \cite{Tubiana2015} \\ 
  $^*$1998 VD31 &         &     &  19.4   &    & \cite{Babadzhanov2008} \\
  $^*$1999 RK45 & 162195  &     & 19.4    &    & \cite{Porubcan2006,Tubiana2015} \\
  $^*$1999 VK12 &         &     &  23.4   &    & \cite{Babadzhanov2008} \\
  $^*$1999 VR6  &         &     &  20.8   &    & \cite{Babadzhanov2008} \\
  $^*$1999 VT25 & 152828  &     &  21.1   &    & \cite{Tubiana2015} \\
  $^*$2001 HB   & 496901  &     &  20.4   &    & \cite{Porubcan2006} \\
  $^*$2001 QJ96 &         &     &   22.1  &    & \cite{Porubcan2006,Brown2010} \\
  $^*$2001 VH75 & 153792  &     &  18.1   &    & \cite{Jenniskens2006,Tubiana2015} \\
  $^*$2002 MX   &         &     &  21.7   &    & \cite{Jopek2011} \\
  $^*$2002 SY50 & 154276  &     &  17.6   & 0.143 $\pm$ 0.173 & \cite{Tubiana2015} \\
  $^*$2003 QC10 & 405212  &     &  18.0   &    & \cite{Porubcan2006,Tubiana2015} \\
  $^*$2002 XM35 &         &     &  23.0   &    & \cite{Porubcan2006} \\
  $^*$2003 SF   &         &     & 19.7    &    & \cite{Porubcan2006} \\
  $^*$2003 UL3  & 380455  &     & 17.8    &    & \cite{Porubcan2006,Babadzhanov2008,Tubiana2015} \\
  $^*$2003 UV11 & 503941  &     & 19.5    & 0.376 $\pm$ 0.075 & \cite{Jopek2011,Dumitru2017} \\
  $^*$2003 WP21 &         &     & 21.5    &    & \cite{Porubcan2006,Babadzhanov2008,Jopek2011} \\
  $^*$2004 WS2  & 170903  &     & 18.0    &    & \cite{Tubiana2015} \\
  $^*$2005 NX39 &         &     & 19.7    &    & \cite{Jopek2011} \\
  $^*$2005 TB15 &         &     & 19.5    &    & \cite{Jopek2011} \\
  $^*$2005 UY6  & 452639  &     & 18.2    & 0.018 $\pm$ 0.025 & \cite{Jopek2011} \\
  $^*$2006 SO198 &        &     & 23.9    &    & \cite{Jopek2011} \\
  $^*$2007 RU17 &         &     & 18.1    &    & \cite{Brown2010,Jopek2011,Dumitru2017} \\
  $^*$2007 UL12 &         &     & 21.1    &    & \cite{Brown2010,Jopek2011} \\
  $^*$2010 TU149 &        &    & 20.7     & 0.025 $\pm$ 0.015 & \cite{Dumitru2017} \\
  \hline
  1937 UB   & 69230   & Hermes   & 17.5 &    & \cite{Asher1993} \\
  1982 TA   & 4197    & Morpheus & 14.9 & 0.37 & \cite{Asher1993,Babadzhanov2001}\\  
  1987 KF   & 4341    & Poseidon & 15.9 & 0.054 $\pm$ 0.091 & \cite{Asher1993,Babadzhanov2001}\\  
  1988 VP4  & 5731    & Zeus     & 15.5 & 0.031 $\pm$ 0.009 & \cite{Asher1993,Babadzhanov2001}\\
  1989 DA   & 168318  &          & 18.9 &    & \cite{Asher1994} \\
  1990 HA   & 217628  & Lugh     & 16.6 &    & \cite{Asher1993,Babadzhanov2001} \\
  1991 BA   &         &          & 28.6 &    & \cite{Asher1993,Asher1994,Babadzhanov2001}\\ 
  1991 GO   &         &          & 20.0 &    & \cite{Asher1993,Babadzhanov2001} \\
  1991 TB2  & 408752  &          & 17.0 &    & \cite{Babadzhanov2001} \\
  1993 KA2  &         &          & 29.0 &    & \cite{Asher1994,Babadzhanov2001} \\   
  1994 AH2  & 8201    &          & 15.7 & 0.154 $\pm$ 0.042 & \cite{Asher1994,Babadzhanov2001} \\
  1995 FF   &         &          & 26.5 &    & \cite{Babadzhanov2001} \\
  1996 SK   & 297274  &          & 16.8 & 0.234 $\pm$ 0.040 & \cite{Babadzhanov2001} \\
  5025 P-L  &  306367 & Nut      & 15.5 &    & \cite{Olsson1988,Asher1993,Babadzhanov2001} \\
  \hline
  $^*$2019 AN12 &   &  & 25.4 &    & TSC \\
  $^*$2019 BJ1  &   &  & 24.9 &    & TSC \\
  $^*$2019 GB   &   &  & 26.9 &    & TSC \\
  $^*$2019 JO5  &   &  & 23.7 &    & TSC \\
  $^*$2019 RV3  &   &  & 19.7 &    & TSC \\
  $^*$2019 TM6  &   &  & 20.5 &    & TSC \\
  $^*$2019 UN12 &   &  & 21.9 &    & TSC \\
  $^*$2019 WJ4  &   &  & 28.4 &    & TSC \\
  $^*$2019 WB5  &   &  & 24.2 &    & TSC \\
  $^*$2019 YM   &   &  & 26.0 &    & TSC \\
  $^*$2020 AL2  &   &  & 25.6 &    & TSC \\
  $^*$2020 BC8  &   &  & 20.0 &    & TSC \\
  $^*$2020 DV   &   &  & 26.7 &    & TSC \\
  $^*$2020 DX2  &   &  & 23.5 &    & TSC \\
  $^*$2020 JV   &   &  & 21.4 &    & TSC \\
  \hline
  \hline
  \end{tabular}}
  \caption{\label{tab:TC_asteroids} Non-exhaustive selection of near-Earth asteroids (NEAs) suspected to be associated with the Taurid complex or the Taurid Swarm Complex (TSC). The last column of the table provides references of works associating these objects with the TC or characterizing their spectral properties \citep[e.g.][]{Popescu2014,Tubiana2015}. Estimates of the absolute magnitude (H) and the geometric albedo were obtained from the Jet Propulsion Laboratory Small-Body Database Browser (\url{https://ssd.jpl.nasa.gov/sbdb.cgi}, accessed in June 2021). The bodies analyzed in this study are indicated with an asterisk ($^*$).}
 \end{table*}

\begin{table*}
	\begin{tabular}{rrrrrrrrrr}
		a (AU) & e & i ($\degree$) & $\Omega$ ($\degree$) & $\omega$ ($\degree$) & $\Delta$a (AU) & $\Delta$e & $\Delta$i ($\degree$) & $\Delta\Omega$ ($\degree$) & $\Delta\omega$  ($\degree$)\\
		\hline
		\hline
		& & & & & & & & & \\[-0.2cm]
		2.254 & 0.8450 & 5.496 & 39.95 & 119.96 & 0.0094 & 0.025 & 0.57 & 2.87 & 2.85 \\
		& & & & & & & & & \\[-0.2cm]
		\hline  
		\hline
	\end{tabular} 
	\caption{Osculating elements and 1-$\sigma$ ranges for 2019-2020 TC asteroid selection.}
	\label{tab:TC2019}
\end{table*}

Following the approach of \cite{Steel1991}, \cite{Porubcan2004} searched the IAU meteor datacenter for clusters of meteoroids that may be associated with Taurid stream filaments. They identified 15 filaments, each containing at least 4 meteoroids, associated with  the N and S branches of the complex. Using the Southworth-Hawkins D-criterion, the authors selected 91 NEAs in the Asteroid Orbital Elements Database\footnote{Ted Bowell, Lowell Observatory - \url{http://alumnus.caltech.edu/~nolan/astorb.html}} that had orbital similarity with one of the Taurid filaments (D$\leq$0.3). 
To claim any relationship between these candidates and the meteoroid streams, \cite{Porubcan2004} required that the orbit of the selected asteroids and the Taurid streams remain similar during the past. After integrating the motion of these bodies over 5000 years, the number of plausible Taurid parents dropped to 10. The most convincing associations were  2003 QC10, 2004 TG10, 2003 UL3 and 2002 XM35. Other possible candidates for TC membership were found for 1999 RK45, 2001 HB, 2001 QJ96, 2003 SF, 2003 WP21 and 2005 TF50. 

Following a comparable procedure, \cite{Jopek2011} identified two groups of Taurid meteors in the IAU meteor datacenter, consistent with the Southern and Northern Taurids. For these meteors, the authors identified 12 possible parent bodies based on orbital similarity at the current epoch: comet 2P/Encke and 9 NEAs (2003 WP21, 2004 TG10, 2005 TB15, 2005 UR, 2006 SO198, 2007 RU17, 2007 UL12, 2002 MX, 2003 UV1, 2201 Oljato, 2005 NX39, 2005 TF50 and 2005 UY6). Among these bodies, 7 asteroids had previously been suggested to belong to the TC in other studies, while several associations with asteroids identified by \cite{Asher1993} \cite{Asher1994}, \cite{Babadzhanov2001}, \cite{Porubcan2006} and \cite{Babadzhanov2008} were not found. 

Other TC asteroids have been proposed in the literature based on their orbital similarity with Taurid meteors. For example, using the D-criterion of \cite{Drummond1981} with a threshold of 0.1, \cite{Brown2010} suggested that asteroids 2007 UL12, 2007 RU17 and 2001 QJ96 could be linked to meteoroid streams of the TC recorded by the Canadian Meteor Orbit Radar \citep{Brown2008,Brown2010}. Similarly, \cite{Dumitru2017} identified plausible associations of several Taurid meteoroid streams with NEAs 2004 TG10 and 2010 TU149, and possibly with 2003 UV11 and 2007 RU17. In other work, the NEAs 2005 UR and 2005 TF50 were found to have orbits close to those of two bright fireballs observed over Poland in 2015 \citep{Olech2016} ; the quality of the computed fireball orbits, however, has been questioned by \cite{Spurny2017}. 

A number of asteroids of several hundred meters in diameter were also linked to a new branch of resonant Taurid meteoroids recorded by \cite{Spurny2017} in 2015.
The authors found a noticeable orbital element correlation between  four asteroids, namely 2004 TG10, 2005 TF50, 2005 UR and 2015 TX24, and the 2015 Taurid resonant branch. In particular, NEA 2015 TX24, 2005 UR, and 2005 TF50 were suggested as being related to the resonant Taurid branch \citep{Spurny2017}.

\subsection{Spectral classification of TC candidates}

If NEAs within the TC have a common origin, it would be expected that they have compositional affinities. To examine this question, the reflectance spectra of several km-sized asteroids proposed as being associated with the TC were analyzed by \cite{Popescu2014}. The authors compared spectra of (2201) Oljato, (4183) Cuno, (4486) Mithra, (5143) Heracles, (6063) Jason, (16960) 1998 QS52 and (269690) 1996 RG3. Among these six bodies, only 1996 RG3 had spectral properties similar to cometary material and thus consistent with an association with comet 2P/Encke. In contrast, the other five asteroids were found to be more similar to the S taxonomic complex; their spectra were akin to Ordinary Chondrite meteorites of high petrologic grade, suggesting a thermally evolved surface inconsistent with comets \citep{Popescu2014}.

Building on this earlier work, \cite{Tubiana2015} examined the spectral properties of potential TC parents of smaller sizes. \cite{Tubiana2015} collected the spectra of 2P/Encke, 1998 QS52, 1999 RK45, 2003 QC10, 2003 UL3, 1999 VT25, 2001 VH75, (6063) Jason, 1998 SS49, 2004 WS2 and 2002 SY50 and laboratory measurement of selected CM chondrites in the wavelength range 400-900 nm. All the spectra obtained were featureless for $\lambda<$800 nm, preventing a clear link between 2P/Encke, the 10 NEAs and CM chondrites. As discussed in \cite{Popescu2014}, most of the NEAs were found to have spectra  similar to the S taxonomic group. While the spectral slope of 2P/Encke was measured to be similar to 1998 QS52, 1999 RK45, 2003 UL3, 2001 VH75, 1984 KB, 1998 SS49, 2004 WS2, and
2002 SY50 for $\lambda<$800 nm, the fit was poor above this range where the spectra also have bad SNR. \cite{Tubiana2015} conclude that there is not strong evidence of a compositional association between 2P/Encke and the NEAs observed, while also acknowledging that a firm conclusion about such a link is difficult to draw from their spectral data alone.  

Thus the picture which emerges from available reflectance spectra is one in which association with 2P/Encke is rare to non-existent for NEAs purported to be in the TC based on present orbital affinities alone. This underscores the serious problem of interlopers in the 2P/Encke orbital phase space as noted by \cite{Valsecchi1999}. It also emphasizes the need to examine past orbital evolution in search of potential associations as an added filter in selecting TC candidates. 

\subsection{Our sample selection}

To explore the dynamical evolution of the TC, we restrict our analysis to those NEAs most likely to belong to the complex. Most of the potential Taurid asteroids reported in previous studies were selected on the basis of empirical D-criterion comparisons between the current  osculating orbital elements of NEAs and the orbits of Taurid meteors (cf. section \ref{sec:litterature}). Since catalogs of orbital elements undergo continuous improvements with time, we have restricted our analysis here to the latest studies, that were conducted using updated catalogs of asteroids and comets. Therefore, in this work, we will explore as possible TC members the parent asteroids suggested by \cite{Porubcan2006}, \cite{Babadzhanov2008}, \cite{Brown2010}, \cite{Jopek2011}, \cite{Olech2016}, \cite{Dumitru2017} and \cite{Spurny2017}.

In addition, the large asteroids for which reflectance spectra were obtained by \cite{Popescu2014} and \cite{Tubiana2015} were also added to our sample, despite their lack of spectral similarity with 2P/Encke. These objects were included to investigate the possibility of random dynamical associations between bodies of different origins. 

In June 2019, the Earth passed unusually close (within five degrees of mean anomaly) of the centre of the Taurid resonant swarm \citep{clawiebro19}, increasing the probability of NEO surveys detecting TC members. Though these newly-discovered asteroids have relatively poorly-known orbital elements, we wanted to include any potential new TC members in this analysis. To make this selection, we used the simulations of \cite{clawiebro19} to determine an average resonant TC orbit, and selected all 2019-2020 apparition bodies within ten standard deviations and include them (see Table~\ref{tab:TC2019}). This resulted in the addition of fifteen asteroids to our sample: 2019 AN12, 2019 BJ1, 2019 GB, 2019 JO5, 2019 RV3, 2019 TM6, 2019 UN1, 2019 WB5, 2019 WJ4, 2019 YM, 2020 AL2, 2020 BC8, 2020 DV, 2020 DX2 and 2020 JV.

\begin{figure}
  \centering
  \includegraphics[width=0.48\textwidth]{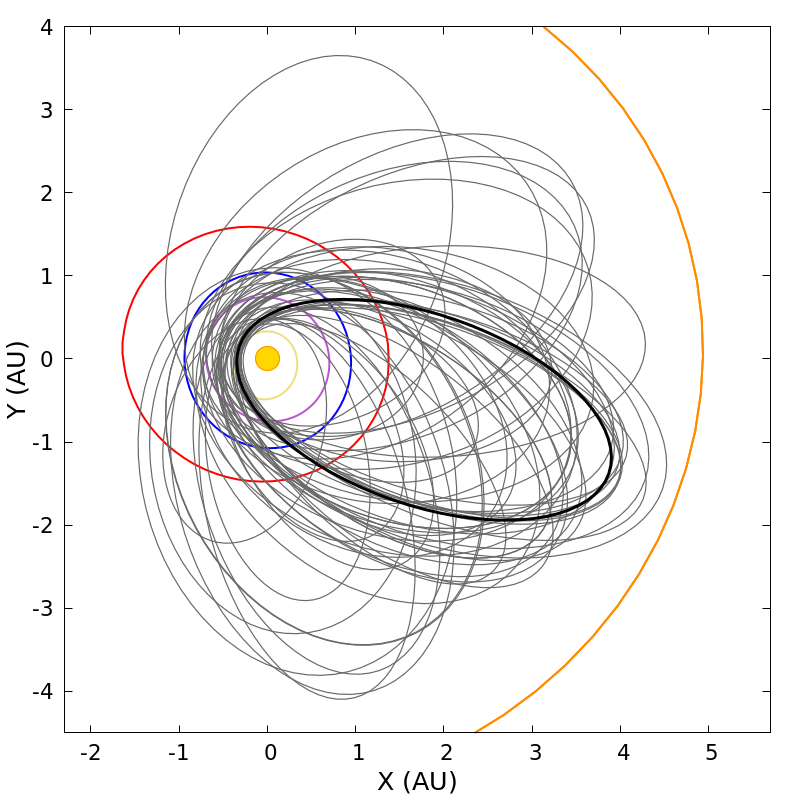}
  \caption{Projection onto the ecliptic plane of the current orbits of bodies in this study. The NEAs orbits are represented in grey, while 2P/Encke is drawn in black. The orbits of the planets from Mercury to Jupiter are plotted in colour.}
  \label{fig:orbits}
\end{figure}

We also investigate the dynamical evolution of the most commonly cited parent body of the Taurid meteoroids, namely comet 2P/Encke.  In total, our selection of potential Taurid members includes 51 NEAs and one known comet, 2P/Encke. The selected objects for our study are summarized in Table \ref{tab:TC_asteroids}. The projection of their current orbit into the ecliptic plane is presented in Figure \ref{fig:orbits}.  As our selection of the most promising TC candidates is motivated by previous publications, some recently discovered NEAs with orbital elements compatible with the TC may not be included in Table \ref{tab:TC_asteroids}. However, we believe that our sample of 52 bodies, while non-exhaustive, is large enough to explore the dynamics of the TC members.

\begin{figure*}
    \centering
    \includegraphics[trim={0.1cm 0.0cm 0.1cm 0.5cm}, clip, width=.99\textwidth]{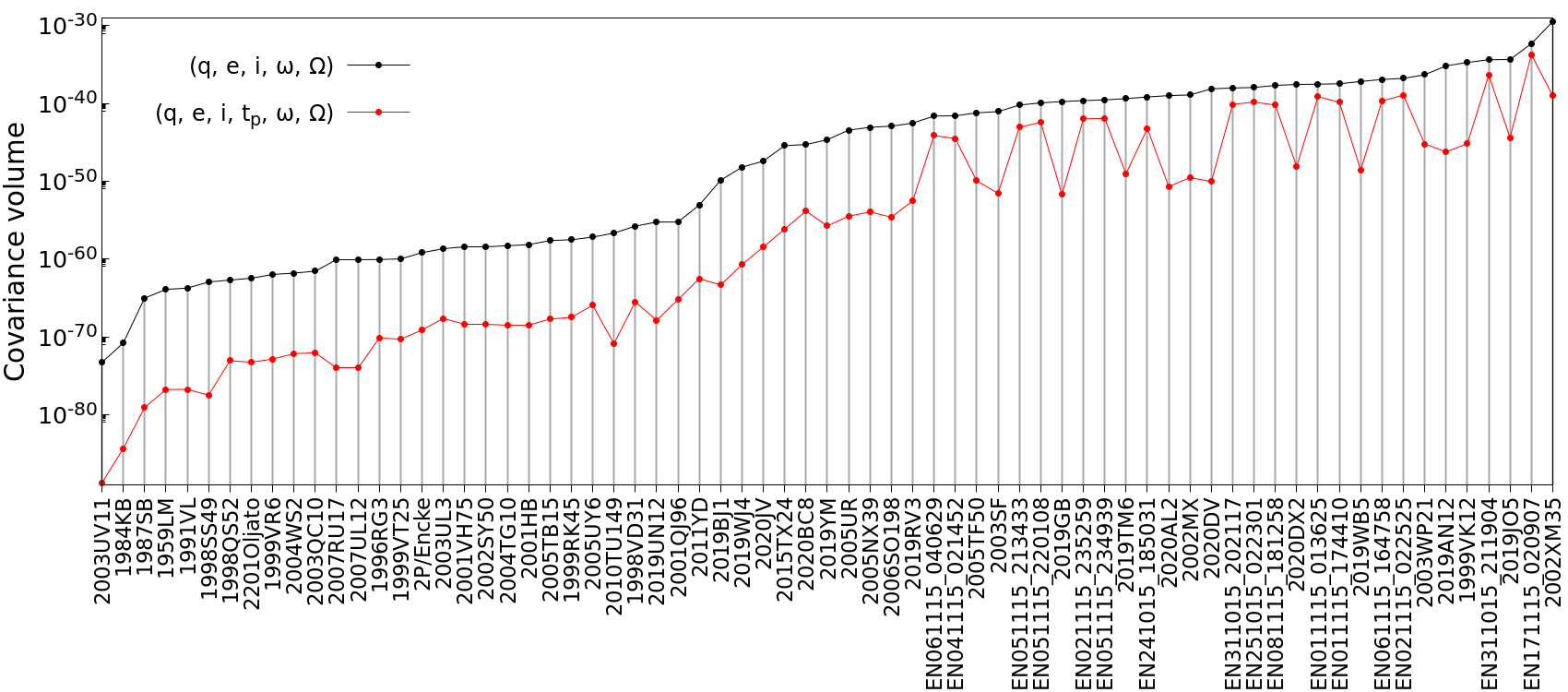}
    \caption{Volume of the hyper-ellipsoid of dimension 5 and 6 defined by the variance-covariance matrix of the bodies integrated in \{e, q (AU), $\Omega$ ($\degree$), $\omega$ ($\degree$), i ($\degree$)\} (in black) and \{e, q (AU), ${t_p}$ (d), $\Omega$ ($\degree$), $\omega$ ($\degree$), i ($\degree$)\} (in red).}
    \label{fig:covariances_volume}
\end{figure*}

\section{Reconnaissance analysis} 

To determine the general dynamical characteristics of our sample objects, we first performed some coarse simulations for all candidates. As discussed later in this section, we subsequently narrow our focus to the subset of NEAs showing the greatest orbital affinities in the recent past. In particular, we are interested in assessing whether or not past orbital associations could be consistent with a past fragmentation event as suggested by the coherent catastrophism hypothesis \citep{Asher1994}.  

The identification of such an event requires backward integration of our objects to demonstrate that one or more of them converged to the same location in the past, with characteristics consistent with a splitting or break-up. However, the precision of our orbital information and the chaotic nature of the dynamics restrict our ability to fully reach this goal. Instead, we will look for a past epoch at which the Minimum Orbit Intersection Distance (MOID) and the relative velocity $V_r$ at the MOID between any two bodies in our sample are both small. These times at which the MOID and $V_r$ reach a minimum are those we will examine more closely for evidence of a possible break-up.

\subsection{Integration}\label{sec:integration}

Each body is integrated over a period of 20 000 years in the past and the future, using a 15$^\text{th}$ order RADAU integrator with a precision parameter LL of 12 \citep{Everhart1985}. The external integration step is fixed to one day, and the position of the body is recorded every year. The force model includes the gravitational attraction of the Sun, the Moon, and the eight planets of the solar system, as well as general relativistic corrections. 

The Yarkovsky effect is expected to have only a small influence on the asteroids' motion relative to our scales of interest over 20 000 years \citep{Bottke2006a} and is therefore neglected in this work. In contrast, non gravitational forces (NGF) due to 2P/Encke's outgassing may have strongly influenced its past evolution \citep[e.g.,][]{Steel1996b,Levison2006}, and are included. Since we have no knowledge of 2P/Encke's NGF outside the limited time span covered by modern observations, we chose to integrate the motion of the comet twice; once without NGF (which we called solution 2P0), and once with constant NGF parameters (solution 2P1). In the second case, the NGF were modelled following \cite{Marsden1973}, with the parameters $A_1$, $A_2$ and $A_3$ characterizing the current activity of the comet listed Table \ref{tab:IC}. 

We note that, if some of the asteroids in our sample are actually extinct comet nuclei, we can not rule out the existence of additional non gravitational forces acting upon the bodies in the past. In such a case, their dynamical history might not be accurately represented by our integration. However, since we have no knowledge about these forces existence, duration and intensity, we have not considered them in our analysis. 

We used as starting conditions the orbital solutions provided by the Jet Propulsion Laboratory (JPL)\footnote{\url{https://ssd.jpl.nasa.gov/sbdb.cgi}}. The orbital elements of the bodies integrated are reported in Table \ref{tab:IC}, Appendix \ref{appendix:IC}.  In order to account for the uncertainty of the initial orbits provided, we created one hundred clones of each body following the multivariate normal distribution defined by the solution's covariance matrix. The motion of each clone was then integrated over at least 20 000 years, each clone representing a possible track for the body's evolution. 

\subsection{Measurement error}\label{sec:errors}

The initial dispersion of each NEAs' clones is highly variable and dependent on the measurement errors associated with the JPL's orbital solution. The solutions available for asteroids observed along a short arc of their orbits are generally less accurate, resulting in larger variances and covariances.

In order to easily compare the quality of the orbital solutions (and therefore the effective phase space volume over which clones are distributed), we compute the volume of the $n$-dimensional hyper-ellipsoid defined by the variance-covariance matrix as: 

\begin{equation}
    V_n=\frac{2\pi^\frac{n}{2}}{n\Gamma(\frac{n}{2})}\prod_{k=1}^{n}\lambda_k
\end{equation}

with $\lambda_k$ the lengths of the ellipsoid's axes, which correspond in our case to the eigenvalues of the covariance matrix, and $\Gamma$ the gamma function defined as $\Gamma(x)=\int_{0}^{\infty}t^{x-1}e^{-t}dt$. An estimate of the volumes computed from the covariance matrix in \{e, q, $\Omega$, $\omega$, i, n=5\} and \{e, q, ${t_p}$, $\Omega$, $\omega$, i, n=6\} (AU, days, $\degree$) is presented in Figure \ref{fig:covariances_volume}. 

While the absolute ordinate value is not very meaningful, Figure \ref{fig:covariances_volume} does allow us to distinguish in a relative sense between accurate (that is, with small hyperellipsoid volume like 2003 UV11) and imprecise (large hyperellipsoid volume like 2002 XM35) NEA orbits. The presence of Taurid fireballs in the figure, identified by the prefix EN, is discussed in Section \ref{sec:fireballs}.

\subsection{Clone analysis}\label{sec:method}

To identify a common origin among the potential TC asteroids, we compared the orbital evolution of all the possible pairs of our 52 near-Earth objects, that is a total of 1326 pairings. For each pair of objects, we computed the MOID and relative velocity $V_r$ of the bodies every year  over time. The MOID, measuring the minimum distance separating the closest points of two orbits, was favoured over the actual physical distance because the uncertainty in our knowledge of the shape of the orbits is much less than the uncertainty in the exact location of the asteroids in their orbit. 

If two or several asteroids originated from the fragmentation of a common parent body, they are expected to present small MOID values and reduced relative velocities at the epoch of separation. In the case of a fragmentation like those seen to occur occasionally in comets, the initial dispersion velocity is expected to be small, reaching at most a few meters per second \citep{Boehnhardt2004}. In contrast, the relative velocity of fragments resulting from a collision between NEAs can reach much higher values, of order $\sim$km/s \citep{Hyodo2020}. 

In order to not miss any possible epoch of common origin, we initially adopt a large MOID and $V_r$ threshold. This allows us to pre-select the subsample of our candidate asteroids which merit more careful study. We use as our first filter a MOID of 0.05 AU and a relative velocity below 1 km/s. Higher relative velocities are unlikely for large fragments even in a crater-producing impact event, as the fraction of ejected mass falls dramatically above this threshold \citep{Hyodo2020}.
To compare the trajectories of two asteroids (hereafter called body \#1 and body \#2), each one represented by a hundred orbits, we proceed as follows. 

At each epoch, we compute the number $N_1$ of clones of body \#1 which approach at least one clone of body \#2 within the MOID and $V_r$ thresholds detailed above. The evolution of $N_1$ with time thus illustrates the epochs when the clones of body \#1 are close to those of body \#2. However, at a given time step, several clones of body \#1 may approach the same clone of body \#2. This situation occurs more often when the clones are highly dispersed, which increases the probability of having a small fraction of the clones approaching another body purely by chance. 

To identify such circumstances, we also compute the number $N_2$ of clones of body \#2 approaching at least one clone of body \#1.  Comparably high values of $N_1$ and $N_2$ therefore improve our confidence in a real link between the two bodies, rather than a random association driven by the uncertainty in the NEA's orbit.

\subsection{Results}

Since the purpose of this work is to identify epochs of possible close approaches between TC candidates, we do not provide in this section a detailed description of the individual evolution of each asteroid integrated. Instead, more generic considerations about the NEAs orbital distribution and clone dispersion are highlighted. For complementary information about the bodies past and future evolution, the reader is referred to the supplementary Appendix B, available online.  

\subsubsection{Orbital considerations}

\begin{figure}
    \centering
    \includegraphics[width=.49\textwidth]{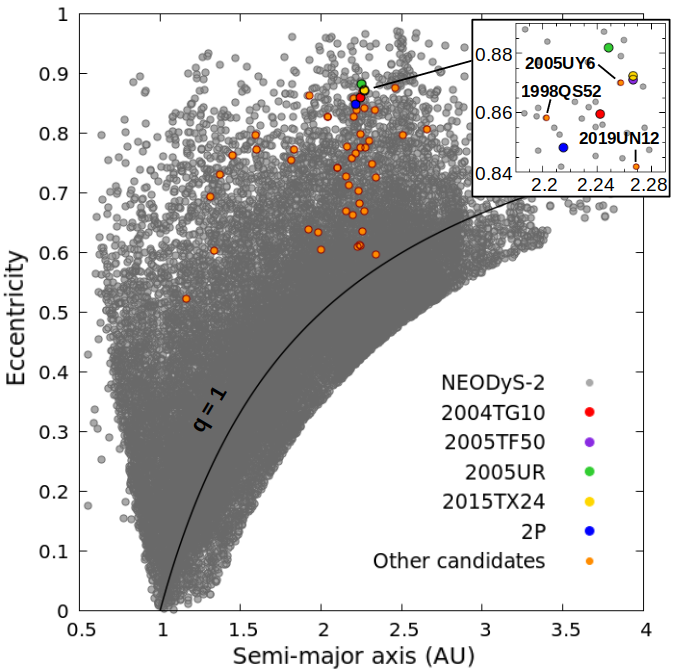}
    \caption{Semi-major axis and eccentricity of TC candidates (in colour) compared with all NEAs listed in the NEODyS-2 catalog (in grey). The inset of the figure shows a small cluster of TC candidates with ($a,e$) orbital elements similar to comet 2P/Encke.}
    \label{fig:ae}
\end{figure}

Figure \ref{fig:ae} depicts the current distribution of our sample in the semi-major axis and eccentricity plane. The objects' orbital elements are compared with those of the NEAs listed in the NEODyS-2 catalog \footnote{\url{https://newton.spacedys.com/neodys/}}. TC candidates are presented in colour in the figure, while the other NEOs are drawn in grey. We note that the NEAs associated with the complex tend to lie towards the highest eccentricities, where the NEA population density drops. In addition, we observe some clustering in ($a,e$) of several asteroids around the current location of comet 2P/Encke (cf. zoomed portion in the upper right part of the Figure). 

The existence of a cluster of our TC candidates around 2P/Encke is not surprising, since TC candidates were commonly identified in previous studies on the basis of their similarity with the comet's orbit (see Section \ref{sec:litterature}). However, we note that this subgroup also contains several asteroids proposed to be directly linked to meteor activity observed on Earth. Among them, we identify the NEAs 2004 TG10, 2005 TF50, 2005 UR and 2015 TX24 (in red, purple, green and yellow respectively) that might be associated with the branch of resonant Taurid meteoroids observed by \cite{Spurny2017}. Asteroid 2019 UN12 was discovered during the 2019-2020 observation campaign targeting unknown large objects belonging to the Taurid resonant swarm. We note that 2005 UY6 is also currently trapped into the 7:2 MMR with Jupiter (cf. supplementary Appendix C). 

The last NEA in this $(a,e)$ cluster is 1998 QS52, the only one of this group for which spectral observations are available. The asteroid was found to belong to the Sq or Sr taxonomic class \citep{Binzel2004,Popescu2014,Tubiana2015}, making a compositional association with 2P/Encke unlikely.

\subsubsection{7:2 mean motion resonance}

Since the 7:2 mean motion resonance with Jupiter is known to play a major role in the evolution of Taurid meteoroid streams \citep{Asher1991,Asher1993b,Asher1998,McBeath1999,Dubietis2007,Spurny2017}, we give special attention to it here. In supplementary Appendix C, we identify the objects for which a significant fraction of their clones (at least 20\%) are found to evolve within the 7:2 MMR during our backward integrations. 

We find that NEAs 2005 UY6, 2015 TX14, 2005 TF50, 2003 UL3, 2019 UN12 and 2005 UR to have likely remained inside the resonance for the past several thousand years. At least 30\% of the clones created for each of these NEAs resided in the resonance over the course of the past 20 000 years. In particular, between 70\% and 100\% of the clones of 2005 UY6 and 2015 TX24 stayed in the resonance during the whole integration period.

We observe that 2019 BJ1 is clearly trapped in the 7:2 MMR at the present epoch, with 100\% of its clones identified as resonant. However, after 5000 years of backward integration, more than 80\% of its clones left the resonance. With about 20\% of their clones in the resonance in the past 2000 to 10 000 years, we suspect NEAs 2020JV, 2019 AN12 and 2019 RV3 to be strong candidates as having also been trapped in the 7:2 MMR. Such objects are particularly germane for our analysis, as the resonance likely increases the residence time of these asteroids in a specific orbital region of orbital elements. Any past fragmentation event occurring in or near the 7:2 MMR is therefore more likely to be identifiable longer into the future.

\subsubsection{Clone dispersion}

\begin{figure}
    \centering
    \includegraphics[width=.49\textwidth]{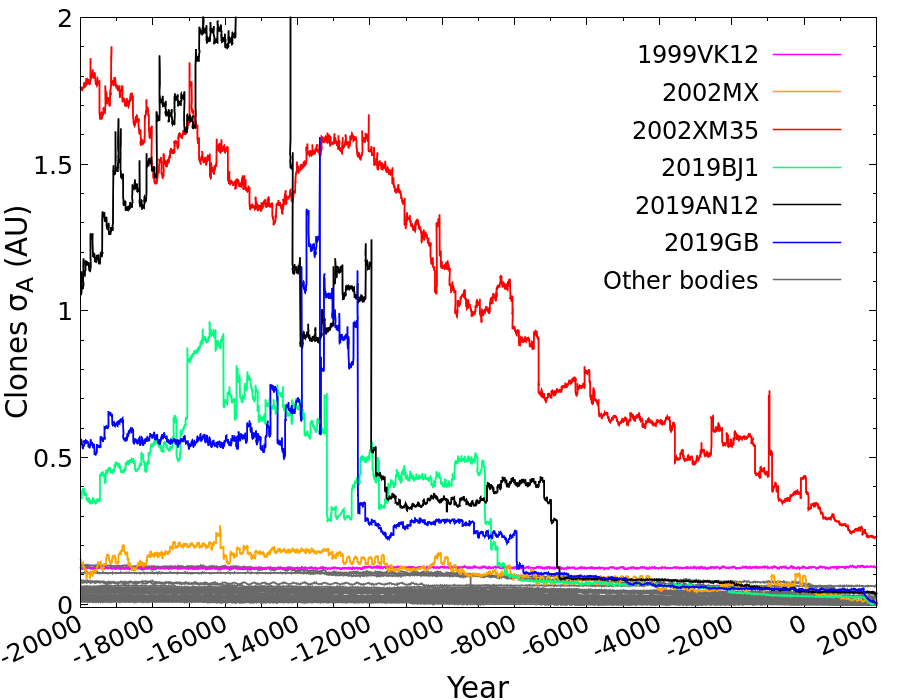}
    \caption{Standard deviation of the semi-major axis of the clones associated with various bodies as shown in the legend.  One hundred clones were created for each body and integrated 22 000 years in the past. }
    \label{fig:sigma_sma}
\end{figure}

Despite the range of measurement uncertainties among our NEA orbits (cf. Section \ref{sec:errors}), we observe that the orbital uncertainty of most objects investigated remains small over the integration period of 20 000 years. This is verified by analyzing the dispersion for each orbital element, of the hundred clones integrated for each body. For example, Figure \ref{fig:sigma_sma} shows the evolution of the clones' dispersion in semi-major axis $a$, characterized by the standard deviation $\sigma_a$. Most of the NEAs selected maintain a $\sigma_a$ level below 0.1 AU for the 20 000 years of integration (in grey in the Figure), implying that their semi-major axis is reasonably stable over this period.

However, due to large measurement uncertainties there is considerable orbital spread for three asteroids in our sample: 1999 VK12, 2002 MX and 2002 XM35, while close encounters with planets led to sudden and significant dispersion of the clones beyond about 6500 BCE for 2019 AN12, 2019 BJ1 and 2019 GB. The same conclusions are drawn when measuring these clones' dispersion in eccentricity, perihelion distance and inclination. Therefore, the search for a common origin between these six asteroids and other NEAs in our sample must be performed with particular care. 

\begin{figure*}
    \centering
    \includegraphics[width=\textwidth]{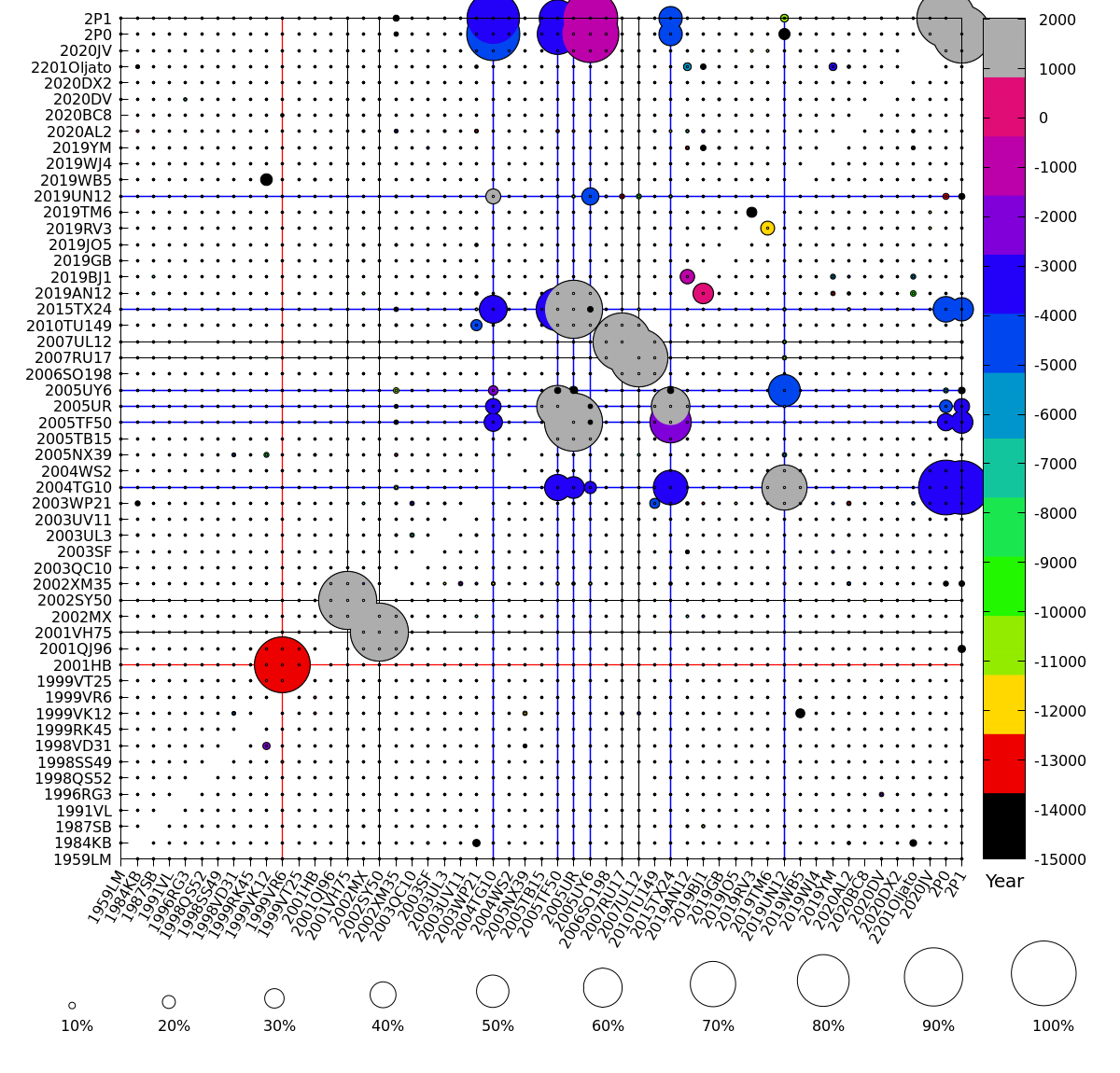}
    \caption{Possible rapprochements (close-approaches) between osculating orbits of TC candidates over the period 20 000 BC to 2000 CE. Labels on the abscissa and ordinate indicate the bodies integrated. Circles at position (body \#1, body \#2) indicate the percentage of clones of body \#1 approaching at least one clone of body \#2 with a MOID smaller than 0.05 AU and a relative velocity below 1 km/s. The size of the circle is proportional to the maximum percentage of clones retained for body \#1 (in abscissa) compared to body \#2 (in ordinate) over the integration period. The colour of the circle indicates the epoch when this highest percentage of clones retained is attained. The colour range has been restricted to the period 15 000 BCE to 2000 CE for clarity. Coordinates with no rapprochements are still given a small black dot to help guide the eye in reading the figure.}
    \label{fig:map_past}
\end{figure*}

\subsubsection{Rapprochement (possible close approach) epochs} \label{sec:rapprochement_all}

Figure \ref{fig:map_past} presents a high-level summary of the possible dynamical associations between all bodies in our study from the current epoch back to 20 000 BCE. The abscissas and ordinates indicate the name of the objects, and circles of variable sizes and colours represent the number of clones of each body (in abscissa) approaching at least one clone of the object in ordinate. The size of the circle is proportional to the maximum number of clones meeting our MOID-$V_r$ criteria over the integration period, at an epoch roughly indicated by the colour bar in the Figure. The colour bar was restricted to the period 15 000 BCE to 2000 CE for clarity, even though the objects' evolution have been analyzed until 20 000 BCE. 

We will refer to an epoch when several clones of each body meet the MOID-$V_r$ threshold as a {\it rapprochement} or orbital convergence. Indeed, if our analysis does not directly compare all the orbital elements of the bodies examined (e.g., with a variation of the $D$ criterion mentioned in Section \ref{sec:litterature}), low MOIDs {\it and} small relative velocities imply a similarity between the orbits compared, indicating a possible orbital convergence between these objects. 

This is however different from a {\it close approach} (that is, true physical proximity, which  could --but does not have to-- occur at these times) and we want to clearly distinguish these two possibilities. We will use these rapprochements as proxies for epochs when fragmentation could occur, and of course the bodies must have been physically close at that time if a break-up did indeed occur. However, because orbital uncertainties most strongly affect our knowledge of the particles' location along their orbit, we do not require a close approach between clones to declare a rapprochement, only that the MOID-$V_r$ thresholds be met.

As an example, we examine the rapprochement between 2P0 (i.e., 2P integrated without NGF) and 2005 TF50 as displayed in Figure \ref{fig:map_past}. The maximum number of clones of 2005 TF50 approaching 2P0 is indicated by the circle at the coordinates $x=$2005 TF50 (near the middle of the $x$-axis), and $y=$ 2P0 (near the top of the $y$-axis). The large blue circle at these coordinates indicates that around 3000 - 4000 BCE, close to 60\% of the 2005 TF50 clones approached at least one clone of 2P0 within our threshold (MOID below 0.05 AU and a relative velocity lower than 1 km/s). In contrast, by examining the location mirrored across a diagonal line with slope = 1, we can see at the coordinates (2P0, 2005 TF50) a much smaller blue circle. This indicates that only about 20\% of the clones of 2P0 approached a clone of 2005 TF50 around the same epoch. 

Grey in Figure \ref{fig:map_past} represents proximity at the current time. Thus, from the large grey circles we immediately see that several pairs of NEAs currently evolve on low MOID-$V_r$ orbits, namely \{2001 VH75, 2002 SY50\}, \{2004 TG10, 2019 UN12\}, \{2005 UR, 2005 TF50\}, \{2005 UR, 2015 TX24\} and \{2007 UL12, 2007 RU17\}. The grey circles in the upper right corner also verify the obvious correlation between 2P0 and 2P1. We also note that the 2P0 and 2P1 produced very similar results in Figure \ref{fig:map_past}, indicating that the inclusion of NGF during the integration period  does not significantly change the results. For the rest of this section then, we will not discriminate between the two solutions, but refer to the comet as 2P. 

Figure \ref{fig:map_past}
allows us to make some initial and rather sweeping conclusions. 
\begin{itemize}
    \item 
 The majority of asteroids show no epochs when our MOID and $V_r$ thresholds are met; therefore we can exclude them as having originated from a low-speed fragmentation of one of our other sample members over this time frame. This is particularly important because they are included in our sample because of their orbital similarity to the TC. They effectively serve as a control sample, demonstrating that not all orbits in/near the TC show the same rapprochements.
 \item We do not find any rapprochements between those NEAs known to have non-cometary spectral characteristics. 
 \item There is a preponderance of dark blue circles corresponding to a set of  rapprochements around 3000 - 4000 BCE. We note that 2P and 2004 TG10 show a rapprochement to each other, and with 2005 TF50, 2005 UR and 2015 TX24 at this time. Because of this clustering of significant mutual rapprochements, we identify these NEAs as the most likely to be related. For the rest of this study, we will refer to this sub-group of five objects as the "G5"; these form the focus of the rest of our study of possible Taurid Complex NEAs.  
\end{itemize}

Other than the grey circles in Figure \ref{fig:map_past} (representing orbits intersecting at the current epoch, and which we expect due to observational selection), and the blue circles of 3000 - 4000 BCE, we also identify a rapprochement epoch of the asteroid 2005 UY6 with 2004 TG10, 2019 UN12 and maybe 2P over a more uncertain time frame (1000 - 4000 BCE). Other associations, involving a smaller fraction of clones, are noticeable in Figure \ref{fig:map_past} (e.g., between 2019 AN12 and 2019 BJ1, or 2019 RV3 and 2019 TM6). The pair \{1999 VR6, 2001 HB\} illustrates the situation where  a high fraction of clones of 1999 VR6 encounter a single clone of 2001 HB around 13 000 BCE.

The set of rapprochements from 3000 - 4000 BCE is the most dramatic feature of Figure \ref{fig:map_past}. So, from this preliminary investigation, we identify a possible dynamical linkage between 2P and asteroids 2004 TG10, 2005 TF50, 2005 UR and 2015 TX24 5000 to 6000 years ago. All these bodies might be related to several of the Taurid complex meteor showers; they were in particular the only NEAs found to be close to the resonant branch of meteoroids observed in 2015 by \cite{Spurny2017}. Among the 52 bodies integrated, these 4 NEAs and comet 2P/Encke are therefore the most promising candidates to examine when searching for a potential fragmentation event within the TC over the past 20 ka. In the following section, we will focus on this group of 5 bodies (hereafter G5) to investigate the hypothesis of a common origin between them. The inclusion of additional NEAs like 2005 UY6 and 2019 UN12, approaching some G5 members around the same epoch, will also be detailed in Section \ref{sec:G5}.

 \section{In-depth analysis: G5 evolution} \label{sec:G5}
 
 Having determined that the G5 group, all of which were previously also mentioned in the analysis of the 2015 Taurid fireball outburst by \cite{Spurny2017}, represent our best candidates on the basis of dynamics for some common origin through fragmentation, our analysis will be largely confined to these asteroids from this point forward. 2P/Encke is the largest object of the G5 group, with a diameter of 4.8 km \citep{Lamy2004}. Next in size, 2004 TG10 has an estimated diameter of 1.32$\pm$0.61 km \citep{Nugent2015}. In contrast, 2005 TF50, 2005 UR and 2015 TX24 are smaller, being about 300 m, 170 m\footnote{\url{https://neo.ssa.esa.int/}} and 250 m \citep{Masiero2017} respectively, as determined by their H magnitude and some albedo assumption. The current orbits of comet 2P/Encke and 2004 TG10 are well determined, while the orbits of 2015 TX24, 2005 UR and 2005 TF50 are more uncertain (cf. Figure \ref{fig:covariances_volume}).  
 
A competing hypothesis for the grouping of the G5 members is that they are coincidental approaches caused by the small number of clones generated for each body or the length of the epoch investigated; additional simulations were conducted to explore this possibility and are detailed below. First, we generated and integrated an additional set of 1000 clones per body (5000 total), to determine if small number statistics was a factor. This larger sample of clones did not materially affect our results, and so for simplicity, we only present in upcoming sections the results obtained with the original 5$\times$100=500 clones.

As a second check, we extended the analysis of the G5 to a period of $\pm$30 000 years. In particular, extending the integrations into the future allow an assessment of the likelihood of rapprochements occurring on a purely dynamical basis, since a fragmentation origin is impossible in this case. These results are detailed in Section~\ref{sec:G5_orbit}. 

Thirdly, to explore the limits of our analysis, an additional independent set of 500 clones was integrated for 100 000 years in the future and the past. The output of this simulation is discussed in Section \ref{sec:G5_precession}.
 
 \subsection{Orbital evolution} \label{sec:G5_orbit}

\begin{figure}
	\includegraphics[width=.49\textwidth]{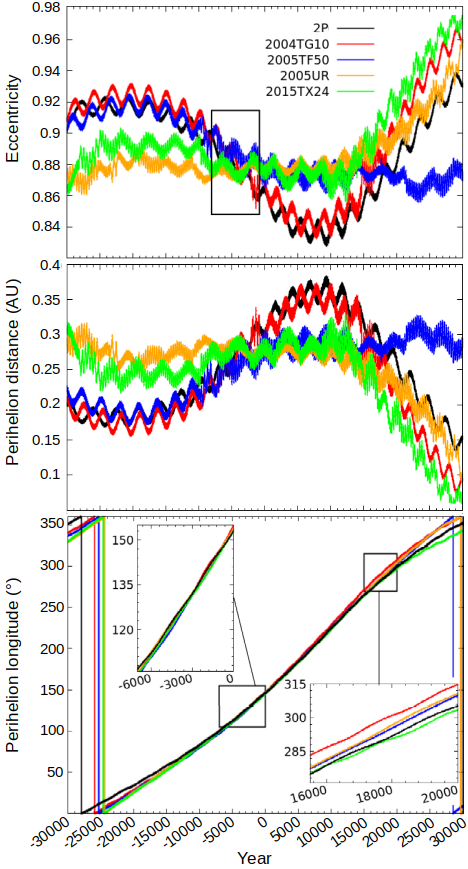}\\
	\caption{Eccentricity (top), perihelion distance (middle) and longitude of perihelion ($\varpi$) (bottom) evolution of the nominal clones created for 2P/Encke (in black), 2004 TG10 (in red), 2005 TF50 (in blue), 2005 UR (in orange) and 2015 TX24 (in green).}
	\label{fig:G5_qe}
\end{figure}

 Figure \ref{fig:G5_qe} presents the evolution of the eccentricity, perihelion distance and longitude of perihelion of the G5 nominal orbits between 30 000 BCE and 30 000 CE. The variations of the NEAs' inclination, argument of perihelion $\omega$ and longitude of the ascending node $\Omega$ are presented in supplementary Appendix B, Figure B2.

 From these figures, we observe that the evolution of each object is dominated by short and long-term oscillations in ($e$, $q$, $i$) caused by the Kozai and secular resonances influencing the bodies' motion. A description of these mechanisms constraining the secular evolution of 2P/Encke is detailed for example in \cite{Valsecchi1995}. Over the integration period, the aphelion of each body remains smaller than 4.95 AU, protecting the G5 from any close encounters with Jupiter. However, this situation could change after 30 000 CE due to an increase of the bodies' eccentricity, especially in the case of asteroids 2004 TG10 and 2015 TX24. 

 Figure \ref{fig:G5_qe} shows the evolution of the G5 nominal clones. This might be different from the real history of the G5 objects, since each clone represents only one possible evolution of one of these bodies over the integration period. However,  Fig~\ref{fig:G5_qe} allows us to quickly understand why the preliminary analysis of Section \ref{sec:rapprochement_all} indicates a possible close approach between the G5 5000 to 6000 years ago; namely that their orbital elements become quite similar at this time. 
 
 Indeed, we observe that the orbits of 2005 TF50, 2005 UR and 2015 TX24 are similar between 3000 BCE and 14 000 CE, while the evolution of 2004 TG10 is more similar to that of 2P/Encke. Around 3500 BCE however, the orbits of these five bodies approach each other (cf. black square in Figure \ref{fig:G5_qe}). This rapprochement in ($q,e$) is also correlated with a similarity in their longitudes of perihelion (bottom panel, red arrow). We note that the nominal orbit of 2005 TF50, close to 2005 UR and 2015 TX24 between 3000 BCE and 14 000 BCE, is similar to those of 2004 TG10 and 2P/Encke before this rapprochement epoch. 
 
 Thus we see an overall convergence in the orbital elements of the nominal orbits of the G5 at this epoch. This rapprochement around 3500 BCE could support the hypothesis of a common genetic origin among the G5 group or it could be simple coincidence. To better analyse the latter possibility, as a control, we examine the future evolution of these bodies. Any rapprochement which occurs in the future cannot be the result of fragmentation, and provides us with a look into the purely dynamical behaviour of these orbits. Indeed, Figure \ref{fig:G5_qe} displays another epoch of rapprochement around 14 000 CE, though the convergence of the five nominal orbits is less narrowly confined in time in this case. The existence of this possible rapprochement in the future needs to be investigated in more detail with the clones generated for each body, but underscores the fact that we cannot strictly rule out subtle dynamical effects. We will examine them further in the next section.

 \subsection{Clone analysis} \label{sec:G5_rapprochements}
 
 \subsubsection{Rapprochement epoch}
 
 \begin{figure*}
  \centering
  \includegraphics[width=0.99\textwidth]{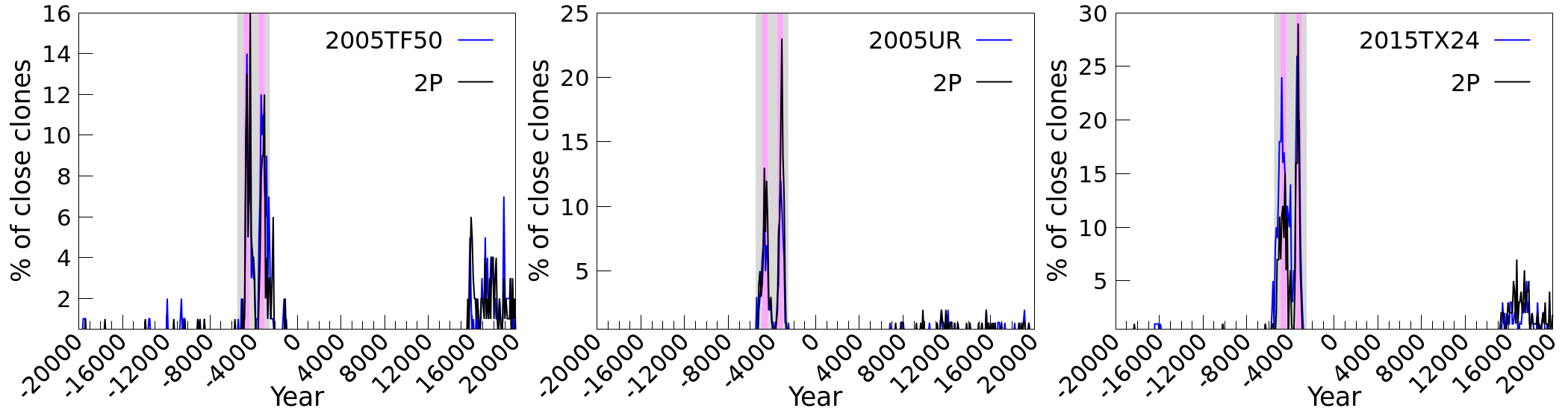}\\
  \includegraphics[width=0.99\textwidth]{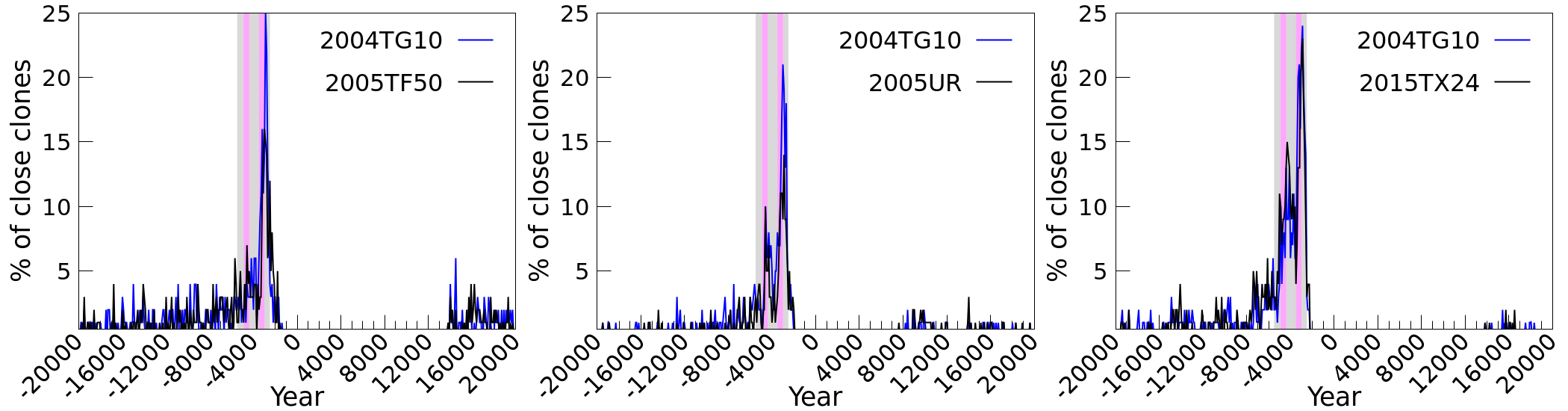}\\
  \includegraphics[width=0.99\textwidth]{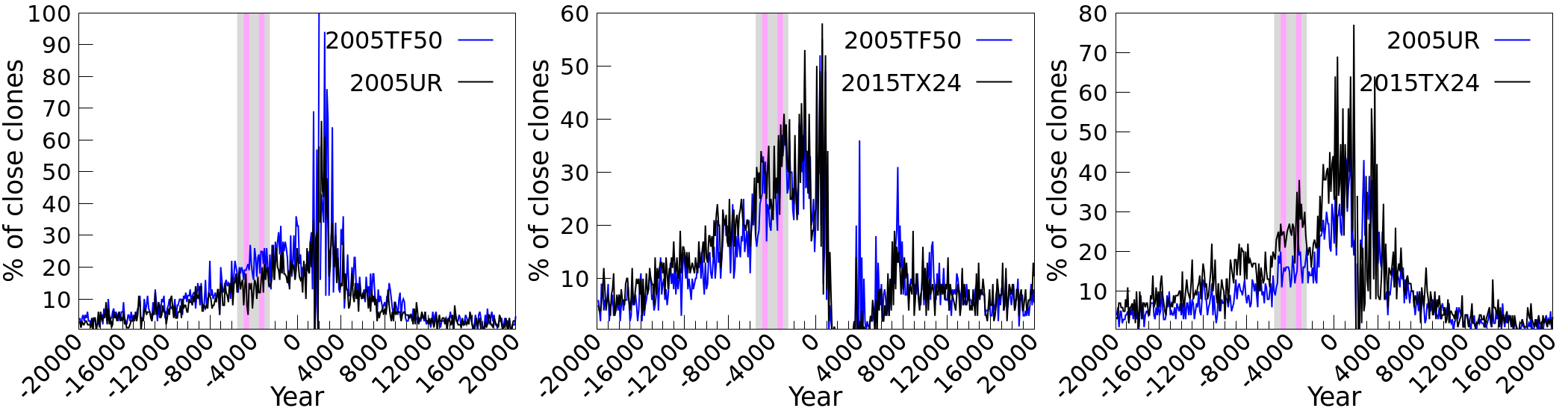}\\
  \includegraphics[trim={0.0cm 0.0cm 49.6cm 0.0cm}, clip, width=.33\textwidth]{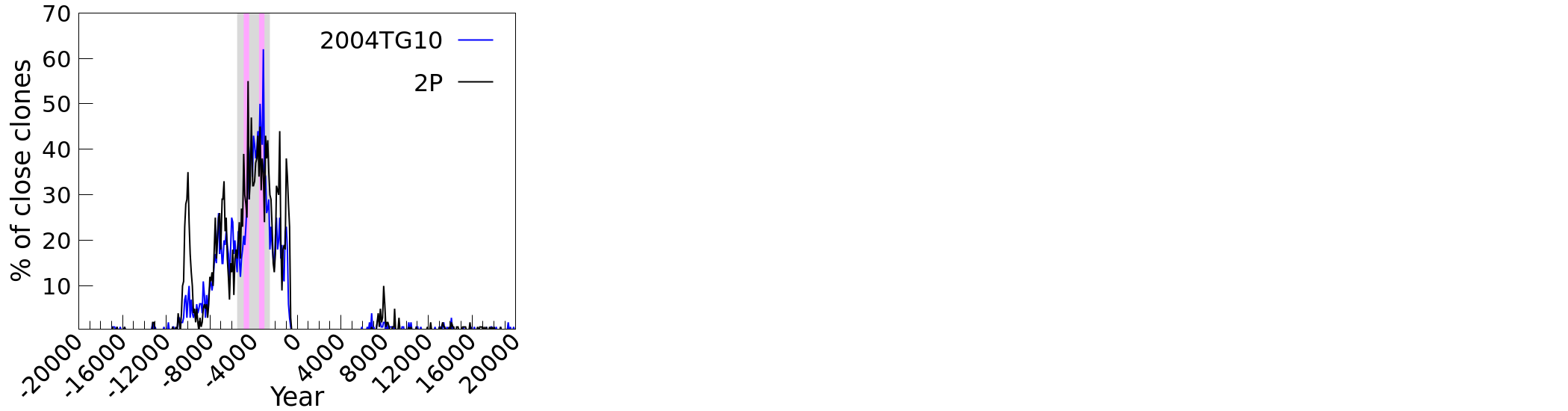}\\
   \caption{Percentage of clones of body \#1 (in blue) and body \#2 (in black) approaching at least a clone of the other object with a MOID smaller than 0.05 AU and a relative velocity below 500 m/s. The grey shaded areas encompasses the period 2500 BCE to 5500 BCE, and the two magenta areas the years 3000 BCE to 3500 BCE and 4400 BCE to 4900 BCE respectively. }
   \label{fig:Main5_rapprochement}
 \end{figure*}
 
 To investigate all the G5 rapprochements in more detail, we tighten the MOID-$V_r$ thresholds.  We follow the methodology described in Section \ref{sec:method}, considering a limiting MOID of 0.05 AU but a more restrictive $V_r$ threshold of 500 m/s. The evolution of $N_1$ and $N_2$, presented as a percentage, is illustrated in Figure \ref{fig:Main5_rapprochement}. 
 
 Each panel of Figure \ref{fig:Main5_rapprochement} represents the percentage of clones of two G5 objects approaching each other between 20 000 BCE and 20 000 CE. For example, the top left panel in the figure displays the percentage of clones of 2P (in black) approaching at least one clone of 2005 TF50 within the MOID and $V_r$ thresholds stated above. The figure also presents the percentage of clones of 2005 TF50 (in blue) approaching at least one clone of 2P with the same restrictions. For any combination of two bodies presented in Figure \ref{fig:Main5_rapprochement}, we remark that the blue and black curves follow a similar trend, with only small variations in the magnitude of the respective peaks. This reflects the fact that for the pairs investigated, a similar fraction of clones of each body is involved in the rapprochements observed. We can conclude that the main peaks in the figure are therefore not due to several clones of a G5 body approaching only a single clone of another object; and more importantly, that the rapprochements observed are statistically robust.
 
 The top panels of Figure \ref{fig:Main5_rapprochement} present the percentage of clones of asteroids 2005 TF50, 2005 UR and 2015 TX24 approaching 2P over the integration period. The second row analyzes the rapprochements of 2005 TF50, 2005 UR and 2015 TX24 with 2004 TG10, and the third row, possible combinations between 2005 TF50, 2005 UR and 2015 TX24. Finally, the last row illustrates the rapprochement between 2004 TG10 and 2P/Encke. The grey shaded areas in the Figure are for reference, and encompass the period 2500 BCE to 5500 BCE, while the two magenta vertical lines circumscribe the years 3000 BCE to 3500 BCE and 4400 BCE to 4900 BCE respectively. 
 
 In the top panels, we consistently observe a double peak around 3200 BCE and 4700 BCE, with 15 to 30\% of clones meeting our MOID and $V_r$ thresholds. The percentage increases and decreases sharply, indicating a possible dynamical approach between \{2005 TF50, 2005 UR, 2015 TX24\} and 2P over a restricted time period. In the second row of panels in Figure \ref{fig:Main5_rapprochement}, a similarly high and sharp peak is visible around 3200 BCE, confirming the rapprochement between \{2005 TF50, 2005 UR, 2015 TX24\} and 2004 TG10 at this epoch. An even higher ($\sim$60\%) and wider peak is observed when comparing 2004 TG10 and comet 2P/Encke at the same period (shown in the bottom panel). If these objects are fragments of each other, the break-up could only have occurred during one or two relatively narrow windows of time.

 In contrast, we see an absence of a similarly well-determined, sharp feature when comparing the possible approaches between 2005 TF50, 2005 UR and 2015 TX24. The percentage of clones meeting the threshold is generally high around the current epoch, as expected for bodies with similar orbits (cf. Section \ref{sec:rapprochement_all}). The number of clones decreases towards the bounds of the integration period, reflecting the slow divergence between the NEA orbits observed in Figure \ref{fig:G5_qe}. As a result, if one of these object split from another, the time frame for this break-up is much less constrained.
 
 Figure~\ref{fig:Main5_rapprochement} provides us with two important results:
 \begin{itemize}
 \item It supports the conclusions drawn in sections \ref{sec:rapprochement_all} and \ref{sec:G5} of a real rapprochement between 2P, 2004 TG10 and NEAs \{2005 TF50, 2005 UR, 2015 TX24\} around a common epoch of 3200 BCE and, to a lesser extent, around 4700 BCE. For the sake of brevity, in the rest of the text, we will refer to the two possible orbital convergences of the G5 members around 3200 BCE and 4700 BCE as the $R_1$ and $R_2$ rapprochements, respectively. 
  \item  In contrast, no sharp enhancement of the percentage of clones meeting the MOID-$V_r$ criteria appears in the future. This result, that the G5 bodies do not show future (and so necessarily purely dynamically-driven) close approaches to the same degree as they do in the past, provides us with another important control sample.
  It suggests that the rapprochements seen in the past are not purely coincidental, though it cannot be viewed as conclusive.
 \end{itemize}
 
  \subsubsection{Sensitivity analysis}

 To test the robustness of the $R_1$ and $R_2$ rapprochements, we rerun the analysis with different MOID and $V_r$ thresholds. We first reduce the limiting MOID to values of 0.01, 5$\times10^{-3}$, 1$\times10^{-3}$, 5$\times10^{-4}$, 1$\times10^{-4}$, 5$\times10^{-5}$ and 1$\times10^{-5}$ AU, while keeping our $V_r$ threshold to 500 m/s. The impact of these new MOID thresholds on the percentage of clones of 2004 TG10, 2005 TF50, 2005 UR and 2015 TX25 approaching 2P/Encke in the past is presented in supplementary Appendix D, Figure D1.
 
 The MOID threshold has only a small impact on the existence of the peaks, though smaller MOIDs reduce the overall number of clones retained. The percentage of clones retained is very similar for MOID thresholds between 5$\times10^{-4}$ AU and 5$\times10^{-2}$ AU. At 5$\times10^{-5}$ AU, the two peaks are still visible, but with only 3 to 12 clones having a relative velocity below 500 m/s. Thus the existence of the peaks is robust against changes in the MOID threshold. To ensure good number statistics however, we retain our initial MOID threshold of 0.05 AU for the rest of our analysis. 
 
 As a second check, we maintain our limiting MOID at 0.05 AU but consider different $V_r$ thresholds of 100, 200, 300, 500, 800 and 1000 m/s (cf. supplementary Appendix D, Figure D2). For a limiting $V_r$ of 1 km/s, we obtain a result similar to Figure \ref{fig:Main5_rapprochement} but with twice the number of clones for each peak. When reducing the $V_r$ threshold, the curve is decreased but maintains its overall shape. The minimum $V_r$ value which still has at least two clones contributing to the peak(s) is between 200 and 300 m/s. 
 
 For a low limiting $V_r$ of 100 to 200 m/s, we still clearly see the $R_1$ rapprochement between the G5 NEAs ; in contrast, the $R_2$ peak disappears at these low relative velocities (cf. Figure D2 in supplementary Appendix D). We therefore conclude that the orbital convergence of the G5 around 3200 BCE is more robust than around 4700 BCE. 
 
Though the percentage of clones approaching a given body depends on the MOID and $V_r$ thresholds considered, the $R_2$ and especially the $R_1$ rapprochement remains visible even down to levels where just a few clones remain. When examining the best pair of clones between any two bodies we obtain MOID values as low as 10 to 100 km (intersecting orbits for all practical purposes) and $V_r$ reaching down to almost 100 m/s depending on the NEAs examined. It is plausible that lower MOID and $V_r$ values could be reached by filling the phase space with more than the hundred clones per body in our work. As a result, we conclude that these rapprochements are consistent with fragmentation events, though whether they are consistent with few m/s break-up speeds is unclear.
 
 In summary, the analysis of the G5 is suggestive of a rapprochement of 2P/Encke, 2004 TG10 and NEAs 2005 TF50, 2005 UR, 2015 TX24 around a common epoch of 3200 ($R_1$) and/or 4700 BCE ($R_2$). One interpretation of these results is that these objects are the remnants of a larger parent body that fragmented in the recent past. It is also possible that the rapprochement could be caused by purely dynamical processes that concentrated the G5 orbits a few millennia ago. In the next sections, we explore two additional possible causes of the $R_1$-$R_2$ events, namely orbital precession, and the effects of the 7:2 resonance.
 
 \subsection{Alternative dynamical explanations for the rapprochements} 
 
 \subsubsection{Orbital precession}\label{sec:G5_precession}
 
\begin{figure}
     \centering
     \includegraphics[width=.5\textwidth]{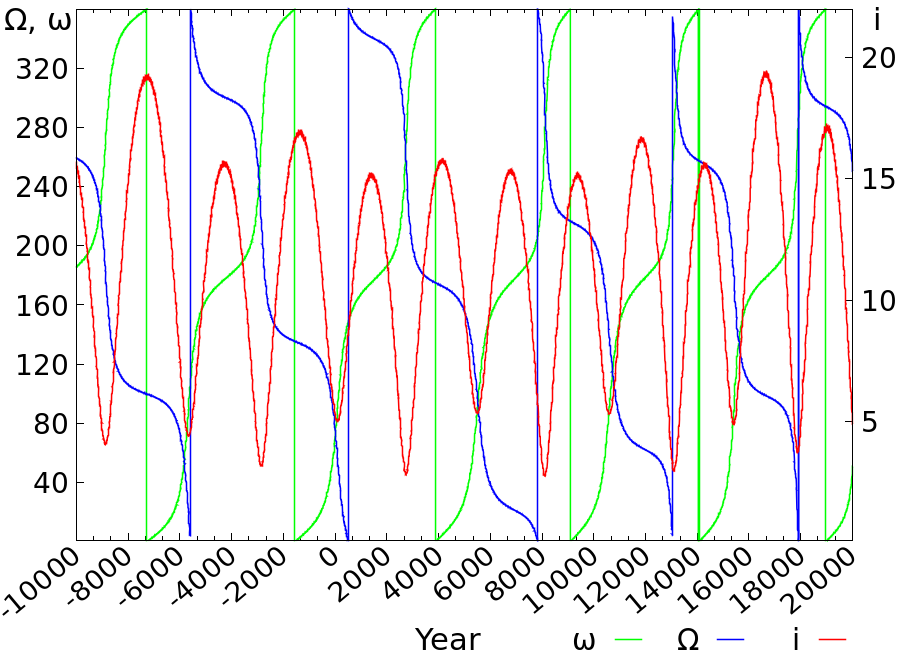}
     \caption{Inclination, argument of perihelion and the longitude of the ascending node of comet 2P/Encke, integrated from the nominal solution of Table \ref{tab:IC} without non-gravitational forces.} 
     \label{fig:2P_angles}
 \end{figure}

 One possible entirely dynamical explanation for the apparent G5 MOID-V$_r$ correlation is the precession cycle of the bodies' angular elements. Indeed, orbital precession will periodically increase and  decrease the MOID of two bodies over time, whether or not the two share a common origin. In addition, we see in Figure \ref{fig:2P_angles} that the period of circulation of 2P/Encke's $\omega$ and $\Omega$ varies from 5000 years to 6000 between 10 000 BCE and 20 000 CE: thus the $R_1$ rapprochement occurred almost exactly one precession cycle of 2P/Encke ago. 
 
  From Figure B2 in supplementary Appendix B, we find the approximate precession periods of 2004 TG10, 2005 TF50, 2005 UR and 2015 TX24 to be 4000-6000, 5000-7000, 7000-8000 and 6000-8000 years respectively over the integration period. While the precession rates of the G5 orbits vary with time, and differ slightly from each other these might still explain the existence of a favourable approach epoch around 3200-4700 BCE. 
 
 As 2005 TF50, 2005 UR and 2015 TX24 are currently on similar orbits (cf. Figure \ref{fig:map_past}), we would expect these orbits to be in a comparable configuration one precession cycle ago if they precess at similar rates. It is therefore plausible that if 2P/Encke evolves so as to reach a small MOID with one of these NEAs, it will do the same with the two other asteroids. This could explain why 2005 TF50, 2005 UR and 2015 TX24 show a similar percentage of clones approaching 2P/Encke or 2004 TG10 in Figure \ref{fig:Main5_rapprochement}. However, this does not suffice to explain the rapprochement between the much larger NEA 2004 TG10 and 2P/Encke at the same epoch, since these two objects are not moving on close orbits at the present day (cf. bottom panel of Figure \ref{fig:Main5_rapprochement}).  
 
 If the G5 approach was uniquely due to orbital precession, we could reasonably expect a similar favourable configuration further in the past or in the future. From Figure \ref{fig:Main5_rapprochement}, we see that no such configuration is visible further in the past (between 20 000 BCE and 6000 BCE). There are some indications of a favourable rapprochement in the future (15 000 - 20 000 CE, see Figure \ref{fig:Main5_rapprochement}); however, this correlation is more modest and extended in time than the peak(s) observed around 3200-4700 BCE. We therefore conclude that the $R_1$-$R_2$ rapprochement is not simply a trivially recurring feature of the dynamics. 
 
\begin{figure}
     \centering
     \includegraphics[width=.5\textwidth]{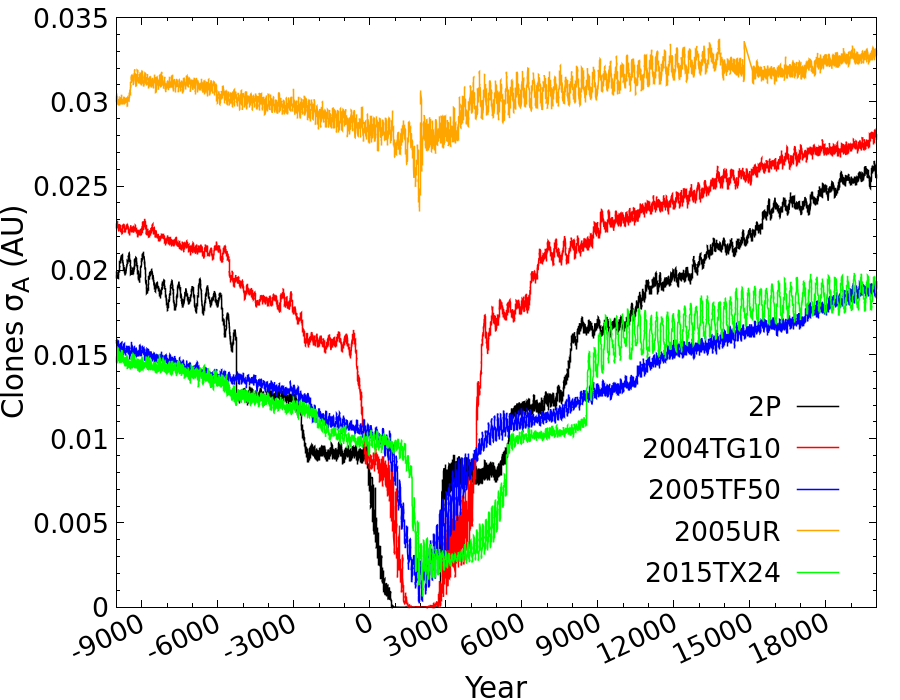} 
     \caption{Standard deviation of the semi-major axis of the clones associated with 2P/Encke (in black), 2004 TG10 (in red), 2005 TF50 (in blue), 2005 UR (in orange) and 2015 TX24 (in green) between 10 000 BCE and 20 000 BCE.}
     \label{fig:sma_dispersion}
 \end{figure}

  From Figure \ref{fig:G5_qe}, we do see that the G5 nominal orbits converge to some extent in the future. Could the future rapprochement of 15000-20000 BCE be broader and less distinct simply because of increased orbital diffusion due to the longer integration time? To examine the relative significance of the enhancement of particles around 15 000 CE versus the $R_1$-$R_2$ rapprochement, we examine the evolution of the standard deviation of the clones semi-major axis $\sigma_a$ (which among all the elements is most sensitive to measurement uncertainties) in Figure \ref{fig:sma_dispersion}. The value of $\sigma_a$ increases with time during both forward and backward integration of the clones as expected. However, the values of $\sigma_a$ at 3200-4700 BCE and at 15 000 CE differ only by a factor of order unity. Thus we conclude that our simulations are correctly capturing the relative strengths of the past and future rapprochements, and that the broadness and weakness of the future event is not simply due to more diffusion due to a longer integration time. Given this result we would expect the past rapprochements to be similar to those we see in the future in magnitude and duration if simply due to dynamics. 
 
 To explore possible rapprochement epochs beyond $\pm$30 000 years we performed an additional simulation of the G5 evolution, generating a hundred clones for each body and integrating them  100 000 years into the past and in the future. The result of the close approach analysis using the methodology described in Section \ref{sec:method}, is presented in supplementary Appendix E, Figure E1. No additional peaks in the percentage of clones meeting the MOID and relative speed thresholds are seen, beyond 20 000 years from today. 
 
Taking all these results together, we can say that we have evidence for a rapprochement between some asteroids of the TC around 3200-4700 BCE. Using forward simulations of the same asteroids as a 'control sample' we find a much weaker signal of rapprochement, but we cannot from this alone completely exclude the possibility of coincidence, or the effects of observational selection or orbital uncertainty coupled with phase space diffusion. Though precession might explain recurring similar geometrical configurations of \{2005 TF50, 2005 UR, 2015 TX24\} with 2P/Encke or 2004 TG10, it does not suffice to clarify why all these bodies approach each other at a common epoch in the past. A quick analysis of the clones' orbital elements highlighted no specific orbital characteristic (e.g., maxima in eccentricity, specific $\omega$ and $\Omega$ values) of the clones during the $R_1$-$R_2$ rapprochement. 

One important fact to note however, is that the $R_1$-$R_2$ rapprochement is not seen for the majority of asteroids presumed to be in the TC. By examining Figure \ref{fig:map_past}, it is clear that most of the TC candidates show no rapprochements at all over the times we examined. Yet these objects were all selected as possible TC members in other studies and on the basis that their orbital elements resemble those of the TC. These asteroids effectively serve as another broad "control sample", demonstrating more clearly that the vast majority of objects with TC-like orbits do not undergo a G5-level of MOID-$V_r$ close-approach; in fact it is quite rare.

 \subsubsection{Resonance}\label{sec:G5_resonance}
 
 Another possible mechanism for the $R_1$-$R_2$ rapprochement is the 7:2 MMR with Jupiter. Figure C2 in supplementary Appendix C shows that a significant fraction of the clones created for asteroids 2005 TF50, 2005 UR and 2015 TX24 spend several millennia within the 7:2 MMR. If the measurement uncertainties on the asteroids' initial orbit are high, one could reasonably expect that the clones created will drift around the orbital space allowed by the resonance. The past rapprochement between these bodies could in this case simply be due to the fact that 2P/Encke and 2004 TG10 were closer to the resonance a few millennia ago. 
  
  \paragraph{2015 TX24}
 
 To explore this hypothesis, we investigated in particular the possible association between 2015 TX24 and 2P/Encke. We examined the sensitivity of the $R_1$-$R_2$ rapprochement to the initial orbit of 2015 TX24. The asteroid was selected among the other members of G5 because most of its clones remain in the 7:2 MMR during the integration period (cf. Figure C2). In addition, errors associated with 2015 TX24's orbital elements are smaller than those measured for 2005 TF50 and 2005 UR (see Figure \ref{fig:covariances_volume}).
 
 We first estimated the semi-major axis extent of the 7:2 MMR, for an eccentricity corresponding to the current orbit of 2015 TX24 (e=0.872). For this purpose, we generated several thousand clones from the asteroid's nominal solution, exaggerating the uncertainty in the orbital elements to fill the resonance. After integration of the clones over several millennia and analysis of the resonant particles (cf. supplementary Appendix C), we estimate that the semi-major axis extent of the 7:2 MMR at this eccentricity ranges from 2.235 AU to 2.286 AU. Our result matches the values of 2.23 AU to 2.28 AU for the 7:2 computed by \cite{Asher1991} for an eccentricity of 0.85.   
 
Next, we conducted several new simulations of 2015 TX24's evolution, generating one hundred clones using its covariance matrix but slightly changing the initial orbit of the asteroid in $a$, $e$, and $i$. A single orbital element was modified for each simulation, the other ones remaining fixed to the original values summarized in Table \ref{tab:IC}. In total, we generated 13 sets of 100 clones with values of semi-major axis ranging from 2.214 AU to 2.31 AU,  8 sets of clones with eccentricities varying between 0.8651 and 0.8790 and 8 sets of clones with inclinations ranging from 2$\degree$ to 10$\degree$. All the clones were integrated for 10 000 years into the past and their orbital evolution compared with the ones created for 2P/Encke using the methodology presented in Section \ref{sec:method}. The percentage of clones approaching 2P/Encke in the past for several variations of 2015 TX24's initial orbit is shown in supplementary Appendix D, Figure D3. 
 
 We finally estimated the maximum deviation in $\Delta_a$, $\Delta_e$ and $\Delta_i$ from the original orbit of 2015 TX24 which still preserves the magnitude and timing of the $R_1$ and $R_2$ rapprochements. All the simulations performed with $0.8702\leq e \leq 0.8732$ ($\Delta_e \simeq 0.03$), $2.2382\leq a \leq 2.2764$ AU ($\Delta_a \simeq 0.04$ AU) and $4\degree \leq i \leq 8 \degree$ ($\Delta_i \simeq 4\degree$) showed the 3200BC rapprochement with 2P/Encke, while simulations starting beyond these $a$, $e$ and $i$ limits showed no close approach with 2P/Encke during the 10 000 years of integration.
 
 We observe that the initial uncertainty on 2015 TX24's orbital elements ($\sigma_e=2e^{-4}$, $\sigma_a=3e^{-3}$, and $\sigma_i=2e^{-3}$) is much smaller than the deviations determined from this sensitivity analysis of $\Delta_e \simeq 0.03$, $\Delta_a \simeq 0.04$ AU and $\Delta_i \simeq 4\degree$. The existence and robustness of the  $R_1$-$R_2$ rapprochement between 2015 TX24 and 2P/Encke is therefore not limited by the accuracy of the asteroid's initial solution. 
 
 We also note that the semi-major axis range $2.2382\leq a \leq 2.2764$ AU necessary to produce a rapprochement with 2P/Encke is close to the bounds of the 7:2 MMR determined previously (2.235 AU to 2.286 AU). This raises the possibility that all the clones approaching the comet around 3200 BCE belong to the the 7:2 MMR, while simulations where most of the clones of outside the resonance never approach 2P/Encke within the MOID and $V_r$ criteria selected. Thus the dynamical association of 2015 TX24 with 2P at 3200 or 4700 BCE appears to encompass a bounded phase space within the 7:2 MMR.

 \paragraph{Resonant clones} 
 
  \begin{figure}
     \centering
     \includegraphics[width=.49\textwidth]{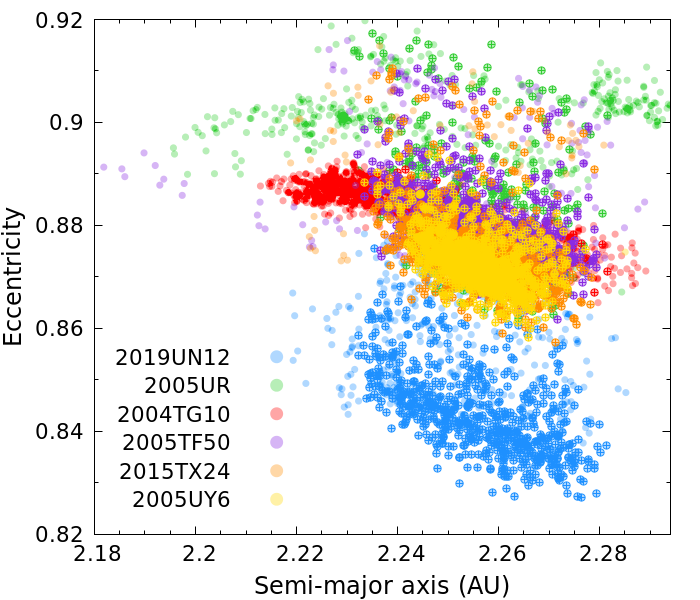}\\
     \includegraphics[width=.49\textwidth]{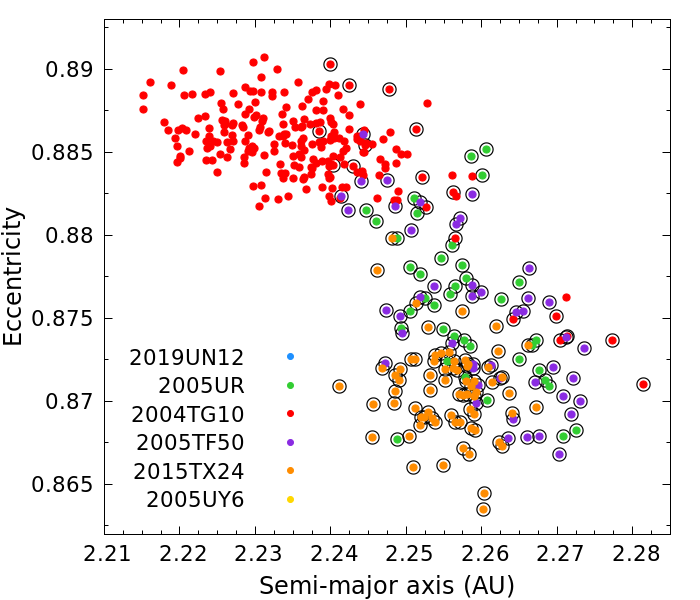}\\
     \caption{Semi-major axis and eccentricity of the clones created for 2004 TG10, 2005 TF50, 2005 UR, 2015 TX24, 2005 UY6 and 2019 UN12 between 3000 BCE and 3500 BCE (top). Particles trapped in the 7:2 MMR with Jupiter are indicated by a circled cross. In the bottom, only the clones approaching at least one clone of comet 2P/Encke within the threshold (MOID $<0.05$ AU; $V_r<$500 m/s) are represented. Resonant clones are circled in black.  }
     \label{fig:AE_clones}
 \end{figure}
 
 To investigate if the clones within a rapprochement are necessarily resonant, we plot in Figure \ref{fig:AE_clones} a compilation of the semi-major axis and eccentricity of all the clones of 2004 TG10, 2005 TF50, 2005 UR, 2015 TX24, 2005 UY6 and 2019 UN12 between 3000 BCE and 3500 BCE, that is during the $R_1$ rapprochement (top panel of the Figure). The particles trapped in the 7:2 MMR with Jupiter are represented by circles with inset crosses, while the non resonant particles are shown by semi-transparent circles. 
 
 Examining the top panel of the figure we see that most of the clones for all bodies (except 2004 TG10) are in the 7:2 MMR around 3200 BCE, as expected from the percentage of resonant clones presented in Figure C2, Appendix C. Individual plots of the clones' semi-major axis and eccentricity of each G5 member at the same epoch are also provided in supplementary Appendix F, Figure F1.
 
 In the bottom panel of Figure \ref{fig:AE_clones}, we plot only those clones approaching 2P/Encke within our threshold (MOID $< 0.05$ AU; $V_r < 500$ m/s). We observe that all the clones retained for 2005 TF50, 2005 UR and 2015 TX24 are located in the resonance. However, this is not the case for 2004 TG10, which has most of its clones meeting our rapprochement criteria outside the 7:2 MMR. And we note that, with the exception of 2P/Encke, the orbital solution integrated for 2004 TG10 is the most accurate of all the NEAs belonging to the G5 (cf. Figure \ref{fig:covariances_volume}).   So the presumption that the resonance alone increases the percentage of clones of 2005 TF50 2005 UR, 2015 TX24 approaching 2P/Encke (say, by constraining their evolution to a reduced orbital space) would still not fully explain the convergence of the comet with 2004 TG10 at the same time. 
 
 We also observe that the clones of 2005 UY6 and 2019 UN12 also wander all around the resonance extent in semi-major axis and eccentricity, and yet neither of these bodies approached 2P/Encke around 3000-3500 BCE (a possible close approach of NEA 2005 UY6 occurs with 2004 TG10 around 2 500 BCE, but with only a small number of clones (cf. Figure E2 in supplementary Appendix E). 
 
 We also note that the NEAs 2003 UL3, 2019 BJ1, 2019 RV3, 2019 AN12 and 2020 JV do not approach {\it any} other asteroid of our sample (see Figure \ref{fig:map_past}), despite the fact that a significant fraction ($15-20\%$) of their clones remain in the 7:2 MMR for the past few millennia (cf. Figure C2). We therefore conclude that being in the 7:2 MMR on a TC-like orbit is not a sufficient condition to approach 2P/Encke and 2004 TG10 in the past.
 
\subsubsection{Summary: Precession and resonance as alternatives} 

From the analysis described in sections \ref{sec:G5_precession} and \ref{sec:G5_resonance}, we have circumstantial evidence that the close approaches represented by the $R_1$-$R_2$ rapprochement are not trivial manifestations of the dynamics. But it is difficult to firmly determine whether this event is related to a fragmentation of a larger body using these asteroidal bodies alone. To examine this hypothesis further, we introduce another set of objects potentially linked with 2P/Encke and/or TC NEAs: fireballs detected from the 2015 Taurid resonant outburst. 
 
 \section{Taurid fireballs} \label{sec:fireballs}
 
 In 2015, the cameras of the European Fireball Network detected an increase in the number of Taurid fireballs compared to previous years \citep{Spurny2017}. Between October 25 and November 17 113 fireballs (among a total of 144 observed) were detected having a well-defined orbital structure within the broader Southern Taurid stream. The fireballs shared similar orbital characteristics, including comparable orientation of the lines of apsides and very similar semi-major axes, notably lying inside the limits of the 7:2 MMR  with Jupiter (i.e., between 2.23 and 2.28 AU).
 
 The fact that the semi-major axes computed for all the fireballs (except two) had a narrow range of values testifies to the exceptional accuracy of the meteor data. For most fireballs recorded, \cite{Spurny2017} cite a formal precision in the radiant determination $<$0.05$\degree$, and $\leq$0.1 $km.s^{-1}$ in the fireball's pre-atmospheric velocity. As a consequence of this accuracy, these observations of the 2015 new branch of fireballs confirmed the theory of \cite{Asher1993b} and \cite{Asher1998}, who had proposed that enhanced Taurid activity was caused by meteoroids trapped in the 7:2 MMR. Since increased meteor rates are not observed every year, it also supports the idea that these meteoroids are not spread along the whole orbit, but concentrated around the resonance centre with an extension of about $\pm$ 30-40$\degree$ in mean anomaly \citep{Asher1993b,Asher1998}.  
 
 In addition to radiants and orbits, \cite{Spurny2017} measured the physical properties of the meteoroids associated with the resonant swarm, using estimates of their photometric mass and end height as a proxy for strengths using the PE criterion \citep{Ceplecha1976}. The PE criterion was able to discriminate between soft cometary material (type IIIb) and denser stony meteorite-like \citep[type I, see][]{Ceplecha1988} objects. \cite{Spurny2017} noted a significant heterogeneity in the physical properties of the meteoroids, which had photometric masses ranging from 0.1 g to 1300 kg and PE values covering all the categories of meteoroid classification. In particular, they found a strong anti-correlation between meteoroid mass and PE criteria, with the largest PE values (hence strongest material) associated with small masses. A later study concentrated on physical properties \citep{Borovicka2020} concluded that the majority of Taurid material has a low mechanical strength; nevertheless, stronger material exists as well, in the form of inclusions or as separate cm-sized bodies. The quality of the fireball orbital measurements, coupled with information about the meteoroids' mass and physical strength, make the data of \cite{Spurny2017} particularly suitable for our dynamical exploration of the Taurid complex. 

 \subsection{Data processing} \label{sec:fireball_data}
 
 In this section, we analyse the orbital history of 16 representative Taurid fireballs out of those observed by the European Fireball Network in 2015, 15 of them belonging to the resonant meteoroid branch (cf. list in Table \ref{tab:IC_fireballs}) and one to the Southern Taurid meteor shower (EN241015\_185031) for comparison. For each station observing a fireball, the original measurements comprise two data sets. The first one contains selected observations of the position of the meteor on the camera focal plane, allowing a precise computation of the meteoroid's position. The second set of measurements consist of less precise measurements across the meteor track, but having better time registration, which makes them more suitable for the velocity determination. For each Taurid, the best orbital solution of the fireball, computed as described in \cite{Spurny2017}, was also available. 
 
 To explore the past evolution of the fireballs and the TC NEAs in a self-consistent manner, we must first compute the covariance matrix of each fireball orbit before performing any integration. The formal uncertainties obtained by fitting the meteor trajectory are much smaller than the real uncertainties associated with the fireball's position and velocity, and are not fully representative of the measurement uncertainties necessary for the clone analysis. Discussion of covariance measurements for meteor orbits are uncommon \citep{JansenSturgeon2019}, but have been given in \cite{Dmitriev2015} and \cite{vida2017, vida2020a}. 
 
\subsubsection{Covariance computation} \label{sec:covariance_computation}

To estimate the covariance matrix of the fireball measurements, we use the Western Meteor Physics Laboratory (WMPL) library developed by \cite{Vida2019}. Among several common functions used for meteor physics, this Python library offers several trajectory determination solvers, including the observational Monte Carlo approach we selected to recompute the fireballs' positions\footnote{The WMPL is published as open source on the GitHub web page \url{https://github.com/wmpg/WesternMeteorPyLib}}. 

With this solver, the angular difference between the meteor measurements and the fitted trajectory is used as a direct estimate of the uncertainty $\sigma$. Several Monte Carlo (MC) runs are then generated as follows: first, Gaussian noise is added to the original observations using the value of $\sigma$ determined by examining the variance of the sight lines from a straight-line trajectory, and then a new trajectory solution is computed from the noise-added data. For each trajectory solution, the dynamics of the meteor as observed for each station is computed. When all the MC runs are performed, the solver keeps the trajectory solution that provides the most consistent dynamics of the meteor as seen from all stations. 

Since the trajectory determined in each run represents a possible solution within the measurement uncertainty, all the runs can be used as an ensemble to estimate the overall uncertainty of the trajectory determination. The Monte Carlo solver of the WMPL library can therefore be used to compute the covariance matrix associated with the trajectory solution of the fireballs, and hence the corresponding covariance of their orbit computation. 

While the WMPL solver offers the best fully automated trajectory computation currently available, the solution obtained is still probably of lower quality than that obtained with the detailed processing performed in \cite{Spurny2017}, in particular with regard to the initial velocity which is computed via a forward modelling method using an ablation code. However this detailed 'manual' processing does not easily produce an associated covariance matrix.  To provide the best estimate of the original state vector {\it and} its covariance, we consider in our analysis the orbital solution determined as described in \cite{Spurny2017} (hereafter called the EN solution), but use the covariance matrix computed with the Monte Carlo (MC) solver. 

For each fireball, we generate 1000 MC runs using the original measurements described in \ref{sec:fireball_data}, and record the apparent radiant and initial velocity computed for each run. Then, the radiants and velocities are re-centred on the EN solution. The centring is performed by subtracting the average solution estimated from the 1000 runs, and by adding the difference to the EN apparent radiant and initial velocity. Then, for these new initial states of the meteor, we compute the orbits following \cite{Vida2019}. These orbits are finally used to compute the covariance matrix of the fireball's osculating elements. 

\subsubsection{Discussion}

 \begin{figure}
     \centering
     \includegraphics[width=.48\textwidth]{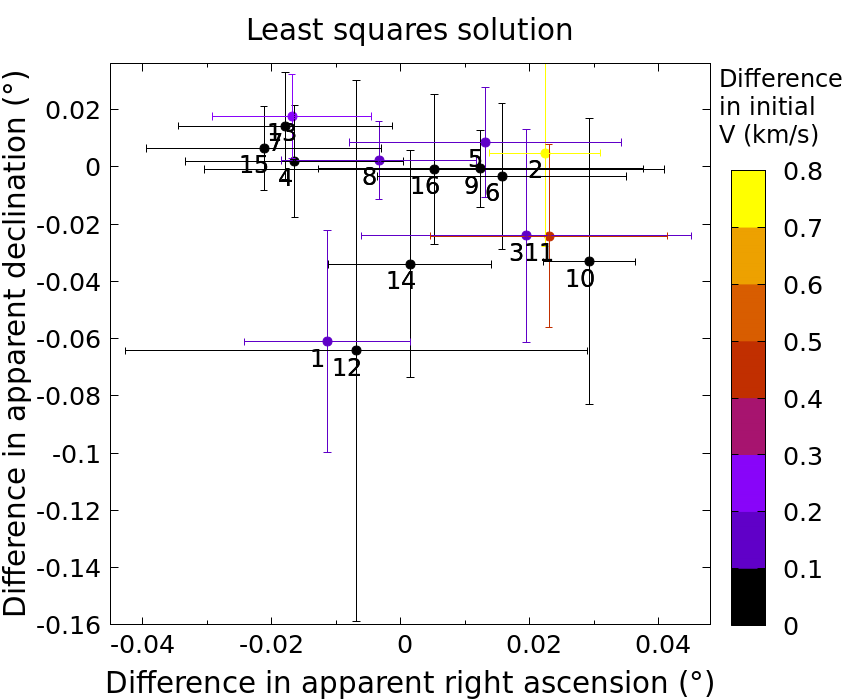}\\
     \includegraphics[width=.48\textwidth]{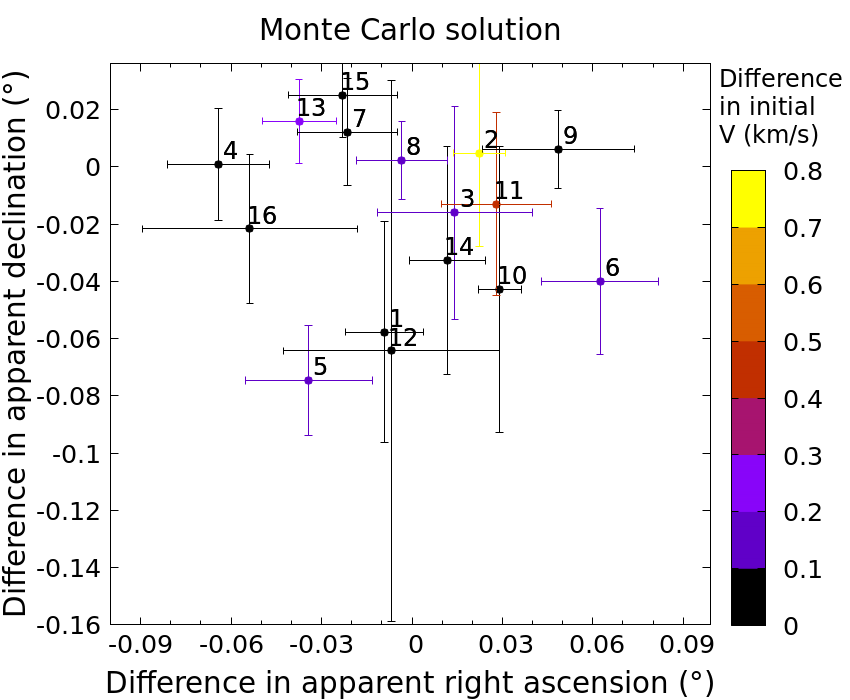}\\[0.2cm] 
    \includegraphics[width=.48\textwidth]{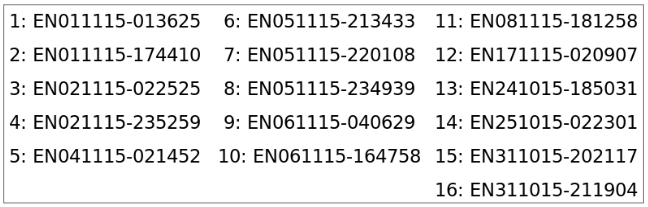}
     \caption{Difference in apparent radiant (right ascension and declination) and initial velocity (colour bar) of the fireballs state vectors computed with the line-of-sights method (top) or Monte Carlo solver (bottom) of the WMPL when compared with the EN solution. Each fireball is labelled by number and referenced by name following Table \ref{tab:IC_fireballs}. See the text for details. }
     \label{fig:error_fireballs}
 \end{figure}

 \begin{figure*}
	\centering
	\includegraphics[width=\textwidth]{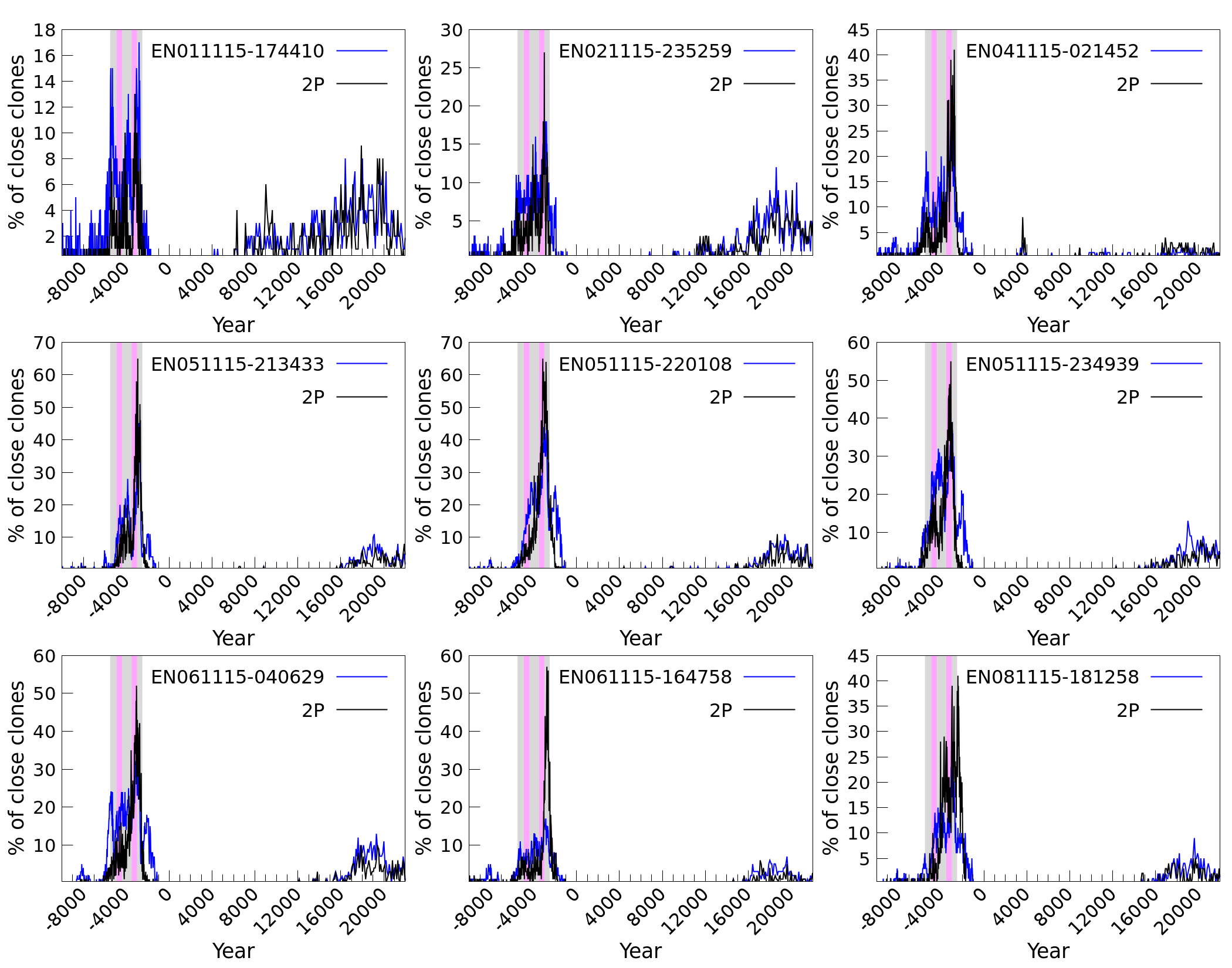}
	\caption{Percentage of the clones for each of the 2015 Taurid fireballs (in blue) and comet 2P/Encke (in black) approaching at least one clone of the other object with a MOID smaller than 0.05 AU and a relative velocity below 1 km/s. The grey shaded areas encompasses the period 2500 BCE to 5500 BCE, and the two magenta areas bracket the years 3000 BCE to 3500 BCE and 4400 BCE to 4900 BCE respectively.}
	\label{fig:fireballs_with_peak}
\end{figure*}

\begin{figure*}
	\centering
	\includegraphics[width=\textwidth]{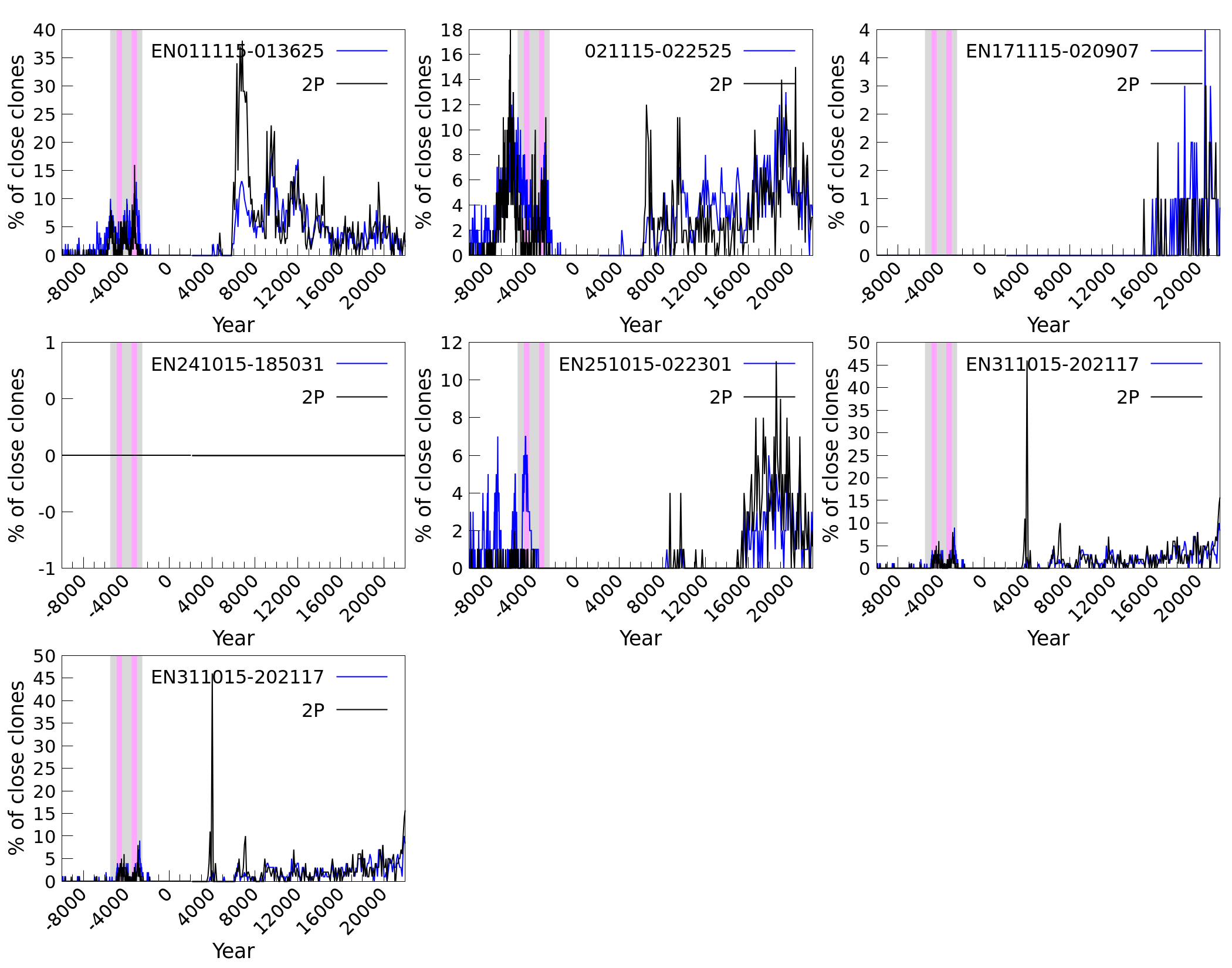}
	\caption{Percentage of the clones for each of the 2015 Taurid fireballs (in blue) and comet 2P/Encke (in black) approaching at least a clone of the other object with a MOID smaller than 0.05 AU and a relative velocity below 1 km/s. The grey shaded areas encompasses the period 2500 BCE to 5500 BCE, and the two magenta areas the years 3000 BCE to 3500 BCE and 4400 BCE to 4900 BCE respectively.}
	\label{fig:fireballs_without_peak}
\end{figure*}

The transposition of the MC covariance matrix to the EN solution is meaningful only if the two trajectory solvers produce similar results. To investigate the agreement of both solutions, we illustrate in Figure \ref{fig:error_fireballs} the difference in apparent radiant and initial velocity between the two solvers. The top panel represents the difference obtained when computing the trajectory using the line-of-sight method of \cite{Borovicka1990} versus the WMPL library. We see that for most fireballs, the difference in apparent right ascension is below 0.04$\degree$, and the divergence in declination is below 0.04$\degree$ (and below 0.06$\degree$ for all fireballs). The absolute difference in initial velocity is generally low, below 200 m/s, except for fireball \#2 and fireball \#11. However, since the initial velocities are not computed in the exact same way and sometimes at slightly different altitudes, we consider the agreement good for most fireballs. 

The bottom panel of Figure \ref{fig:error_fireballs} illustrates the divergence obtained between the EN solution and the best solution determined from all the MC runs. This time, the difference in radiant is  slightly higher ($<0.08\degree$ in right ascension, and in declination) which is not surprising since the solution offering the most consistent dynamics of the meteor as seen from all stations is not necessarily the solution that best fits the meteor's positional measurements \citep{Vida2019}. The differences in the meteor's initial velocity between the two methods is consistent with the previous description. We therefore believe that the differences observed in the initial state vector of the meteor when considering the MC or the EN solution remain small enough to meaningfully associate the covariance matrix computed with the MC solver with the EN solution. 

The uncertainty in the fireball's radiant determined with the Monte Carlo solver is two to four times higher than the EN uncertainties provided by \cite{Spurny2017}, which is not surprising since the published uncertainties refer to the trajectory fit formal errors. In contrast, the uncertainty on the initial velocity determined with the MC solver appeared to be smaller than the real uncertainty measured by the EN solution by a factor 2. In fact the EN velocity error is not just a formal error but takes into account the uncertainties of meteoroid mass and density, which affect the deceleration in the early part of atmospheric trajectory.\footnote{Note that EN velocity errors used in this work were revised in comparison with those published in \citet{Spurny2017}.} In order to account for the worst possible accuracy of the orbital solution, we decided to scale the velocity estimates obtained with the MC so the standard deviation of all the velocity values replicates the EN uncertainty.

The volume of the n-dimensional hyper-ellipsoid defined by the covariance matrix in ($e,q,\Omega,\omega,i$, n=5) and ($e,q,t_p,\Omega,\omega,i$, n=6) is shown along with the volumes computed for the NEAs in Figure \ref{fig:covariances_volume}. Units of the different orbital elements are degrees, AU and days as appropriate. The relative covariance volume illustrates the comparatively good quality of the fireballs' orbits compared to many of the NEAs linked to the TC. When excluding the uncertainty of the location of the meteoroid along its orbit (represented by the time of perihelion passage $t_p$), the covariance volume of the fireballs is comparable to that of several NEAs, including 2005 TF50. The accuracy of the fireball orbits decreases when the perihelion time, $t_p$, is taken into account, reflecting the comparatively poor knowledge of the actual location of the fireball in its orbit. In this comparison the fireball covariances are lower than for any NEA investigated. However, since our close approach analysis relies on the MOID between two bodies, errors related to the location of the objects on their orbits are of reduced importance.  We therefore consider the quality of the fireballs orbits and the associated covariances sufficient to perform a clone analysis as was previously done for the NEAs. 
 
 \subsection{Results} \label{sec:fireballs_results} 

 In order to avoid the strong gravitational influence of the Earth and the Moon at the beginning of the simulation (the meteoroid orbits are computed in the vicinity of the Earth), we have integrated the initial orbit back 60 days into the past, following the approach suggested by \cite{Clark2011}. At this epoch, a hundred clones of each orbit were then created using the associated covariance matrix and these were integrated over the period 10 000 BC to 22 000 AD. This integration window was chosen to cover the possible rapprochement epochs observed for the G5 around 3200-4700 BCE and 15 000-18 000 CE. The analysis of the clones of each fireball approaching any body of the G5 was performed as described in Section \ref{sec:method}. The fireball osculating elements and corresponding Julian date used for the starting solution are summarized in Table \ref{tab:IC_fireballs}. 
 
 The percentage of clones of each fireball trapped in the 7:2 MMR through time is presented in Appendix C, Figure C3. We observe that at least 30\% of the clones of all the fireballs remain in the resonance over the integration period, with the exception of our comparator Southern Taurid EN241014\_185031 that evolves outside the resonance. We thus confirm the result of \cite{Spurny2017}, that is that these 15 members of the new Taurid branch are probably concentrated in the 7:2 MMR. 
 
 The percentage of clones approaching comet 2P/Encke with a MOID smaller than 0.05 AU and a relative velocity below 1 km/s during the integration period is illustrated in Figures \ref{fig:fireballs_with_peak} and \ref{fig:fireballs_without_peak}. The higher relative velocity threshold of 1 km/s was selected because of the  larger uncertainty generally associated with the fireballs' initial orbit, resulting in more significant dispersion of the clones and a lower number of particles retained. However, results obtained with our previous limiting velocity of 500 m/s reproduce the characteristics of Figures \ref{fig:fireballs_with_peak} and \ref{fig:fireballs_without_peak}, just with poorer statistics (cf. supplementary Appendix E, Figures E3 and E4).

 In Figure \ref{fig:fireballs_with_peak}, we note that at least 9 fireballs out of the 16 integrated show a close approach with 2P/Encke between 3500 BCE and 3000 BCE, as observed for the G5 NEAs ($R_1$ rapprochement). The peak is particularly notable for the fireballs recorded on November 4, 5 and 6 ($>$30\%) and is still noticeable for the one observed on November 8 ($\sim$20\%). A peak around 3000 BCE is also perceptible for fireball EN011115\_174410 and EN021115\_235259, but with a smaller number of clones for each body ($\sim$10-15\%). 
 
 In Figure \ref{fig:fireballs_without_peak}, we see that fireballs observed on October 25, 31 and November 1, 17, 2015 do not approach 2P/Encke in the past, or at least not with a significant number of clones. Less than $<$15\% of the clones of fireballs EN011115\_013625 and EN021115\_022525 approach 2P/Encke at any epoch between 10 000 BCE and 2000 BCE, and between 7000 CE and 22 000 CE. Not a single clone of our comparator Southern Taurid EN241014\_185031 approaches the comet during the integration period with our MOID and $V_r$ criteria. A large peak in the number of 2P/Encke clones approaching fireball EN311015\_202117 around 4500 CE illustrates the situation when many clones of the comet ($>$40\%) approach only a few clones of the fireballs ($<$2\%). This peak is therefore probably not characteristic of a real close approach between the two bodies, but is more plausibly a random association caused by the clones' dispersion.

 \begin{figure}
 	\centering
 	\includegraphics[width=.48\textwidth]{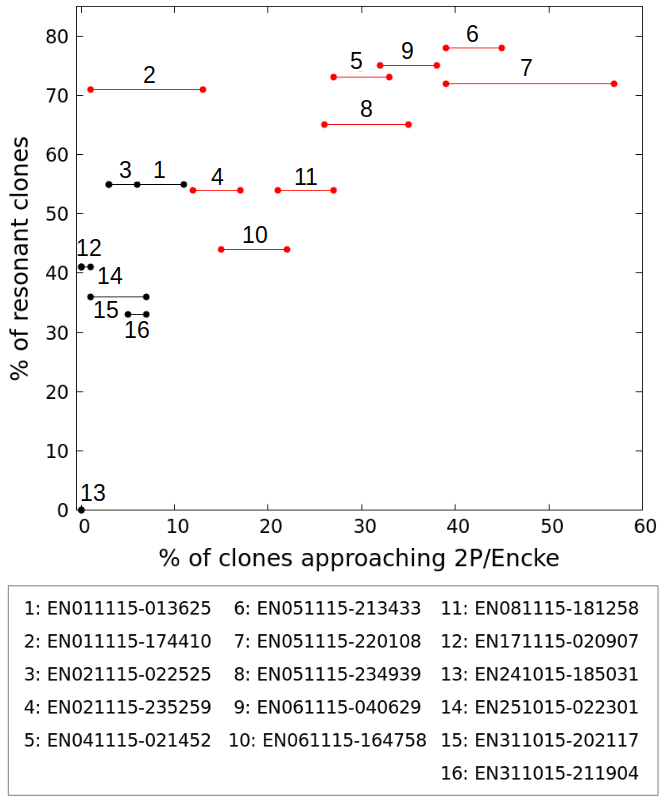}
 	\caption{The percent of resonant clones for each of the 2015 Taurid fireballs as a function of the maximum percent of clones approaching 2P/Encke for each fireball retained during the close approach epoch around 3000 BCE. The range in percentage of clones retained for each body is represented by the line connecting the solid circles. The fireballs shown in Figure \ref{fig:fireballs_with_peak} are highlighted in red.}
 	\label{fig:percent_fireballs_approaching_2P}
 \end{figure}
 
 When comparing the orbital evolution of the fireballs' clones with the other G5 members, we do not observe any significant rapprochement between the fireballs and the four NEAs in the past or the future. For EN021115\_022525 only, we observe a possible rapprochement of the fireball with 2004 TG10 around the $R_1$ and $R_2$ epochs (cf. Figure E5 and E6 in supplementary Appendix E). However, the small number of particles (<4) limits the robustness identification of such a rapprochement. 

 We remark that all the Encke-approaching fireballs discussed above were observed on Earth during a short period of about six days. The orbital elements of these meteors are indeed similar to each other (cf. Table \ref{tab:IC_fireballs}), but do not present specific particularities in eccentricity, semi-major axis or angular elements. There is also no obvious correlation between a significant fraction of the clones approaching 2P/Encke and the meteoroid mass, PE-criterion or fireball classification type.
 
However, several alternatives may explain the time-concentration of fireballs showing an $R_1$ rapprochement. One possibility relates to the quality of the fireball's initial orbit. We observe that fireballs which show a  significant $R_1$ rapprochement (like EN041115\_021452, EN051115\_213433, EN051115\_220108, EN051115\_234939 and EN061115\_040629) also possess the best determined orbit in 2015 (cf. Figure \ref{fig:covariances_volume}). This raises the possibility that some fireballs of Figure \ref{fig:fireballs_without_peak} do not show a clear $R_1$ rapprochement, not because such rapprochement did not occur, but because of the larger uncertainty of their initial orbit.
 
 Also, we see a correlation between the magnitude of the $R_1$ rapprochement and the percentage of clones trapped into the 7:2 MMR around the same epoch. In Figure \ref{fig:percent_fireballs_approaching_2P}, we present the percentage of clones of each fireball approaching 2P/Encke as a function of the number of resonant clones around 3200 BCE. Despite the scatter of the plot, there is a potential correlation between these two quantities, especially when excluding the fireball \#2 (EN011115\_174410). If real, this correlation confirms once again the importance of the 7:2 MMR in the dynamics of the fireballs and G5 members. However, it is still unclear if the resonance is the main driver for approaching 2P/Encke around 3200 BCE, or if this orbital rapprochement increases instead the probability of being located at the resonance at this epoch.

 \section{Forward modelling: Implications for TC formation} \label{sec:TC_formation}

In the previous sections, backward integrations of the orbits of TC candidates were examined, and behaviour compatible with a break-up event within the complex several thousand years ago were found. In this section, we consider a simple forward model, where hypothetical material originating from such a break-up is integrated forward in time, to examine whether such an event is consistent with some or all of the Taurid meteor showers observed at Earth. 

The Taurid meteoroid streams produce at least four meteor showers, that are observed on Earth every year. Because the streams have a very low inclination, the Earth moves within the complex for a large fraction of its orbit, with activity from the showers occurring over a large solar longitude (SL) range ($\sim$120$\degree$). Major showers associated with the complex are the Northern Taurids (NTA, SL=207-258$\degree$), the Southern Taurids (STA, SL=167-238$\degree$), the Daytime $\beta$-Taurids (BTA, SL=75-116$\degree$) and the Daytime $\zeta$-Perseids (ZPE, SL=59-103$\degree$). The Daytime May Arietids (SMA, SL=44-75$\degree$) may also belong to the complex, along with other additional minor meteor showers \citep{Asher1991,Jenniskens2006}. 

If a progenitor object underwent a large break-up between 3000 BCE and 4000 BCE, it is reasonable to presume a significant fraction of meteoroids released during this event might have contributed to the formation of the Taurid stream complex observed today. 

To explore this possibility more quantitatively, we performed several simulations of meteoroids released during a break-up event around 3200 BCE, at the maximum of the $R_1$ rapprochement. Since we do not know the exact orbit of the progenitor, we ejected a meteoroid stream from each of the G5 orbits 5200 years ago. Each simulation involved the generation of 6000 to 9000 particles, of masses between $4\times10^{-9}$ kg and 4 kg, that were integrated until the present epoch as described in \cite{Egal2019} or \cite{Egal2020}.

For 2P/Encke and 2004 TG10, we ejected three sets of 3000 meteoroids generated using different ejection velocities. These velocities were selected to characterize different ejection scenarios. In the first case (scenario 1), the meteoroids are released from the parent body with a fixed ejection velocity of 2 m/s, approximating the separation velocity of cometary fragments \citep{Boehnhardt2004}. The second set of simulations emulates meteoroids ejected via regular cometary activity (scenario 2).
 The ejection velocity of these meteoroids was generated using the cometary dust ejection model of \cite{Crifo1997}, with ranges between 1 m/s and 125 m/s. Finally, if the fragmentation was caused by a collision with another object, the meteoroid dispersion velocity could reach values above a few hundred meters per second \citep{Hyodo2020}. To investigate this hypothesis, we also simulated a third set of meteoroids, ejected from the parent comet orbit with a fixed velocity of 1 km/s (scenario 3). All the meteoroid trails simulated with the different ejection velocities models were integrated until the present epoch. The distributions of nodal locations that cross the ecliptic plane in 2015 are presented in Figure G1 in Appendix G.
 
\begin{figure}
   \includegraphics[width=0.48\textwidth]{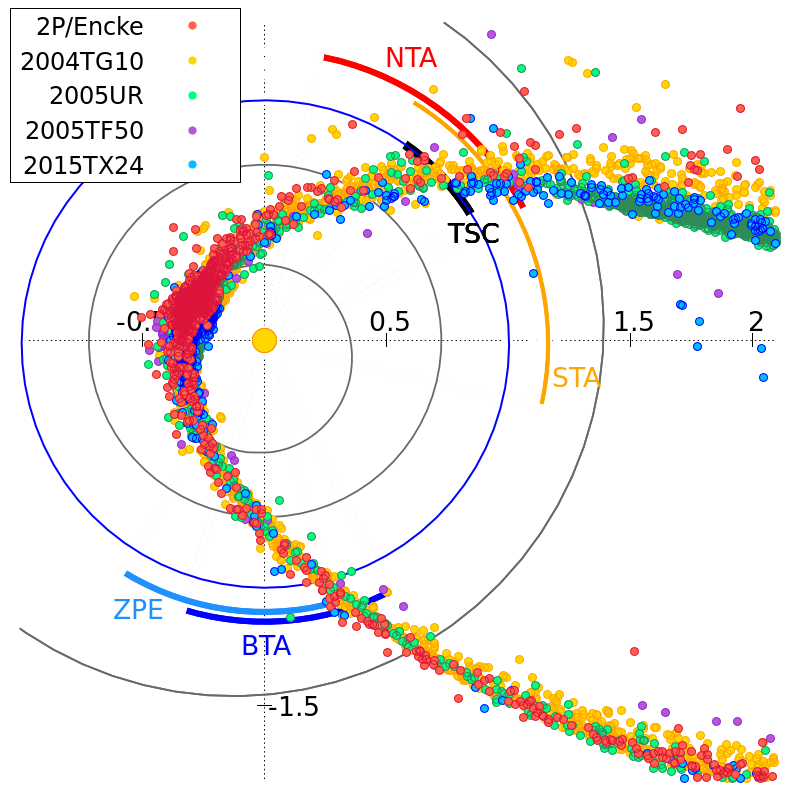}\\
   \includegraphics[width=0.48\textwidth]{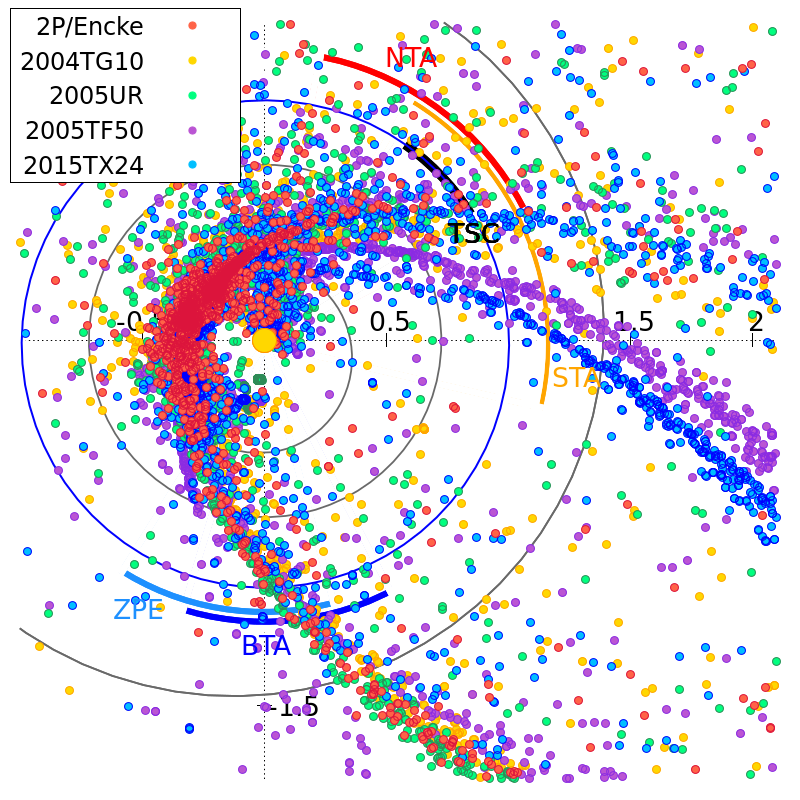}\\
   \caption{Distribution of nodal locations for model meteoroids crossing the ecliptic plane in 2015 based on ejection scenarios 2 (upper panel) and 3 (lower panel) for the G5 member's orbit circa 3200 BCE (see text for details). The nodes are colour coded by the meteoroids parent body. The time periods where the annual Taurid meteor showers (NTA, STA, BTA and ZPE) are active are indicated by the coloured arcs of circle close to the Earth's orbit (in blue). The solar longitude range where the Taurid resonant swarm was observed in 2015 by \protect\cite{Spurny2017} is illustrated by the black circle arc.} \label{fig:ejection_in_2015}
 \end{figure}

 We found that scenarios 1 and 2 produced very similar results for 2P/Encke and 2004 TG10 (cf. Figure Figure G1), so we only generated two sets of 3000 meteoroids (6000 in total) corresponding to scenarios 2 and 3 for NEAs 2005 TF50, 2005 UR and 2015 TX24. The distribution of the nodal location of meteoroids crossing the ecliptic plane in 2015, when the Taurid resonant branch was observed by \cite{Spurny2017}, is presented in Figure \ref{fig:ejection_in_2015}. The top panel of the figure corresponds to scenario 2; the bottom panel, scenario 3. The nodes of the particles are coloured as a function of their parent body. The epoch of visibility of the main annual Taurid meteor showers, namely the NTA, STA, BTA and ZPE, are shown by the coloured arcs of circles in the Figure. 
 
We observe that the nodes of meteoroids ejected with low initial velocity forms a well defined structure in the ecliptic plane having an elliptical shape crossing the Earth's orbit at two different times during the year. Meteoroids released circa 3200 BCE from G5-like orbits should contribute to the meteor activity observed in November and during the first weeks of July. While these meteoroids might contribute to the core activity of the Northern, Southern and $\beta$-Taurids, the narrow dispersion of the model nodal locations for each parent body fails to match the long duration of these showers. Though a combination of ejecta from several members of the G5 may better approximate the overall duration of the TSC, such a distribution remains comparatively narrow.  It seems likely that older ejections from 2P/Encke (or an earlier progenitor on a similar orbit) would be necessary to explain the full nodal extent of the contemporary Taurid complex of streams: low speed material deposited during $R_1-R_2$ rapprochement would not be sufficient to explain the entire TC.   

As expected, the nodal footprint of meteoroids ejected with a higher velocity of 1 km/s is much more dispersed. We observe that meteoroids ejected from 2P/Encke, 2004 TG10 and 2005 UR cross the ecliptic plane at a similar location, while the nodal position of meteoroids ejected from 2005 TF50 and 2015 TX24 are separate, though close to each other. As a result, this higher velocity material could explain longer activity of the NTA, STA and BTA; however such a scenario still would not completely explain the Taurid complex, and especially the TSC. 

Can a low-speed break-up (scenario 2) explain the outburst of resonant TSC fireballs observed in 2015? Many simulated meteoroids do approach Earth's orbit in the correct time interval. However, the number of meteoroids simulated is too small to compare with the observed meteor activity profile. Instead,  we examine the proportion of particles ejected from the G5 bodies that were trapped in the 7:2 MMR in 2015. Based on their semi-major axis and eccentricity alone, we find that several meteoroids simulated from 2P/Encke, 2005 TF50 and 2005 UR entered and remained into this resonance until approaching Earth in 2015. However, about 55 to 95\% of the scenario 2 meteoroids from these three bodies that were observable that specific year showed a semi-major axis smaller than 2.3 AU (e$\sim$0.85), which places them outside the 7:2 MMR. So material ejected from these bodies in 3200 BCE can end up in the resonance, but the majority would have been outside it, inconsistent with observations.

In contrast, we observe that about 78\% and 98\% of the scenario 2 meteoroids ejected from 2004 TG10 and 2015 TX24 exhibit a semi-major axis and an eccentricity in 2015 compatible with the 7:2 MMR. In particular, when comparing the predicted and observed geocentric radiants of the meteoroids in 2015 (cf. Figure G2 in Appendix G), we observe that meteoroids ejected in 2015 TX24 with low velocities contribute the most to the TSC. Ejecta from the other G5 members in our simulations would tend to contribute to the NTA activity in 2015. 

Therefore we conclude that a more-or-less gentle break-up of an Encke-like orbit 5200 years ago could populate the TSC. However, such a break-up, whether due to a fragmentation of the parent comet or to a collision with another object, does not explain on its own the current total activity of the Taurid meteoroid complex. More realistic simulations of the Taurid meteoroid complex are therefore required to confirm these preliminary observations about the TC formation.

 \section{Conclusions}
 
 In this work, we examined the orbital evolution of 51 asteroids and 16 fireballs associated with the Taurid Complex and comet 2P/Encke. The motion of each body was integrated over tens of thousands of years into the past and the future, to explore the possible dynamical linkages between them. In order to account for measurement errors, each NEA and fireball was represented by a sample of 100 clones, generated from its orbital covariance matrix. 
 
 The simulations were examined for epochs of small MOID ($<$0.05AU) and reduced relative velocity ($V_r<$500 m/s, 1 km/s) between objects, corresponding to possible close approaches between objects, and thus possible fragmentation events. Over the 1326 possible combinations of two bodies in our sample of 52 objects, only 12 pairs showed such possible past dynamical associations in the last 20 000 years (cf. Figure \ref{fig:map_past}), pointing toward an orbital convergence of 2P/Encke and NEAs 2004 TG10, 2005 TF50, 2005 UR, 2015 TX24 and possibly 2005 UY6 and 2019 UN12 between 3000 BCE and 4000 BCE. 
 
 Our analysis then focused on a reduced list of five objects (called G5), which included 2P/Encke, 2004 TG10, 2005 TF50, 2005 UR and 2015 TX24. These appeared particularly promising since these bodies had all been associated by other authors with Taurid meteor activity observed on Earth in the past (see references in Table \ref{tab:TC_asteroids}). In particular, the G5 NEAs have all been previously identified as overlapping with the resonant meteoroid branch of the Southern Taurid stream detected by \cite{Spurny2017}. We therefore investigated the G5 group more carefully, and identified a possible rapprochement between 2P/Encke and 2004 TG10, as well as between these two bodies and 2005 TF50, 2005 UR and 2015 TX24 during a few hundred year period around 3200 BCE (and, to a lesser extent, around 4700 BCE). We refer to these possible orbital convergences circa 3200 BCE and 4700 BCE as the $R_1$ and $R_2$ rapprochements. Our analysis of 2005 UY6 and 2019 UN12 did not show any significant rapprochement with the members of the G5. 
 
Extending our analysis to 16 high-quality fireballs observed by \cite{Spurny2017} (comprising 15 members of a new branch of meteoroids trapped into the 7:2 MMR with Jupiter and a comparator non-resonant Southern Taurid meteor) revealed that 9 of the 16 meteoroids integrated do have low MOID and relative velocity with respect to comet 2P/Encke between 3500 BCE and 3000 BCE, reproducing the scenario observed for the G5 asteroids ($R_1$ rapprochement). For the remaining 6 resonant meteoroids the past behaviour is more uncertain while the comparator STA member does not show any past rapprochements with 2P/Encke.

We therefore find compelling evidence that 2P/Encke, several NEAs associated with the TC and some meteoroids belonging to the Southern Taurid stream display a convergence of their orbital elements five to six thousand years in the past. The reason of such a rapprochement between these objects is not clear. 
This phenomenon could be a marker of the fragmentation of a large parent body a few millennia ago. We examined a number of dynamical effects which might provide a different explanation. We found no clear alternative mechanism; nevertheless it is possible that the orbital convergence is a coincidence resulting from purely dynamical processes. The absence of any spectral data for 2004 TG10, 2005 TF50, 2005 UR and 2015 TX24 means that the question cannot be resolved by this means at this time.
	
On the other hand, though the existence of common dynamical effects affecting the orbital evolution of the G5 members is undeniable, that in itself is insufficient to rule out a common origin between these objects about 5 millennia ago. The disruption of a large parent body (due to a collision or a structural fragmentation) might plausibly have resulted in the dispersion of cm to km-sized objects within the TC, giving birth to the G5 members and filling the 7:2 MMR and nearby phase space with meteoroid streams.

To investigate the impact of such breakup event on the formation of the Taurid meteoroid complex, we simulated the ejection of meteoroids from each G5 orbit around 3200 BCE and integrated them up to the present and concluded that a simplified break-up modelling can not reproduce the full extent of the Taurid meteoroid complex, including all features of the TSC observed in 2015. However, material released during this event could contribute to the core of several Taurid showers detected on Earth, and also fill the 7:2 MMR with meteoroids. However, older meteoroid ejection from the parent(s) object(s) or other sources of dust would be needed to explain the duration of the Taurid showers observed today. 

The hypothetical fragmentation of 2P/Encke's parent body around 3200 BCE is compatible with the giant breakup scenario of \cite{Clube1984}, \cite{Asher1993b} and other work. In this scenario, a 50 to 100 km comet was injected into the inner solar system 20 000 to 30 000 years ago ; after a period of normal outgassing during 10 000 years, the parent body is proposed to have suffered a series of fragmentation and splitting events that produced the NEAs and meteoroid streams of the complex. In this scenario we could imagine the G5 formation as one of the last major splitting events of a larger parent body, and one of the last fragmentation of the residuals of the giant progenitor that cumulatively has created the Taurid meteoroid complex. 

Estimating the initial size of such a giant progenitor would require identifying all the subproducts resulting from its fragmentation and outgassing. This would include the Taurid meteoroid swarm, which may itself include hundred-meter bodies yet to be discovered \citep{Spurny2017,Clark2019}. Such estimate is therefore difficult to produce and beyond the scope of this study; however, some considerations can be gleaned from our dynamical analysis. We first observe that among the 52 bodies integrated, only a few TC candidates displayed a possible past orbital convergence with another object in our sample. In addition to the G5 members and NEAs 2005 UY6 and 2019 UN12, three pairs of asteroids presented an opportunity of rapprochement in the past 20 000 years: \{2019 AN12, 2019 BJ1\}, \{2019 RV3, 2019 TM6\}, and \{2010 TU149, 2003 WP1\} (cf. Figure \ref{fig:map_past}).

Among these 13 objects, we identify three km-sized bodies: 2P/Encke (D$\sim$4.8km), 2004 TG10 (D$\sim$1.32$\pm$0.61 km) and 2005 UY6 \citep[D$\sim$2.249$\pm$1.084 km, cf.][]{Masiero2017}. Four asteroids possess a diameter of a few hundred meters (2005 TF50, 2005 UR, 2015 TX24 and 2010 TU149), while the size of the other NEAs cited above is still undetermined. Thus the small number of possibly TC-linked objects and their relatively small size is insufficient to support the diameter of 50-100 km suggested by \cite{Clube1984} for the giant progenitor. An extension of our analysis to other TC candidates \citep[and especially those belonging to the \emph{Hephaistos group} of][]{Asher1993} might help in resolving this discrepancy. However, our analysis of the TC dynamics suggests that this common parent progenitor, if it exists, might have been a comet of more moderate size. 

In any case, a compositional analysis of the G5 members is necessary either to support the hypothesis of a common origin of these objects, or to reject any linkage between them that is not due to purely dynamical processes. While the recovery of, let alone spectral measurements, will be difficult for most G5 members because of short orbital arcs, the reflectance spectra of 2004 TG10 is most urgently needed to test this proposition. The asteroid is arguably the most significant dynamical linkage in the TC other than 2P/Encke, and combines a considerable diameter of 1.32$\pm$0.61 km \citep{Nugent2015} and an accurate orbit (see Figure \ref{fig:covariances_volume}) that should ease its observation. In addition its albedo of 0.018$\pm$0.037 \citep{Nugent2015}, even if uncertain, suggests a compatibility with a cometary nature. 

We therefore strongly encourage future spectral surveys of the TC asteroids to target in priority 2004 TG10, as well as NEAs 2004 TG10, 2005 TF50, 2005 UR and 2015 TX24. Unfortunately however, because of the uncertain ephemeris of 2005 TF50, 2005 UR and 2015 TX24, such a spectral analysis may first require the recovery of these NEAs. 

\section*{Acknowledgements}

Funding for this work was provided in part through NASA co-operative agreement 80NSSC21M0073. PS and JB were supported by grant no. 19-26232X from the Czech Science Foundation. This work was funded in part by the Natural Sciences and Engineering Research Council of Canada Discovery Grants program (Grants no. RGPIN-2016-04433 \& RGPIN-2018-05659). We highly thank Denis Vida for his help and advice regarding the use of the WMPL library and the covariance computation of the observed Taurid fireballs. We also thank the reviewer, Marcel Popescu, for his comments that helped improving this manuscript.

\section*{Data availability statement}
 	
The data underlying this article will be shared on reasonable request to the corresponding author.




\bibliographystyle{mnras}
\bibliography{references} 

\begin{thebibliography}{}
\makeatletter
\relax
\def\mn@urlcharsother{\let\do\@makeother \do\$\do\&\do\#\do\^\do\_\do\%\do\~}
\def\mn@doi{\begingroup\mn@urlcharsother \@ifnextchar [ {\mn@doi@}
  {\mn@doi@[]}}
\def\mn@doi@[#1]#2{\def\@tempa{#1}\ifx\@tempa\@empty \href
  {http://dx.doi.org/#2} {doi:#2}\else \href {http://dx.doi.org/#2} {#1}\fi
  \endgroup}
\def\mn@eprint#1#2{\mn@eprint@#1:#2::\@nil}
\def\mn@eprint@arXiv#1{\href {http://arxiv.org/abs/#1} {{\tt arXiv:#1}}}
\def\mn@eprint@dblp#1{\href {http://dblp.uni-trier.de/rec/bibtex/#1.xml}
  {dblp:#1}}
\def\mn@eprint@#1:#2:#3:#4\@nil{\def\@tempa {#1}\def\@tempb {#2}\def\@tempc
  {#3}\ifx \@tempc \@empty \let \@tempc \@tempb \let \@tempb \@tempa \fi \ifx
  \@tempb \@empty \def\@tempb {arXiv}\fi \@ifundefined
  {mn@eprint@\@tempb}{\@tempb:\@tempc}{\expandafter \expandafter \csname
  mn@eprint@\@tempb\endcsname \expandafter{\@tempc}}}

\bibitem[\protect\citeauthoryear{{Asher}}{{Asher}}{1991}]{Asher1991}
{Asher} D.~J.,  1991, PhD thesis, Oxford Univ. (England).

\bibitem[\protect\citeauthoryear{{Asher} \& {Clube}}{{Asher} \&
  {Clube}}{1993}]{Asher1993b}
{Asher} D.~J.,  {Clube} S.~V.~M.,  1993, \qjras, \href
  {https://ui-adsabs-harvard-edu.ezproxy.obspm.fr/abs/1993QJRAS..34..481A} {34,
  481}

\bibitem[\protect\citeauthoryear{{Asher} \& {Izumi}}{{Asher} \&
  {Izumi}}{1998}]{Asher1998}
{Asher} D.~J.,  {Izumi} K.,  1998, \mn@doi [\mnras]
  {10.1046/j.1365-8711.1998.01395.x}, \href
  {https://ui-adsabs-harvard-edu.ezproxy.obspm.fr/abs/1998MNRAS.297...23A}
  {297, 23}

\bibitem[\protect\citeauthoryear{{Asher}, {Clube}  \& {Steel}}{{Asher}
  et~al.}{1993}]{Asher1993}
{Asher} D.~J.,  {Clube} S.~V.~M.,   {Steel} D.~I.,  1993, \mn@doi [\mnras]
  {10.1093/mnras/264.1.93}, \href
  {https://ui.adsabs.harvard.edu/abs/1993MNRAS.264...93A} {264, 93}

\bibitem[\protect\citeauthoryear{{Asher}, {Clube}, {Napier}  \&
  {Steel}}{{Asher} et~al.}{1994}]{Asher1994}
{Asher} D.~J.,  {Clube} S.~V.~M.,  {Napier} W.~M.,   {Steel} D.~I.,  1994,
  \mn@doi [Vistas in Astronomy] {10.1016/0083-6656(94)90002-7}, \href
  {https://ui.adsabs.harvard.edu/abs/1994VA.....38....1A} {38, 1}

\bibitem[\protect\citeauthoryear{{Babadzhanov}}{{Babadzhanov}}{2001}]{Babadzhanov2001}
{Babadzhanov} P.~B.,  2001, \mn@doi [\aap] {10.1051/0004-6361:20010583}, \href
  {https://ui.adsabs.harvard.edu/abs/2001A%26A...373..329B} {373, 329}

\bibitem[\protect\citeauthoryear{Babadzhanov \& Obrubov}{Babadzhanov \&
  Obrubov}{1992}]{Babadzhanov1992}
Babadzhanov P.~B.,  Obrubov Y.,  1992, Celestial Mechanics and Dynamical, 54,
  111

\bibitem[\protect\citeauthoryear{{Babadzhanov}, {Williams}  \&
  {Kokhirova}}{{Babadzhanov} et~al.}{2008}]{Babadzhanov2008}
{Babadzhanov} P.~B.,  {Williams} I.~P.,   {Kokhirova} G.~I.,  2008, \mn@doi
  [\mnras] {10.1111/j.1365-2966.2008.13096.x}, \href
  {https://ui.adsabs.harvard.edu/abs/2008MNRAS.386.1436B} {386, 1436}

\bibitem[\protect\citeauthoryear{Beech, Hargrove  \& Brown}{Beech
  et~al.}{2004}]{Beech2004a}
Beech M.,  Hargrove M.,   Brown P.~G.,  2004, The Observatory, 124, 277

\bibitem[\protect\citeauthoryear{{Binzel}, {Rivkin}, {Stuart}, {Harris}, {Bus}
  \& {Burbine}}{{Binzel} et~al.}{2004}]{Binzel2004}
{Binzel} R.~P.,  {Rivkin} A.~S.,  {Stuart} J.~S.,  {Harris} A.~W.,  {Bus}
  S.~J.,   {Burbine} T.~H.,  2004, \mn@doi [\icarus]
  {10.1016/j.icarus.2004.04.004}, \href
  {https://ui-adsabs-harvard-edu.ezproxy.obspm.fr/abs/2004Icar..170..259B}
  {170, 259}

\bibitem[\protect\citeauthoryear{{Boehnhardt}}{{Boehnhardt}}{2004}]{Boehnhardt2004}
{Boehnhardt} H.,  2004, in {Festou} M.~C.,  {Keller} H.~U.,   {Weaver} H.~A.,
  eds, , Comets II.
{University of Arizona Press, Tucson}, pp 301--316

\bibitem[\protect\citeauthoryear{{Borovicka}}{{Borovicka}}{1990}]{Borovicka1990}
{Borovicka} J.,  1990, Bulletin of the Astronomical Institutes of
  Czechoslovakia, \href
  {https://ui-adsabs-harvard-edu.ezproxy.obspm.fr/abs/1990BAICz..41..391B} {41,
  391}

\bibitem[\protect\citeauthoryear{{Borovi{\v{c}}ka} \&
  {Spurn{\'y}}}{{Borovi{\v{c}}ka} \& {Spurn{\'y}}}{2020}]{Borovicka2020}
{Borovi{\v{c}}ka} J.,  {Spurn{\'y}} P.,  2020, \mn@doi [\planss]
  {10.1016/j.pss.2020.104849}, \href
  {https://ui-adsabs-harvard-edu.ezproxy.obspm.fr/abs/2020P&SS..18204849B}
  {182, 104849}

\bibitem[\protect\citeauthoryear{Bottke, Vokrouhlick{\'{y}}, Rubincam  \&
  Nesvorn{\'{y}}}{Bottke et~al.}{2006}]{Bottke2006a}
Bottke W.,  Vokrouhlick{\'{y}} D.,  Rubincam D.,   Nesvorn{\'{y}} D.,  2006,
  \mn@doi [Annual Review of Earth and Planetary Sciences]
  {10.1146/annurev.earth.34.031405.125154}, 34, 157

\bibitem[\protect\citeauthoryear{{Brown}, {Weryk}, {Wong}  \& {Jones}}{{Brown}
  et~al.}{2008}]{Brown2008}
{Brown} P.,  {Weryk} R.~J.,  {Wong} D.~K.,   {Jones} J.,  2008, Earth Moon and
  Planets, \href
  {https://ui-adsabs-harvard-edu.ezproxy.obspm.fr/abs/2008EM&P..102..209B}
  {102, 209}

\bibitem[\protect\citeauthoryear{{Brown}, {Wong}, {Weryk}  \&
  {Wiegert}}{{Brown} et~al.}{2010}]{Brown2010}
{Brown} P.,  {Wong} D.~K.,  {Weryk} R.~J.,   {Wiegert} P.,  2010, \mn@doi
  [\icarus] {10.1016/j.icarus.2009.11.015}, \href
  {https://ui.adsabs.harvard.edu/abs/2010Icar..207...66B} {207, 66}

\bibitem[\protect\citeauthoryear{{Ceplecha}}{{Ceplecha}}{1988}]{Ceplecha1988}
{Ceplecha} Z.,  1988, Bulletin of the Astronomical Institutes of
  Czechoslovakia, \href
  {https://ui-adsabs-harvard-edu.ezproxy.obspm.fr/abs/1988BAICz..39..221C} {39,
  221}

\bibitem[\protect\citeauthoryear{{Ceplecha} \& {McCrosky}}{{Ceplecha} \&
  {McCrosky}}{1976}]{Ceplecha1976}
{Ceplecha} Z.,  {McCrosky} R.~E.,  1976, \mn@doi [\jgr]
  {10.1029/JB081i035p06257}, \href
  {https://ui-adsabs-harvard-edu.ezproxy.obspm.fr/abs/1976JGR....81.6257C} {81,
  6257}

\bibitem[\protect\citeauthoryear{Clark \& Wiegert}{Clark \&
  Wiegert}{2011}]{Clark2011}
Clark D.,  Wiegert P.,  2011, \mn@doi [Meteoritics & Planetary Science]
  {10.1111/j.1945-5100.2011.01226.x}, 46, 1217

\bibitem[\protect\citeauthoryear{{Clark}, {Wiegert}  \& {Brown}}{{Clark}
  et~al.}{2019a}]{clawiebro19}
{Clark} D.~L.,  {Wiegert} P.,   {Brown} P.~G.,  2019a, \mn@doi [\mnras]
  {10.1093/mnrasl/slz076}, \href
  {https://ui.adsabs.harvard.edu/abs/2019MNRAS.487L..35C} {487, L35}

\bibitem[\protect\citeauthoryear{{Clark}, {Wiegert}  \& {Brown}}{{Clark}
  et~al.}{2019b}]{Clark2019}
{Clark} D.~L.,  {Wiegert} P.,   {Brown} P.~G.,  2019b, \mn@doi [\mnras]
  {10.1093/mnrasl/slz076}, \href
  {https://ui-adsabs-harvard-edu.ezproxy.obspm.fr/abs/2019MNRAS.487L..35C}
  {487, L35}

\bibitem[\protect\citeauthoryear{{Clube} \& {Napier}}{{Clube} \&
  {Napier}}{1984}]{Clube1984}
{Clube} S.~V.~M.,  {Napier} W.~M.,  1984, \mn@doi [\mnras]
  {10.1093/mnras/211.4.953}, \href
  {https://ui-adsabs-harvard-edu.ezproxy.obspm.fr/abs/1984MNRAS.211..953C}
  {211, 953}

\bibitem[\protect\citeauthoryear{{Crifo} \& {Rodionov}}{{Crifo} \&
  {Rodionov}}{1997}]{Crifo1997}
{Crifo} J.~F.,  {Rodionov} A.~V.,  1997, \mn@doi [Icarus]
  {10.1006/icar.1997.5690}, \href
  {http://cdsads.u-strasbg.fr/abs/1997Icar..127..319C} {127, 319}

\bibitem[\protect\citeauthoryear{Dmitriev, Lupovka  \& Gritsevich}{Dmitriev
  et~al.}{2015}]{Dmitriev2015}
Dmitriev V.,  Lupovka V.,   Gritsevich M.,  2015, \mn@doi [Planetary and Space
  Science] {10.1016/j.pss.2015.06.015}, 117, 223

\bibitem[\protect\citeauthoryear{{Drummond}}{{Drummond}}{1981}]{Drummond1981}
{Drummond} J.~D.,  1981, \mn@doi [\icarus] {10.1016/0019-1035(81)90020-8},
  \href
  {https://ui-adsabs-harvard-edu.ezproxy.obspm.fr/abs/1981Icar...45..545D} {45,
  545}

\bibitem[\protect\citeauthoryear{{Dubietis} \& {Arlt}}{{Dubietis} \&
  {Arlt}}{2007}]{Dubietis2007}
{Dubietis} A.,  {Arlt} R.,  2007, \mn@doi [\mnras]
  {10.1111/j.1365-2966.2007.11488.x}, \href
  {https://ui-adsabs-harvard-edu.ezproxy.obspm.fr/abs/2007MNRAS.376..890D}
  {376, 890}

\bibitem[\protect\citeauthoryear{{Dumitru}, {Birlan}, {Popescu}  \&
  {Nedelcu}}{{Dumitru} et~al.}{2017}]{Dumitru2017}
{Dumitru} B.~A.,  {Birlan} M.,  {Popescu} M.,   {Nedelcu} D.~A.,  2017, \mn@doi
  [\aap] {10.1051/0004-6361/201730813}, \href
  {https://ui-adsabs-harvard-edu.ezproxy.obspm.fr/abs/2017A&A...607A...5D}
  {607, A5}

\bibitem[\protect\citeauthoryear{{Egal}, {Wiegert}, {Brown}, {Moser},
  {Campbell-Brown}, {Moorhead}, {Ehlert}  \& {Moticska}}{{Egal}
  et~al.}{2019}]{Egal2019}
{Egal} A.,  {Wiegert} P.,  {Brown} P.~G.,  {Moser} D.~E.,  {Campbell-Brown} M.,
   {Moorhead} A.,  {Ehlert} S.,   {Moticska} N.,  2019, \mn@doi [\icarus]
  {10.1016/j.icarus.2019.04.021}, \href
  {https://ui.adsabs.harvard.edu/abs/2019Icar..330..123E} {330, 123}

\bibitem[\protect\citeauthoryear{{Egal}, {Wiegert}, {Brown}, {Campbell-Brown}
  \& {Vida}}{{Egal} et~al.}{2020}]{Egal2020}
{Egal} A.,  {Wiegert} P.,  {Brown} P.~G.,  {Campbell-Brown} M.,   {Vida} D.,
  2020, \mn@doi [\aap] {10.1051/0004-6361/202038953}, \href
  {https://ui-adsabs-harvard-edu.ezproxy.obspm.fr/abs/2020A&A...642A.120E}
  {642, A120}

\bibitem[\protect\citeauthoryear{Everhart}{Everhart}{1985}]{Everhart1985}
Everhart E.,  1985, International Astronomical Union Colloquium, 83, 185–202

\bibitem[\protect\citeauthoryear{Hyodo \& Genda}{Hyodo \&
  Genda}{2020}]{Hyodo2020}
Hyodo R.,  Genda H.,  2020, \mn@doi [arXiv] {10.3847/1538-4357/ab9897}, 898, 30

\bibitem[\protect\citeauthoryear{{Jansen-Sturgeon}, {Sansom}  \&
  {Bland}}{{Jansen-Sturgeon} et~al.}{2019}]{JansenSturgeon2019}
{Jansen-Sturgeon} T.,  {Sansom} E.~K.,   {Bland} P.~A.,  2019, \mn@doi
  [Meteoritics and Planetary Science] {10.1111/maps.13376}, \href
  {https://ui-adsabs-harvard-edu.ezproxy.obspm.fr/abs/2019M&PS...54.2149J} {54,
  2149}

\bibitem[\protect\citeauthoryear{{Jenniskens}}{{Jenniskens}}{2006}]{Jenniskens2006}
{Jenniskens} P.,  2006, {Meteor Showers and their Parent Comets}.
Cambridge University Press, Cambridge, UK

\bibitem[\protect\citeauthoryear{Jones}{Jones}{1986}]{Jones1986}
Jones J.,  1986, \mn@doi [Monthly Notices of the Royal Astronomical Society]
  {10.1093/mnras/221.2.257}, 221, 257

\bibitem[\protect\citeauthoryear{{Jopek}}{{Jopek}}{2011}]{Jopek2011}
{Jopek} T.~J.,  2011, \memsai, \href
  {https://ui.adsabs.harvard.edu/abs/2011MmSAI..82..310J} {82, 310}

\bibitem[\protect\citeauthoryear{{Lamy}, {Toth}, {Fernandez}  \&
  {Weaver}}{{Lamy} et~al.}{2004}]{Lamy2004}
{Lamy} P.~L.,  {Toth} I.,  {Fernandez} Y.~R.,   {Weaver} H.~A.,  2004, in
  {Festou} M.~C.,  {Keller} H.~U.,   {Weaver} H.~A.,  eds, , Comets II.
{University of Arizona Press, Tucson}, p.~223

\bibitem[\protect\citeauthoryear{{Levison}, {Terrell}, {Wiegert}, {Dones}  \&
  {Duncan}}{{Levison} et~al.}{2006}]{Levison2006}
{Levison} H.~F.,  {Terrell} D.,  {Wiegert} P.~A.,  {Dones} L.,   {Duncan}
  M.~J.,  2006, \mn@doi [\icarus] {10.1016/j.icarus.2005.12.016}, \href
  {https://ui-adsabs-harvard-edu.ezproxy.obspm.fr/abs/2006Icar..182..161L}
  {182, 161}

\bibitem[\protect\citeauthoryear{Marsden \& Sekanina}{Marsden \&
  Sekanina}{1973}]{Marsden1973}
Marsden B.~G.,  Sekanina Z.,  1973, The Astronomical Journal, 78

\bibitem[\protect\citeauthoryear{{Masiero} et~al.,}{{Masiero}
  et~al.}{2017}]{Masiero2017}
{Masiero} J.~R.,  et~al., 2017, \mn@doi [\aj] {10.3847/1538-3881/aa89ec}, \href
  {https://ui.adsabs.harvard.edu/abs/2017AJ....154..168M} {154, 168}

\bibitem[\protect\citeauthoryear{{McBeath}}{{McBeath}}{1999}]{McBeath1999}
{McBeath} A.,  1999, WGN, Journal of the International Meteor Organization,
  \href
  {https://ui-adsabs-harvard-edu.ezproxy.obspm.fr/abs/1999JIMO...27...53M} {27,
  53}

\bibitem[\protect\citeauthoryear{{Napier}}{{Napier}}{2010}]{Napier2010}
{Napier} W.~M.,  2010, \mn@doi [\mnras] {10.1111/j.1365-2966.2010.16579.x},
  \href
  {https://ui-adsabs-harvard-edu.ezproxy.obspm.fr/abs/2010MNRAS.405.1901N}
  {405, 1901}

\bibitem[\protect\citeauthoryear{{Napier}}{{Napier}}{2019}]{Napier2019}
{Napier} W.~M.,  2019, \mn@doi [\mnras] {10.1093/mnras/stz1769}, \href
  {https://ui-adsabs-harvard-edu.ezproxy.obspm.fr/abs/2019MNRAS.488.1822N}
  {488, 1822}

\bibitem[\protect\citeauthoryear{{Napier}, {Asher}, {Bailey}  \&
  {Steel}}{{Napier} et~al.}{2015}]{Napier2015}
{Napier} B.,  {Asher} D.,  {Bailey} M.,   {Steel} D.,  2015, \mn@doi [Astronomy
  and Geophysics] {10.1093/astrogeo/atv198}, \href
  {https://ui-adsabs-harvard-edu.ezproxy.obspm.fr/abs/2015A&G....56f6.24N} {56,
  6.24}

\bibitem[\protect\citeauthoryear{{Nugent} et~al.,}{{Nugent}
  et~al.}{2015}]{Nugent2015}
{Nugent} C.~R.,  et~al., 2015, \mn@doi [\apj] {10.1088/0004-637X/814/2/117},
  \href
  {https://ui-adsabs-harvard-edu.ezproxy.obspm.fr/abs/2015ApJ...814..117N}
  {814, 117}

\bibitem[\protect\citeauthoryear{{Olech} et~al.,}{{Olech}
  et~al.}{2016}]{Olech2016}
{Olech} A.,  et~al., 2016, \mn@doi [\mnras] {10.1093/mnras/stw1261}, \href
  {https://ui.adsabs.harvard.edu/abs/2016MNRAS.461..674O} {461, 674}

\bibitem[\protect\citeauthoryear{{Olsson-Steel}}{{Olsson-Steel}}{1988}]{Olsson1988}
{Olsson-Steel} D.,  1988, \mn@doi [\icarus] {10.1016/0019-1035(88)90127-3},
  \href
  {https://ui-adsabs-harvard-edu.ezproxy.obspm.fr/abs/1988Icar...75...64O} {75,
  64}

\bibitem[\protect\citeauthoryear{{Popescu}, {Birlan}, {Nedelcu}, {Vaubaillon}
  \& {Cristescu}}{{Popescu} et~al.}{2014}]{Popescu2014}
{Popescu} M.,  {Birlan} M.,  {Nedelcu} D.~A.,  {Vaubaillon} J.,   {Cristescu}
  C.~P.,  2014, \mn@doi [\aap] {10.1051/0004-6361/201424064}, \href
  {https://ui-adsabs-harvard-edu.ezproxy.obspm.fr/abs/2014A&A...572A.106P}
  {572, A106}

\bibitem[\protect\citeauthoryear{{Porub{\v c}an}, {Williams}  \& {Korno{\v
  s}}}{{Porub{\v c}an} et~al.}{2004}]{Porubcan2004}
{Porub{\v c}an} V.,  {Williams} I.~P.,   {Korno{\v s}} L.,  2004, \mn@doi
  [Earth Moon and Planets] {10.1007/s11038-005-2243-5}, \href
  {https://ui.adsabs.harvard.edu/abs/2004EM%26P...95..697P} {95, 697}

\bibitem[\protect\citeauthoryear{{Porub{\v c}an}, {Korno{\v s}}  \&
  {Williams}}{{Porub{\v c}an} et~al.}{2006}]{Porubcan2006}
{Porub{\v c}an} V.,  {Korno{\v s}} L.,   {Williams} I.~P.,  2006, Contributions
  of the Astronomical Observatory Skalnate Pleso, \href
  {https://ui.adsabs.harvard.edu/abs/2006CoSka..36..103P} {36, 103}

\bibitem[\protect\citeauthoryear{{Southworth} \& {Hawkins}}{{Southworth} \&
  {Hawkins}}{1963}]{Southworth1963}
{Southworth} R.~B.,  {Hawkins} G.~S.,  1963, Smithsonian Contributions to
  Astrophysics, \href
  {https://ui-adsabs-harvard-edu.ezproxy.obspm.fr/abs/1963SCoA....7..261S} {7,
  261}

\bibitem[\protect\citeauthoryear{{Spurn{\'y}}, {Borovi{\v c}ka}, {Mucke}  \&
  {Svore{\v n}}}{{Spurn{\'y}} et~al.}{2017}]{Spurny2017}
{Spurn{\'y}} P.,  {Borovi{\v c}ka} J.,  {Mucke} H.,   {Svore{\v n}} J.,  2017,
  \mn@doi [\aap] {10.1051/0004-6361/201730787}, \href
  {https://ui.adsabs.harvard.edu/abs/2017A%26A...605A..68S} {605, A68}

\bibitem[\protect\citeauthoryear{{Steel} \& {Asher}}{{Steel} \&
  {Asher}}{1996a}]{Steel1996b}
{Steel} D.~I.,  {Asher} D.~J.,  1996a, \mn@doi [\mnras]
  {10.1093/mnras/280.3.806}, \href
  {https://ui-adsabs-harvard-edu.ezproxy.obspm.fr/abs/1996MNRAS.280..806S}
  {280, 806}

\bibitem[\protect\citeauthoryear{{Steel} \& {Asher}}{{Steel} \&
  {Asher}}{1996b}]{Steel1996}
{Steel} D.~I.,  {Asher} D.~J.,  1996b, \mn@doi [\mnras]
  {10.1093/mnras/281.3.937}, \href
  {https://ui-adsabs-harvard-edu.ezproxy.obspm.fr/abs/1996MNRAS.281..937S}
  {281, 937}

\bibitem[\protect\citeauthoryear{{Steel}, {Asher}  \& {Clube}}{{Steel}
  et~al.}{1991}]{Steel1991}
{Steel} D.~I.,  {Asher} D.~J.,   {Clube} S.~V.~M.,  1991, \mn@doi [\mnras]
  {10.1093/mnras/251.4.632}, \href
  {https://ui-adsabs-harvard-edu.ezproxy.obspm.fr/abs/1991MNRAS.251..632S}
  {251, 632}

\bibitem[\protect\citeauthoryear{Stohl}{Stohl}{1983}]{Stohl1983}
Stohl J.,  1983, in Asteroids, Comets, and Meteors. \url
  {http://adsabs.harvard.edu/full/1983acm..proc..419S}

\bibitem[\protect\citeauthoryear{{\v{S}}tohl \& Porub{\v{c}}an}{{\v{S}}tohl \&
  Porub{\v{c}}an}{1992}]{Stohl1992}
{\v{S}}tohl J.,  Porub{\v{c}}an V.,  1992, \mn@doi [Symposium - International
  Astronomical Union] {DOI: 10.1017/S0074180900091324}, 152, 315

\bibitem[\protect\citeauthoryear{{Tubiana}, {Snodgrass}, {Michelsen}, {Haack},
  {B{\"o}hnhardt}, {Fitzsimmons}  \& {Williams}}{{Tubiana}
  et~al.}{2015}]{Tubiana2015}
{Tubiana} C.,  {Snodgrass} C.,  {Michelsen} R.,  {Haack} H.,  {B{\"o}hnhardt}
  H.,  {Fitzsimmons} A.,   {Williams} I.~P.,  2015, \mn@doi [\aap]
  {10.1051/0004-6361/201425512}, \href
  {https://ui-adsabs-harvard-edu.ezproxy.obspm.fr/abs/2015A&A...584A..97T}
  {584, A97}

\bibitem[\protect\citeauthoryear{{Valsecchi}}{{Valsecchi}}{1999}]{Valsecchi1999}
{Valsecchi} G.~B.,  1999, in {Svoren} J.,  {Pittich} E.~M.,   {Rickman} H.,
  eds, IAU Colloq. 173: Evolution and Source Regions of Asteroids and Comets.
  p.~353

\bibitem[\protect\citeauthoryear{{Valsecchi}, {Morbidelli}, {Gonczi},
  {Farinella}, {Froeschle}  \& {Froeschle}}{{Valsecchi}
  et~al.}{1995}]{Valsecchi1995}
{Valsecchi} G.~B.,  {Morbidelli} A.,  {Gonczi} R.,  {Farinella} P.,
  {Froeschle} C.,   {Froeschle} C.,  1995, \mn@doi [\icarus]
  {10.1006/icar.1995.1183}, \href
  {https://ui-adsabs-harvard-edu.ezproxy.obspm.fr/abs/1995Icar..118..169V}
  {118, 169}

\bibitem[\protect\citeauthoryear{Vida, Brown  \& Campbell-Brown}{Vida
  et~al.}{2017}]{vida2017}
Vida D.,  Brown P.~G.,   Campbell-Brown M.,  2017, \mn@doi [Icarus]
  {10.1016/j.icarus.2017.06.020}, 296, 197

\bibitem[\protect\citeauthoryear{Vida, Gural, Brown, Campbell-Brown  \&
  Wiegert}{Vida et~al.}{2019}]{Vida2019}
Vida D.,  Gural P.~S.,  Brown P.~G.,  Campbell-Brown M.,   Wiegert P.,  2019,
  \mn@doi [Monthly Notices of the Royal Astronomical Society]
  {10.1093/mnras/stz3160}, 491, 2688

\bibitem[\protect\citeauthoryear{Vida, Gural, Brown, Campbell-Brown  \&
  Wiegert}{Vida et~al.}{2020}]{vida2020a}
Vida D.,  Gural P.,  Brown P.~G.,  Campbell-Brown M.,   Wiegert P.,  2020,
  \mn@doi [Monthly Notices of the Royal Astronomical Society]
  {10.1093/mnras/stz3160}, 491, 2688

\bibitem[\protect\citeauthoryear{{Whipple}}{{Whipple}}{1940}]{Whipple1940}
{Whipple} F.~L.,  1940, Proceedings of the American Philosophical Society,
  \href
  {https://ui-adsabs-harvard-edu.ezproxy.obspm.fr/abs/1940PAPhS..83..711W} {83,
  711}

\bibitem[\protect\citeauthoryear{Whipple}{Whipple}{1967}]{Whipple1967}
Whipple F.,  1967, in Weinberg J.,  ed., The Zodiacal Light and the
  Interplanetary Medium. NASA SP-150, pp 409--426

\bibitem[\protect\citeauthoryear{{Whipple} \& {El-Din Hamid}}{{Whipple} \&
  {El-Din Hamid}}{1952}]{Whipple1952}
{Whipple} F.~L.,  {El-Din Hamid} S.,  1952, Helwan Institute of Astronomy and
  Geophysics Bulletins, \href
  {https://ui-adsabs-harvard-edu.ezproxy.obspm.fr/abs/1952HelOB..41....3W} {41,
  3}

\bibitem[\protect\citeauthoryear{Wiegert, Vaubaillon  \&
  Campbell-Brown}{Wiegert et~al.}{2009}]{Wiegert2009}
Wiegert P.,  Vaubaillon J.,   Campbell-Brown M.,  2009, \mn@doi [Icarus]
  {10.1016/j.icarus.2008.12.030}, 201, 295

\makeatother
\end{thebibliography}


\begin{thebibliography}{}
\makeatletter
\relax
\def\mn@urlcharsother{\let\do\@makeother \do\$\do\&\do\#\do\^\do\_\do\%\do\~}
\def\mn@doi{\begingroup\mn@urlcharsother \@ifnextchar [ {\mn@doi@}
  {\mn@doi@[]}}
\def\mn@doi@[#1]#2{\def\@tempa{#1}\ifx\@tempa\@empty \href
  {http://dx.doi.org/#2} {doi:#2}\else \href {http://dx.doi.org/#2} {#1}\fi
  \endgroup}
\def\mn@eprint#1#2{\mn@eprint@#1:#2::\@nil}
\def\mn@eprint@arXiv#1{\href {http://arxiv.org/abs/#1} {{\tt arXiv:#1}}}
\def\mn@eprint@dblp#1{\href {http://dblp.uni-trier.de/rec/bibtex/#1.xml}
  {dblp:#1}}
\def\mn@eprint@#1:#2:#3:#4\@nil{\def\@tempa {#1}\def\@tempb {#2}\def\@tempc
  {#3}\ifx \@tempc \@empty \let \@tempc \@tempb \let \@tempb \@tempa \fi \ifx
  \@tempb \@empty \def\@tempb {arXiv}\fi \@ifundefined
  {mn@eprint@\@tempb}{\@tempb:\@tempc}{\expandafter \expandafter \csname
  mn@eprint@\@tempb\endcsname \expandafter{\@tempc}}}

\bibitem[\protect\citeauthoryear{{Asher}}{{Asher}}{1991}]{Asher1991}
{Asher} D.~J.,  1991, PhD thesis, Oxford Univ. (England).

\bibitem[\protect\citeauthoryear{{Spurn{\'y}}, {Borovi{\v c}ka}, {Mucke}  \&
  {Svore{\v n}}}{{Spurn{\'y}} et~al.}{2017}]{Spurny2017}
{Spurn{\'y}} P.,  {Borovi{\v c}ka} J.,  {Mucke} H.,   {Svore{\v n}} J.,  2017,
  \mn@doi [\aap] {10.1051/0004-6361/201730787}, \href
  {https://ui.adsabs.harvard.edu/abs/2017A%26A...605A..68S} {605, A68}

\makeatother
\end{thebibliography}



\appendix

\section{Initial conditions} \label{appendix:IC}

\begin{table*}
    \centering
    \resizebox{\textwidth}{!}{
    \begin{tabular}{lccrrrrrrrrr}
   \large Body &\large JD &\large e &\large q (AU) & \large i ($\degree$) &\large $\Omega$ ($\degree$) &\large $\omega$ ($\degree$) &\large $\sigma_e$\hspace{8pt}\mbox{ } &\large $\sigma_q$ (AU) & \large$\sigma_i$ ($\degree$) &\large $\sigma_\Omega$ ($\degree$) & \large$\sigma_\omega$  ($\degree$)\\
    \hline
    \hline
    & & & & & & & & & & & \\[-0.2cm]
    1959 LM    & 2459000.5 & 0.6341 & 0.7255 & 6.70404 & 294.8675 & 236.3490 & 5.848e-16 & 2.251e-15 & 1.096e-11 & 7.307e-10 & 7.356e-10 \\ 
    1984 KB    & 2459000.5 & 0.7662 & 0.5173 & 4.9184 & 169.3965 & 337.1379 & 7.253e-09  & 1.6472e-08 & 1.4185e-06 & 4.5665e-05 & 4.501e-05  \\ 
    1987 SB    & 2459000.5 & 0.6630 & 0.7411 & 3.03970 & 82.2363  & 168.9065 & 4.378e-16 & 2.165e-15 & 4.052e-12 & 3.472e-09 & 3.365e-09 \\ 
    1991 VL    & 2459000.5 & 0.7722 & 0.4176 & 9.03054 & 309.4920 & 227.7858 & 9.269e-16 & 3.086e-15 & 1.189e-11 & 7.383e-10 & 6.882e-10 \\ 
    1996 RG3   & 2459000.5 & 0.6050 & 0.7900 & 3.57169 & 158.1866 & 300.1153 & 8.536e-16 & 3.402e-15 & 5.978e-11 & 9.888e-09 & 1.016e-08 \\ 
    1998 QS52  & 2459000.5 & 0.8583 & 0.3121 & 17.5399 & 260.4250 & 242.9564 & 2.7539e-08 & 6.0635e-08 & 5.6174e-06 & 2.7625e-05 & 2.6143e-05 \\ 
    1998 SS49  & 2459000.5 & 0.6393 & 0.6940 & 10.7637 & 41.4923 & 102.4329 & 3.8306e-08 & 7.3768e-08 & 4.5762e-06 & 1.8719e-05 & 1.8992e-05 \\ 
    1998 VD31  & 2459000.5 & 0.8060 & 0.5150 & 11.5908 & 37.2749  & 123.6114 & 3.393e-13 & 2.385e-12 & 1.442e-09 & 1.869e-09 & 2.888e-08 \\ 
    1999 RK45  & 2456485.5 & 0.7732 & 0.3624 & 5.8937 & 120.0317 & 4.0833 & 4.3675e-07 & 6.9317e-07 & 1.8875e-05 & 0.00015968 & 0.00017501 \\ 
    1999 VK12  & 2451493.5 & 0.7760 & 0.5022 & 9.5141 & 48.9555 & 102.7292 & 0.013103   & 0.0042698  & 0.19118    & 0.024999   & 0.058831   \\ 
    1999 VR6   & 2456103.5 & 0.7581 & 0.5305 & 8.5236 & 213.0010 & 294.0661 & 2.7689e-08 & 6.0245e-08 & 1.1813e-05 & 3.0803e-05 & 3.5565e-05 \\ 
    1999 VT25  & 2459000.5 & 0.5231 & 0.5543 & 5.1483 & 221.9965 & 319.1545 & 9.81e-08   & 1.1393e-07 & 1.5175e-05 & 3.9299e-05 & 4.6095e-05 \\ 
    2001 HB    & 2454840.5 & 0.6939 & 0.4022 & 9.2898 & 195.9979 & 237.7826 & 2.1425e-07 & 2.8413e-07 & 1.9972e-05 & 4.0962e-05 & 3.826e-05  \\ 
    2001 QJ96  & 2456191.5 & 0.7973 & 0.3227 & 5.8560 & 338.8107 & 121.5842 & 4.8301e-06 & 7.664e-06  & 8.6302e-05 & 7.0778e-05 & 0.00079101 \\ 
    2001 VH75  & 2459000.5 & 0.7416 & 0.5428 & 10.6142 & 276.2737 & 244.2313 & 8.2862e-08 & 1.7336e-07 & 1.4392e-05 & 4.4316e-05 & 4.5186e-05 \\ 
    2002 MX    & 2452447.5 & 0.7965 & 0.5105 & 1.9625 & 284.3024 & 237.5817 & 0.0012271  & 0.00034368 & 0.0043164  & 0.019449   & 0.02701    \\ 
    2002 SY50  & 2459000.5 & 0.7416 & 0.5428 & 10.6142 & 276.2737 & 244.2313 & 8.2862e-08 & 1.7336e-07 & 1.4392e-05 & 4.4316e-05 & 4.5186e-05 \\ 
    2002 XM35  & 2452610.5 & 0.8387 & 0.3764 & 3.0633 & 229.9635 & 312.6386 & 0.021409   & 0.010156   & 0.15634    & 0.44092    & 0.17179    \\ 
    2003 QC10  & 2456754.5 & 0.7314 & 0.3688 & 5.0381 & 0.1490 & 120.6716 & 5.2122e-08 & 7.0881e-08 & 5.8868e-06 & 5.8591e-05 & 5.7869e-05 \\ 
    2003 SF    & 2452902.5 & 0.7779 & 0.4803 & 5.7377 & 77.6849 & 31.7699 & 0.0017177  & 0.00060053 & 0.0096907  & 0.11227    & 0.11467    \\ 
    2003 UL3   & 2455783.5 & 0.7982 & 0.4526 & 14.6537 & 153.1540 & 13.0051 & 5.6676e-08 & 1.2313e-07 & 9.5581e-06 & 5.3207e-05 & 6.7675e-05 \\ 
    2003 UV11  & 2455950.5 & 0.7628 & 0.3446 & 5.9267 & 31.9697 & 124.7580 & 1.6716e-08 & 2.4304e-08 & 8.3869e-06 & 5.5045e-06 & 7.0951e-06 \\ 
    2003 WP21  & 2452967.5 & 0.7872 & 0.4888 & 4.3165 & 38.0821 & 123.6485 & 0.01329    & 0.007525   & 0.13534    & 0.103      & 0.1033     \\ 
    2004 WS2   & 2459000.5 & 0.6030 & 0.5306 & 8.2628 & 87.7532 & 115.3459 & 5.7842e-08 & 7.4945e-08 & 3.428e-06  & 3.0602e-05 & 2.9655e-05 \\ 
    2005 NX39  & 2453566.5 & 0.8767 & 0.3031 & 14.1514 & 121.7512 & 38.1503 & 0.00076071 & 0.00023698 & 0.027995   & 0.023406   & 0.029745   \\ 
    2005 TB15  & 2455073.5 & 0.7556 & 0.4432 & 7.2654 & 9.7556 & 138.8093 & 9.4949e-08 & 1.7429e-07 & 1.8159e-05 & 5.105e-05  & 5.3794e-05 \\ 
    2005 UY6   & 2456783.5 & 0.8701 & 0.2932 & 12.1642 & 343.6145 & 180.7515 & 1.0614e-07 & 2.2887e-07 & 2.1819e-05 & 6.1525e-05 & 6.5109e-05 \\ 
    2006 SO198 & 2454009.5 & 0.8632 & 0.2637 & 9.7663 & 10.3858 & 125.3702 & 0.0012063  & 0.0006303  & 0.029545   & 0.00049672 & 0.012531   \\ 
    2007 RU17  & 2459000.5 & 0.8279 & 0.3510 & 9.0810 & 17.3986 & 129.8870 & 1.3955e-07 & 2.7532e-07 & 2.117e-05  & 2.2415e-05 & 3.3496e-05 \\ 
    2007 UL12  & 2455365.5 & 0.8279 & 0.3513 & 9.0740 & 17.54552 & 129.7513 & 1.3363e-07 & 2.7574e-07 & 2.1133e-05 & 2.2559e-05 & 3.3869e-05 \\ 
    2010 TU149 & 2459000.5 & 0.8282 & 0.3782 & 1.9715 & 59.7214 & 91.6957 & 3.9827e-07 & 6.8133e-07 & 1.4177e-05 & 0.00019767 & 0.0001737  \\ 
    2019 AN12  & 2458495.5 & 0.7031 & 0.6636 & 0.2747 & 244.7785 & 319.1332 & 0.0061023  & 0.0016622  & 0.0025561  & 0.191      & 0.18165    \\ 
    2019 BJ1   & 2459000.5 & 0.6699 & 0.7480 & 0.3650 & 148.0145 & 47.9587 & 1.4347e-05 & 3.4682e-06 & 1.8553e-05 & 0.0012072  & 0.0012167  \\ 
    2019 GB    & 2458575.5 & 0.7255 & 0.6422 & 0.3167 & 356.1610 & 111.2738 & 0.00068194 & 0.00019291 & 0.00028175 & 0.0073767  & 0.0081539  \\ 
    2019 JO5   & 2458614.5 & 0.8387 & 0.3578 & 5.0469 & 166.8953 & 297.0251 & 0.0059422  & 0.0025825  & 0.020705   & 0.38857    & 0.37264    \\ 
    2019 RV3   & 2459000.5 & 0.6135 & 0.8655 & 6.9705 & 352.3650 & 125.8658 & 0.000183   & 5.7945e-05 & 0.00083363 & 0.0053286  & 0.013256   \\ 
    2019 TM6   & 2458772.5 & 0.6105 & 0.8682 & 7.8844 & 317.9598 & 160.0473 & 0.0014038  & 0.00021496 & 0.0080342  & 0.052983   & 0.084218   \\ 
    2019 UN12  & 2459000.5 & 0.8419 & 0.3587 & 4.9622 & 226.3733 & 297.8379 & 2.796e-06  & 1.5221e-06 & 2.4215e-05 & 2.1522e-05 & 4.8213e-05 \\ 
    2019 WB5   & 2458815.5 & 0.7282 & 0.5849 & 8.8774 & 91.3470 & 69.4900  & 0.0063041  & 0.0019258  & 0.055501   & 0.051124   & 0.059089   \\ 
    2019 WJ4   & 2459000.5 & 0.6819 & 0.7120 & 10.6840 & 67.9386 & 71.6561  & 0.00056529 & 0.00014003 & 0.0075858  & 8.8627e-05 & 0.00075555 \\ 
    2019 YM    & 2458836.5 & 0.6702 & 0.7110 & 2.2262 & 88.8390 & 70.9899  & 0.00069634 & 0.00019173 & 0.001982   & 0.00076825 & 0.0037177  \\ 
    2020 AL2   & 2458854.5 & 0.7760 & 0.5089 & 3.0577 & 103.6810 & 102.3035 & 0.0029517  & 0.00098471 & 0.013701   & 0.0066257  & 0.0077668  \\ 
    2020 BC8   & 2459000.5 & 0.7480 & 0.5834 & 7.3673 & 318.4827 & 252.7763 & 2.552e-05  & 2.804e-05  & 0.00028641 & 0.0010678  & 0.0025806  \\ 
    2020 DV    & 2458898.5 & 0.5971 & 0.9428 & 0.3718 & 238.5310 & 225.5829 & 0.00296    & 0.0001237  & 0.0015625  & 0.092664   & 0.086664   \\ 
    2020 DX2   & 2458900.5 & 0.6119 & 0.8718 & 9.8584 & 327.3564 & 247.8972 & 0.0041197  & 0.00025579 & 0.046628   & 0.026709   & 0.034924   \\ 
    2020 JV    & 2459000.5 & 0.6363 & 0.8205 & 3.3190 & 282.6038 & 235.6837 & 3.098e-05  & 4.0645e-06 & 0.00018755 & 0.0011263  & 0.0013171  \\ 
    2201 Oljato& 2456571.5 & 0.7126 & 0.6244 & 2.5234 & 75.0078 & 98.1935  & 3.4489e-08 & 7.4987e-08 & 5.1959e-06 & 0.00011779 & 0.00011771 \\ 
    2004 TG10  & 2454041.5 & 0.8594 & 0.3152 & 3.6968 & 212.3165 & 310.0147 & 7.5802e-08 & 1.6871e-07 & 2.3427e-05 & 0.00014631 & 0.0001422  \\ 
    2005 TF50  & 2453663.5 & 0.8711 & 0.2921 & 10.7125 & 0.7351 & 159.7653 & 0.00027346 & 0.00029415 & 0.013431   & 0.019207   & 0.0041268  \\ 
    2005 UR    & 2453668.5 & 0.8817 & 0.2660 & 6.9330 & 20.0281 & 140.4776 & 0.0020397  & 0.0011086  & 0.044026   & 0.04213    & 0.014095   \\ 
    2015 TX24  & 2457307.5 & 0.8722 & 0.2896 & 6.0420 & 33.0112 & 126.9973 & 0.00023245 & 8.433e-05  & 0.0030824  & 0.0024525  & 0.0030072  \\ 
    2P/Encke & 2457097.5 & 0.8483 & 0.3360 & 11.7818 & 334.5678 & 186.5437 & 7.3307e-08 & 1.6324e-07 & 8.5551e-06 & 4.1478e-05 & 4.1911e-05 \\ 
    & & & & & & & & & & & \\[-0.2cm]
    \hline  
    \hline
    \end{tabular}}
    \caption{Osculating elements and 1-$\sigma$ uncertainties of the potential Taurid asteroids examined in our work. Non-gravitational coefficients selected for comet 2P/Encke are $A_1=-2.49\times10^{-11}$, $A_2=-2.69\times10^{-12}$ and $A_3=3.86\times10^{-9}$. This data was retrieved from the Jet Propulsion Laboratory Small-Body Database Browser in April 2021. }
    \label{tab:IC}
\end{table*}

\begin{table*}
    \centering
  \resizebox{\textwidth}{!}{
    \begin{tabular}{ccccccccccrrrrr}
    	    \hline
    	    & & & & & & & & & & & & & &\\[-0.2cm]
    \large Fireball & \large  JD & \large  e & \large  q  &  \large i  & \large  $\Omega$  & \large  $\omega$ & \large  $\sigma_e$ &  \large $\sigma_q$ & \large  $\sigma_i$ & \large  $\sigma_\Omega$ &  \large $\sigma_\omega$  & \large  PE & \large  Mass& \large  Type\\[0.1cm]
     &  &  &\large(AU) &  \large  ($\degree$) & \large ($\degree$) & \large   ($\degree$) & \large &  \large  (AU) & \large ($\degree$) & \large  ($\degree$) &  \large  ($\degree$) &    & \large (kg)  &\\
    \hline
    \hline
    & & & & & & & & & & & & & &\\[-0.2cm]
    EN011115\_013625 & 2457267.566956 & 0.8692 & 0.2973 & 5.4523 & 38.0954 & 120.4350 & 1.432191e-06 & 7.873789e-07 & 0.00174645   & 2.420699e-10 & 0.003584172 & -4.92  &  0.021   &  II   \\
    EN011115\_174410 & 2457268.239006 & 0.8664 & 0.3015 & 5.2057 & 38.7534 & 120.0043 & 8.928923e-07 & 3.604696e-07 & 0.001745804  & 5.150798e-09 & 0.001045232 & -4.62  &   0.0089   &  II   \\
    EN021115\_022525 & 2457268.600986 & 0.8687 & 0.2964 & 5.9180 & 39.1297 & 120.5693 & 9.392982e-07 & 2.75715e-07  & 0.00355279   & 3.783807e-10 & 0.001321566 & -5.57  &   0.2    &    IIIA    \\
    EN021115\_235259 & 2457269.495136 & 0.8612 & 0.3127 & 5.4869 & 40.0240 & 118.6900 & 1.123619e-06 & 3.448058e-07 & 0.0004135622 & 6.757754e-11 & 0.001639336 & -4.68 &    0.0031   &  II  \\
    EN041115\_021452 & 2457270.593666 & 0.8567 & 0.3250 & 5.5567 & 41.1262 & 117.2041 & 5.072631e-07 & 4.443779e-07 & 0.0001851174 & 7.748561e-11 & 0.002772794 & -5.01   &  0.0076   &  II \\
    EN051115\_213433 & 2457272.398996 & 0.8450 & 0.3499 & 5.4537 & 42.9310 & 114.3883 & 1.086207e-06 & 2.092026e-07 & 0.0007907425 & 1.938426e-11 & 0.001201646 & -4.63  &   0.0038  &   II \\
    EN051115\_220108 & 2457272.417456 & 0.8460 & 0.3482 & 5.6481 & 42.9501 & 114.5702 & 7.686731e-07 & 1.976774e-07 & 0.0004270112 & 2.207222e-11 & 0.001353193  &-5.54  &   0.05   &    IIIA  \\
    EN051115\_234939 & 2457272.492816 & 0.8473 & 0.3454 & 5.7606 & 43.0284 & 114.8776 & 1.239752e-06 & 4.449336e-07 & 0.0003865704 & 1.239077e-10 & 0.001340714 & -4.56   &  0.0026  &   I  \\
    EN061115\_040629 & 2457272.671176 & 0.8473 & 0.3456 & 5.8053 & 43.2088 & 114.8465 & 8.639428e-07 & 8.921245e-07 & 7.673269e-05 & 2.961442e-11 & 0.006376054 & -4.89 &    0.01     &  II \\
    EN061115\_164758 & 2457273.200016 & 0.8458 & 0.3458 & 5.4411 & 43.7251 & 114.9068 & 6.672841e-07 & 2.368707e-07 & 0.003560585  & 5.329541e-09 & 0.001322023 & -4.64  &   0.0098    & II \\
    EN081115\_181258 & 2457275.259006 & 0.8377 & 0.3675 & 5.5726 & 45.7954 & 112.3363 & 1.500359e-06 & 1.880955e-07 & 0.001653755  & 6.333231e-10 & 0.000445499 & -5.05   &  0.049  &    II \\
    EN171115\_020907 & 2457283.589666 & 0.8082 & 0.4356 & 4.9554 & 54.1951 & 104.5768 & 5.587592e-06 & 3.013486e-06 & 0.006263584  & 2.62072e-08  & 0.01748613 & -4.93  &   0.0062   &  II  \\
    EN241015\_185031 & 2457260.285086 & 0.8266 & 0.3411 & 5.5229 & 30.8175 & 116.7770 & 8.590657e-07 & 1.597854e-07 & 0.0004497353 & 8.659177e-10 & 0.000297973 & -4.87   &  0.012   &   II \\
    EN251015\_022301 & 2457260.599316 & 0.8956 & 0.2372 & 6.2701 & 31.1425 & 127.7357 & 1.362773e-06 & 6.970074e-07 & 0.002715623  & 1.969699e-10 & 0.002849672 & -5.87 &    0.18   &    IIIB\\
    EN311015\_202117 & 2457267.348116 & 0.8639 & 0.3044 & 5.0745 & 37.8674 & 119.7336 & 1.560010e-06  & 4.758023e-07 & 0.000887986  & 8.314003e-10 & 0.000985649  &-4.43 &    0.007  &    I  \\
    EN311015\_211904 & 2457267.388236 & 0.8741 & 0.2861 & 5.5617 & 37.9098 & 121.7467 & 2.163988e-06 & 2.615601e-07 & 0.004121612  & 1.171538e-09 & 0.02253486 & -5.68  &   0.039 &     IIIB     \\
    & & & & & & & & & & & & & &\\[-0.2cm]
    \hline  
    \hline
    \end{tabular}}
    \caption{Osculating elements and 1-$\sigma$ uncertainties of the 2015 Taurid fireballs integrated based on data from \citep{Spurny2017}. Estimates of the meteoroids mass, PE-criterion and type are also provided. Names of each fireball follow the convention of EN-day-month-year. }
    \label{tab:IC_fireballs}
\end{table*}

Appendices B, C, D, E, F, G \& H are provided as supplementary material and are accessible online. 

\bsp	
\label{lastpage}

\end{document}


\onecolumn
  
  \begin{center}
  \Large \textbf{SUPPLEMENTARY APPENDICES}\\[0.5cm]
  \end{center}

\setcounter{topnumber}{4}
\setcounter{totalnumber}{4}

\newcommand{\fignumprefix}{}
\renewcommand{\thefigure}{\fignumprefix\arabic{figure}}
\renewcommand{\theHfigure}{figure.\thefigure}
\newcommand{\setfignumprefix}[1]{%
	\renewcommand{\fignumprefix}{#1}
	\setcounter{figure}{0}
}

\textbf{APPENDIX B: ORBITAL EVOLUTION}\\ \label{appendix:orb_evolution_all}

\textbf{B1  Semi-major axis and eccentricity of TC candidates}\\

\setfignumprefix{B}
\begin{figure*}[!ht]
  \centering
  \includegraphics[width=0.47\textwidth]{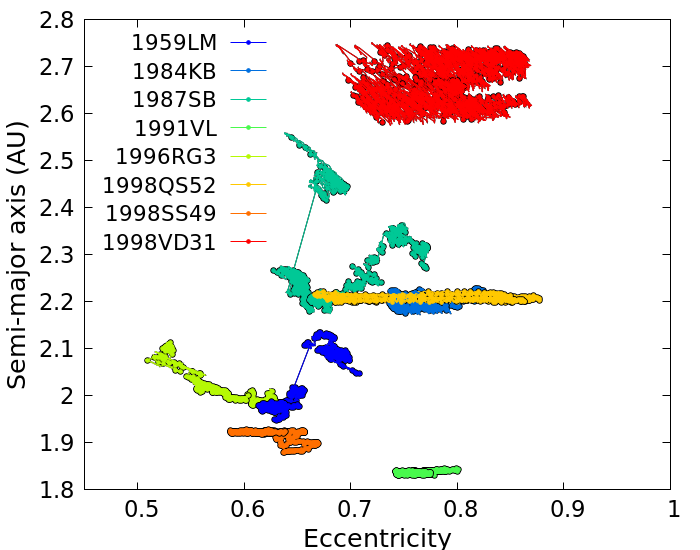}
  \includegraphics[width=0.47\textwidth]{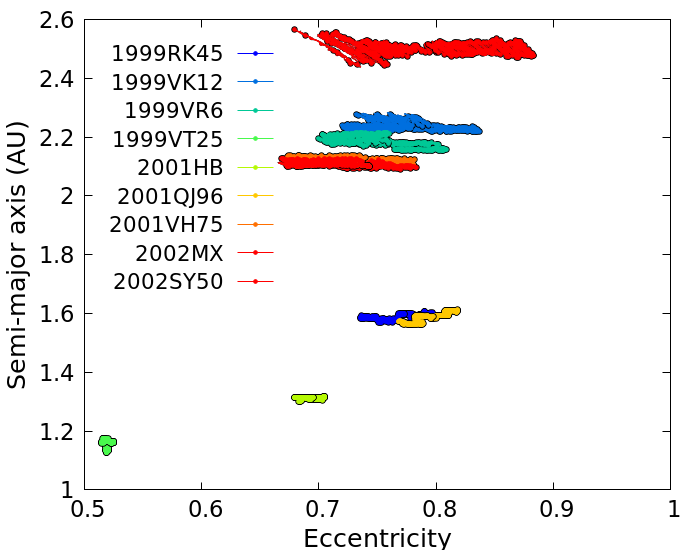}\\
  \includegraphics[width=0.47\textwidth]{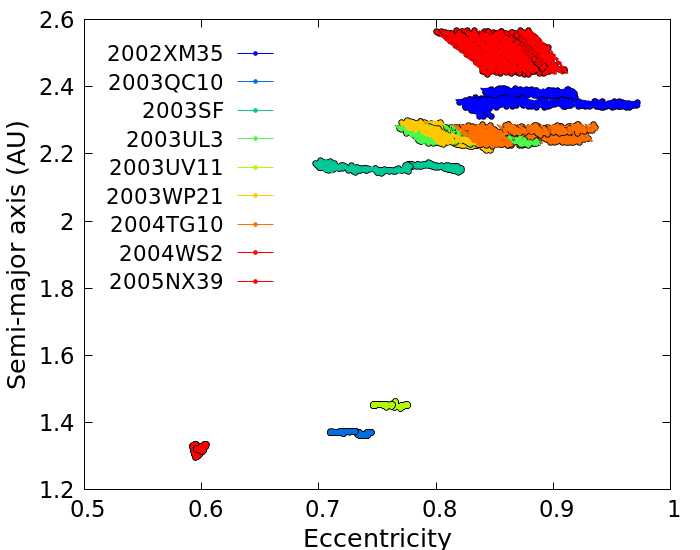}
  \includegraphics[width=0.47\textwidth]{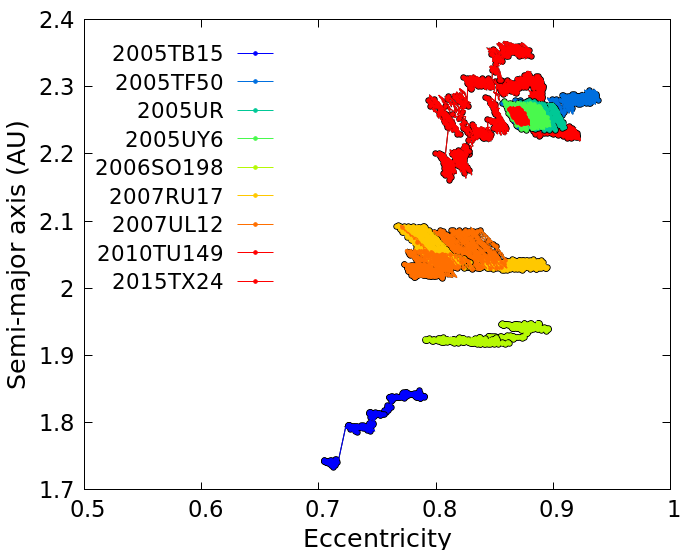}\\
  \includegraphics[width=0.47\textwidth]{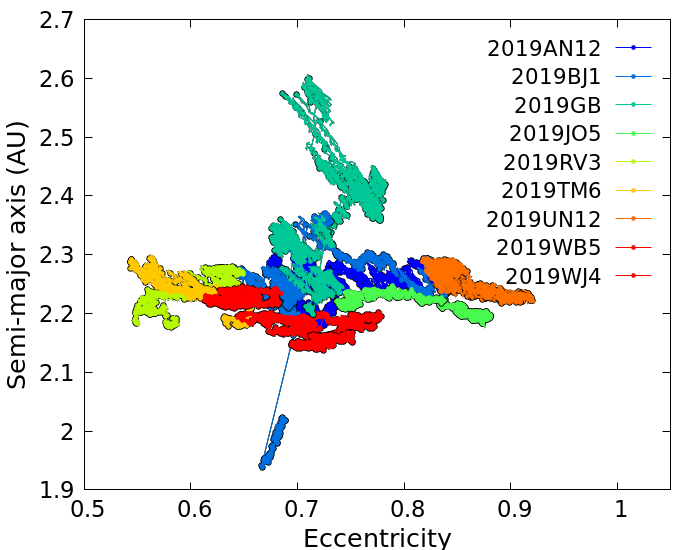}
  \includegraphics[width=0.47\textwidth]{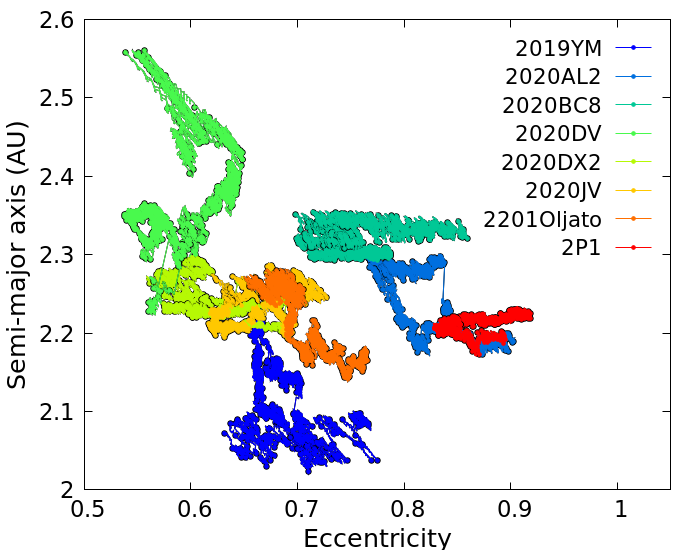}
  \caption{Nominal eccentricity and semi-major axis of the 51 NEAs integrated and comet 2P/Encke over $\pm$20 000 years. The orbital elements of the bodies every year are linked with a solid line, while their value every 100 years is represented by the filled circles along the line.}
  \label{fig:ae_allNEAs}
\end{figure*}

\textbf{B2  Precession cycle of 2004 TG10, 2005 TF50, 2005 UR and 2015 TX24}\\

\begin{figure*}[!ht]
  \centering
  2004 TG10 \hspace{7cm} 2005 TF50\\
  \includegraphics[width=.49\textwidth]{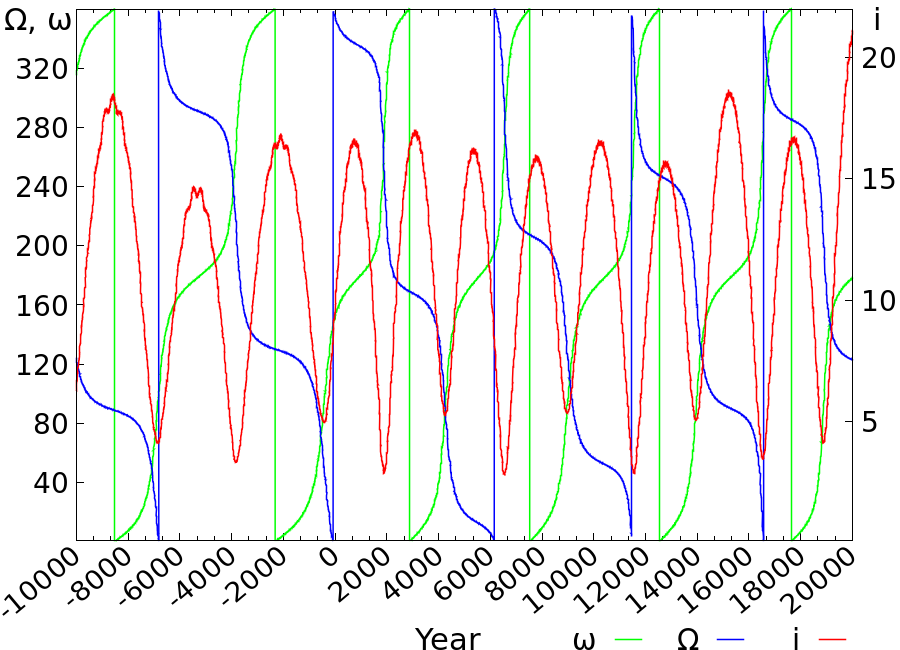}
  \includegraphics[width=.49\textwidth]{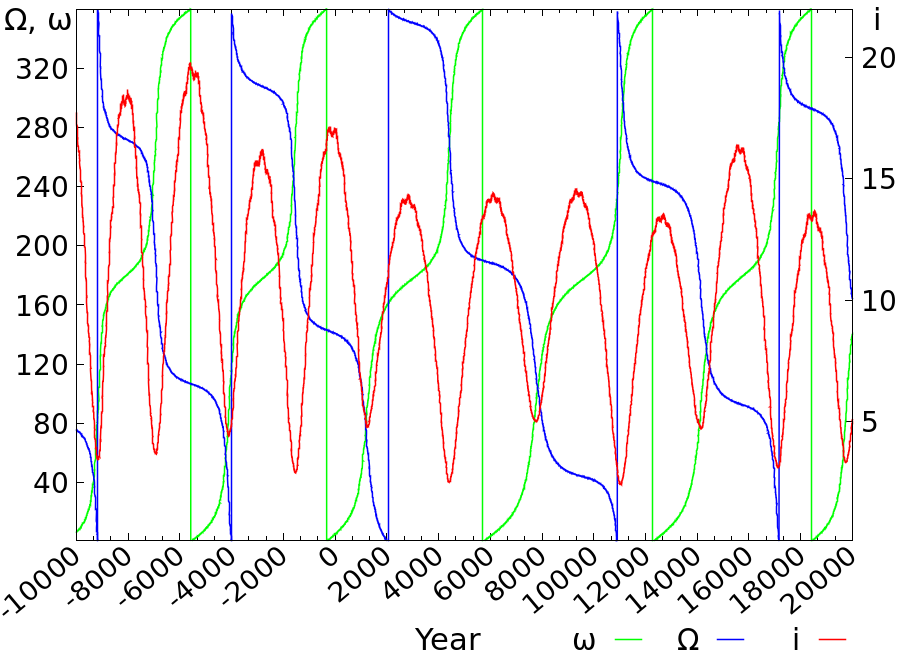}\\[0.2cm]
  2005 UR \hspace{7cm} 2015 TX24\\
  \includegraphics[width=.49\textwidth]{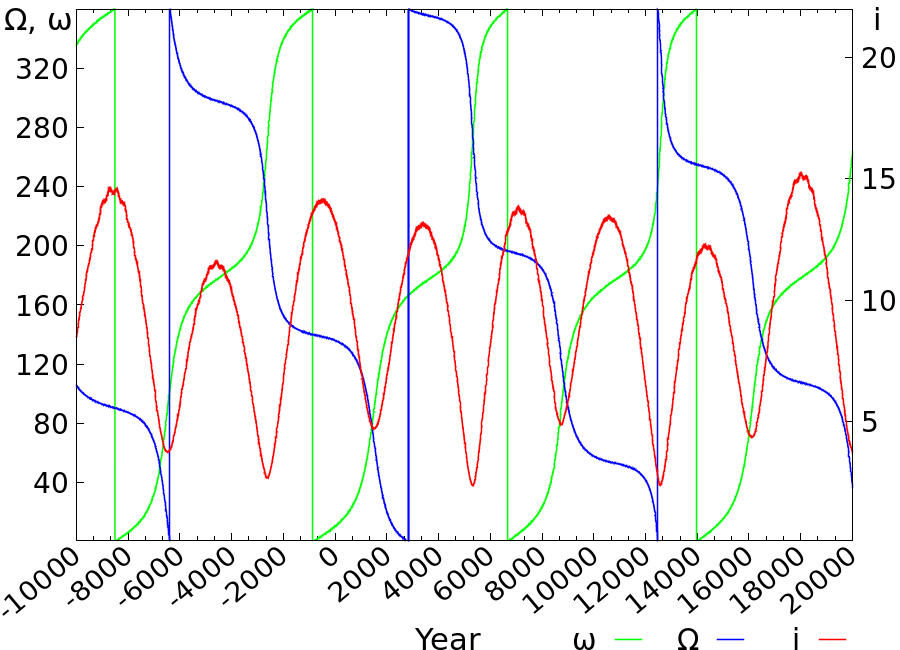}
  \includegraphics[width=.49\textwidth]{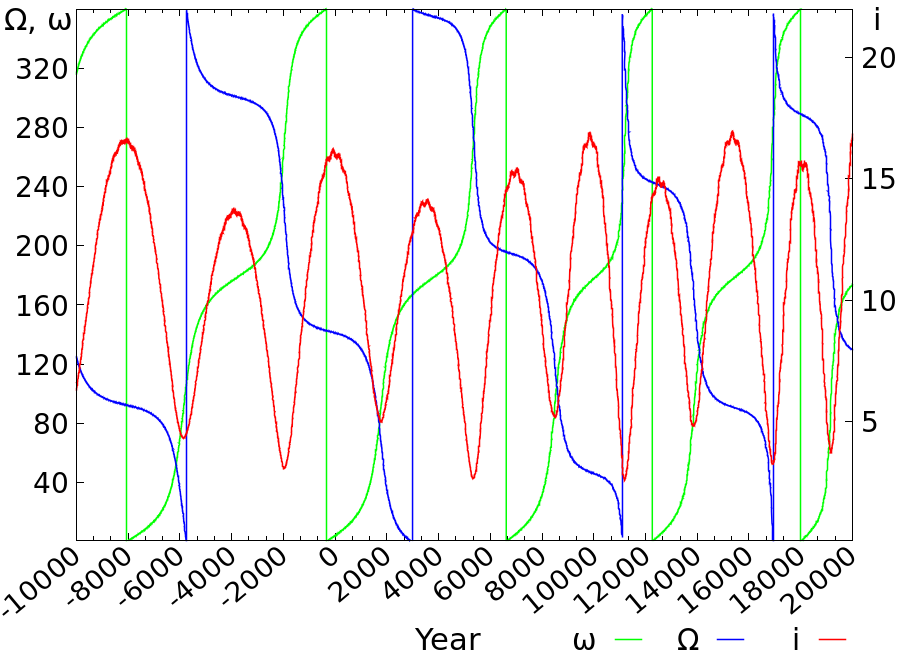}\\
  \caption{Inclination, perihelion argument and longitude of the ascending node of NEAs 2004 TG10, 2005 TF50, 2005 UR and 2015 TX24. The asteroids were integrated from the nominal solutions provided in Table A1 over a period of 30 000 years.}
  \label{fig:All_angles}
\end{figure*}

\newpage
\textbf{APPENDIX C: \text{7:2} MEAN MOTION RESONANCE}\\ \label{appendix:resonant_asteroids}

Mean motion resonances (MMRs) with Jupiter are known to influence the orbital evolution of the objects associated with the Taurid Complex. In particular, the 7:2 MMR is responsible for the formation of a resonant branch of meteoroids, that increases the level of activity of the Taurid meteor shower when approaching Earth. The resonant branch may also include large NEAs, of order a few hundred meters to 1 km in size, that are protected from dynamical diffusion by the resonance over long periods of time. 

\begin{figure}[!ht]
	\centering
	\includegraphics[trim={0cm 0.0cm 0.0cm 0.0cm}, clip, width=\textwidth]{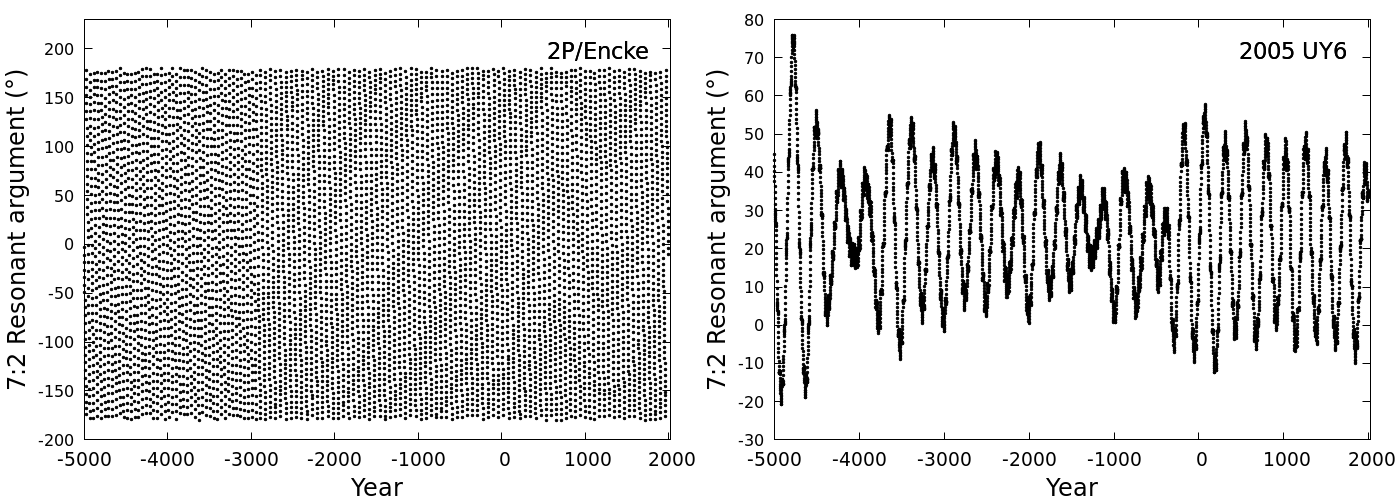}
	\caption{7:2 resonant arguments for comet 2P/Encke and asteroid 2005 UY6 based on integrations of the nominal orbits provided in Table A1. 2005 UY6 has been trapped into the 7:2 MMR with Jupiter over the past millennia, while 2P/Encke evolves outside the resonance. }
	\label{fig:example_resonant}
\end{figure}

Since the 7:2 MMR may play an important role on the evolution of the TC members investigated, we determined the proportion of objects in our sample that were likely trapped in the resonance during several millennia. To that end, for each body integrated, we computed the value of the resonant argument $\sigma$ as described in \cite{Asher1991}:  

\begin{equation}
\sigma =  7\lambda_J - 2\lambda - 5\varpi
\end{equation}

where $\lambda_J$ is the mean longitude of Jupiter, and $\lambda$ and $\varpi$ the mean longitude and  longitude of perihelion of the body integrated. If the body evolves inside the resonance, the resonant angle $\sigma$ librates around 0, with an amplitude between 0 and 180$\degree$. If the amplitude of the oscillations is low, the object is strongly trapped into the resonance. The top panel of figure \ref{fig:example_resonant} illustrates the case of a body not evolving into the 7:2 MMR (i.e., comet 2P/Encke), while the bottom panel represents the evolution of an object strongly trapped in the resonance (namely 2005 UY6). 

\begin{figure}[!ht]
	\centering
	\includegraphics[width=.49\textwidth]{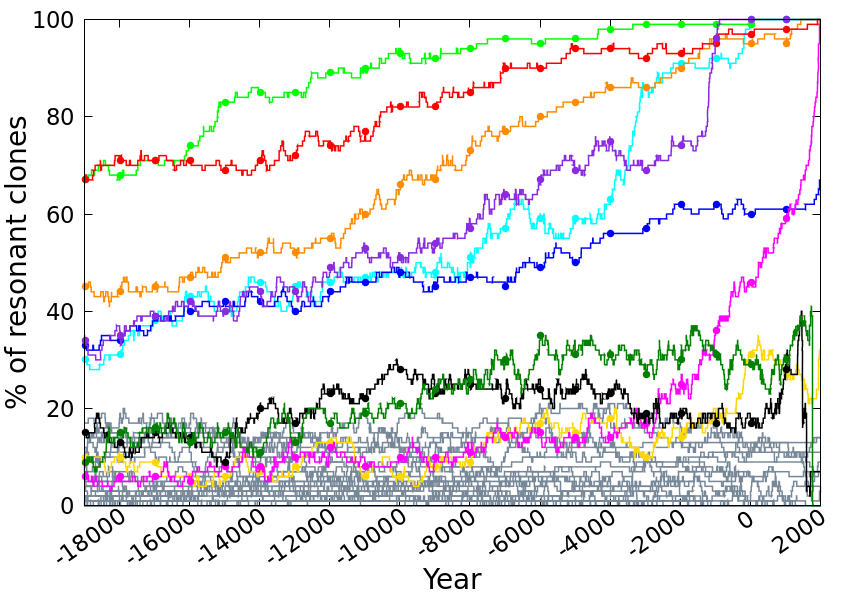}
	\includegraphics[width=.49\textwidth]{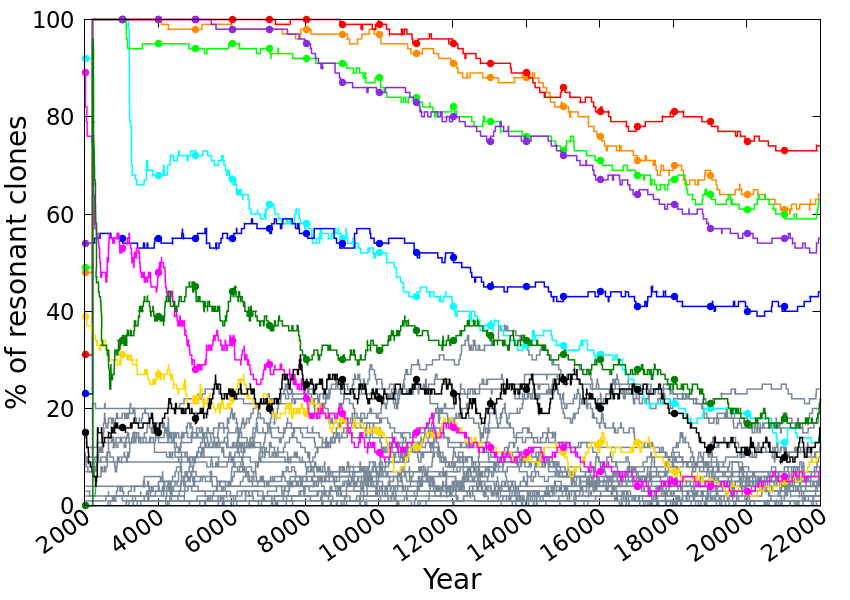}\\[0.2cm]
	\includegraphics[width=.55\textwidth]{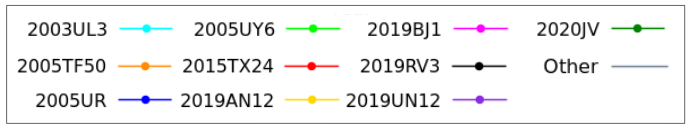}
	\caption{Percentage of clones, for each body, evolving into the 7:2 MMR with Jupiter. Objects with at least 20\% of clones belonging to the resonance in the past are shown in color, while the other ones are drawn in grey.}
	\label{fig:percent_resonant}
\end{figure}

For each body in our sample, we analyzed the evolution of the resonant argument $\sigma$ for every clone created from the body's nominal orbit. We then determined the percentage of clones trapped into the 7:2 MMR over time. Figure \ref{fig:percent_resonant} presents the percentage of resonant clones of each body integrated over the past and future 20 000 years. For clarity reasons, the objects with the highest percentage of resonant clones are represented in color, while objects with less than 20\% of the clones evolving in the 7:2 MMR most of the time are drawn in grey.

\setfignumprefix{C}

From this figure, we see that NEAs 2005 UY6, 2015 TX24, 2005 TF50, 2019 UN12, 2003 UL3 and 2005 UR have at least 50\% of their clones trapped into the resonance during several thousand of years. In particular, most of the clones of 2005 UY6, 2015 TX24, 2005 TF50 and 2019 UN12 stay in the resonance during the whole integration period. In a lower extent, about 20\% of the clones of 2019 AN12, 2019 RV3 and 2020 JV remained in the resonance during the integration period. The recently observed NEA 2019 BJ1 clearly evolves in the 7:2 MMR at the current epoch, but its number of resonant clones decreases significantly after 5000 years of integration. 

\begin{figure}[!ht]
	\centering
	\includegraphics[width=.5\textwidth]{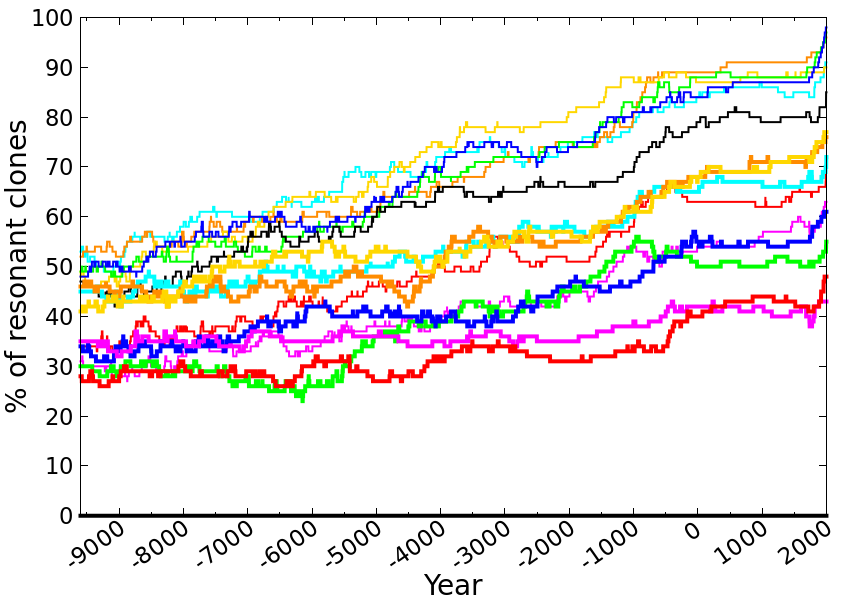}\\[0.2cm]
	\includegraphics[width=.5\textwidth]{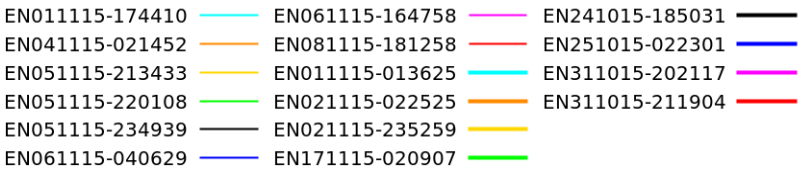}
	\caption{Percentage of clones, for each fireball, evolving into the 7:2 MMR with Jupiter.}
	\label{fig:percent_fireballs}
\end{figure}

In figure \ref{fig:percent_fireballs}, we present the results of the same analysis applied to the fireballs investigated in Section 5. As expected, all the fireballs except the comparator Southern Taurid EN241015\_185031 evolved in the 7:2 MMR at the epoch of observation in 2015. During the backward integration, the percentage of resonant clones decreases regularly for most of the fireballs, reaching a minimum of about 30\% resonant clones around 10 000 BCE. We therefore conclude that all the fireballs remained inside or close to the 7:2 MMR during the integration period, confirming their membership in the Taurid Swarm Complex. 

\onecolumn
\textbf{APPENDIX D: SENSITIVITY ANALYSIS} \label{appendix:sensitivity} \\

\textbf{D1  Different MOID thresholds}\\

\setfignumprefix{D}
\begin{figure*}[!ht]
  \centering
  \includegraphics[width=.49\textwidth]{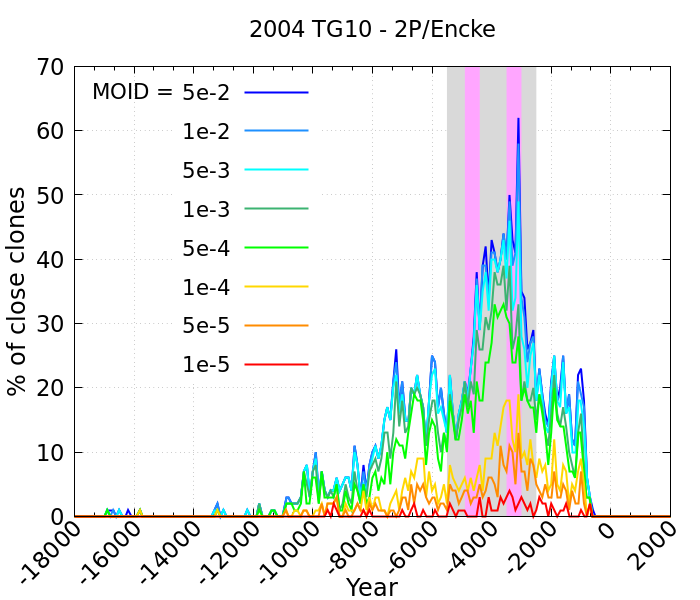}
  \includegraphics[width=.49\textwidth]{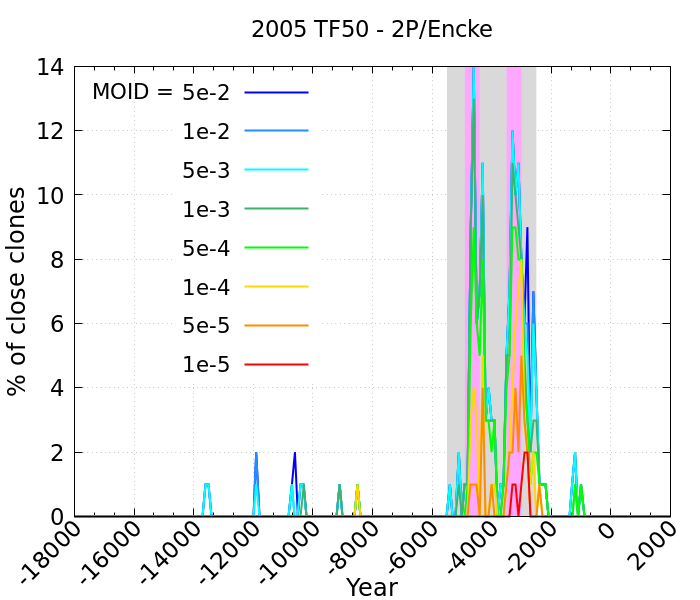}\\
  \includegraphics[width=.49\textwidth]{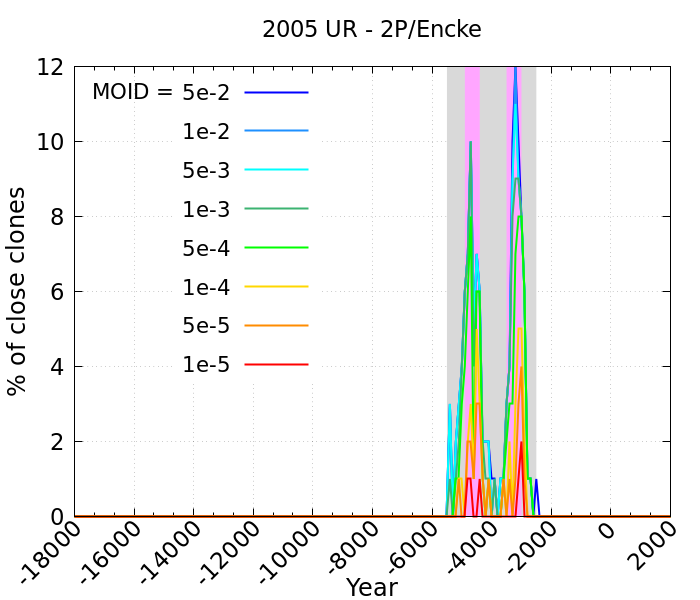}
  \includegraphics[width=.49\textwidth]{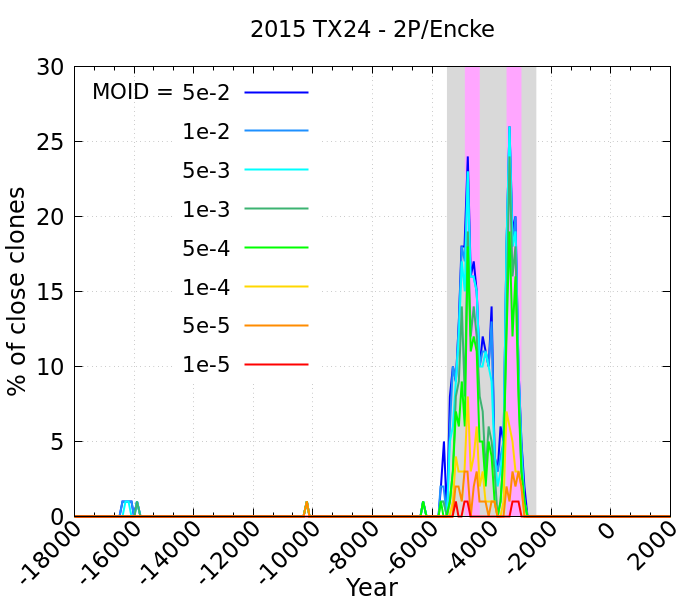}
  \caption{Percentage of clones of NEAs 2004 TG10, 2005 TF50, 2005 UR and 2015 TX24 approaching at least one clone of comet 2P/Encke with a relative velocity below 500 m/s and different MOID thresholds (0.05, 0.01, 0.005, 0.001, 0.0005, 0.0001, 0.00005 and 0.00001 AU). }
  \label{fig:different_MOIDs}
\end{figure*}

\newpage
\textbf{D2  Different relative velocity thresholds}\\

\begin{figure*}[!ht]
  \centering
  \includegraphics[width=.49\textwidth]{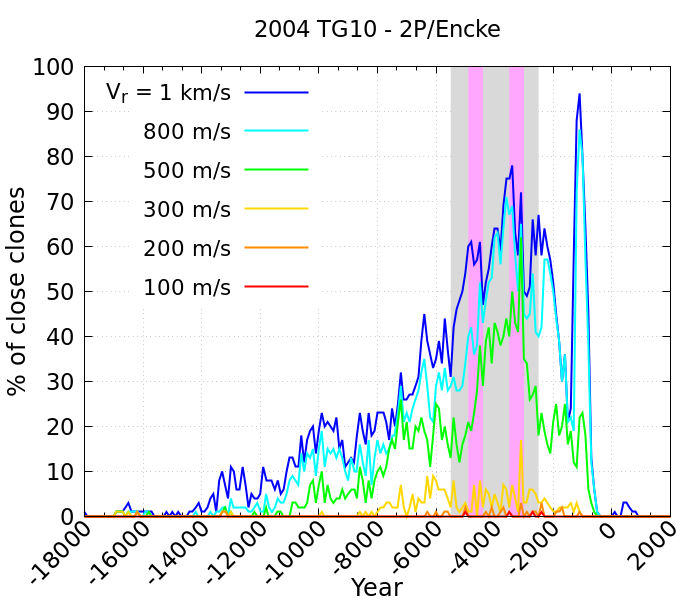}
  \includegraphics[width=.49\textwidth]{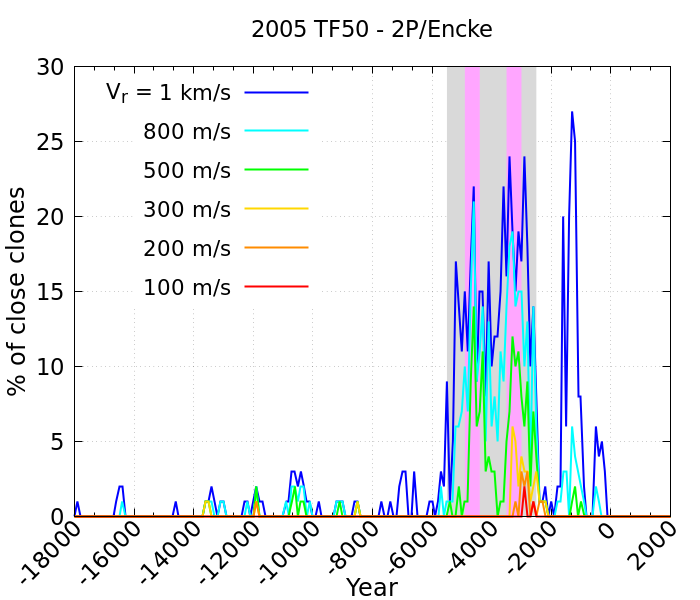}\\
  \includegraphics[width=.49\textwidth]{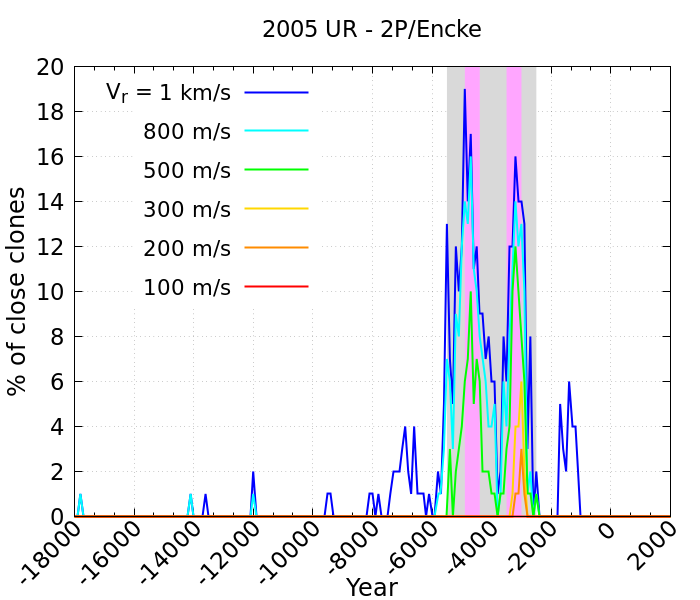}
  \includegraphics[width=.49\textwidth]{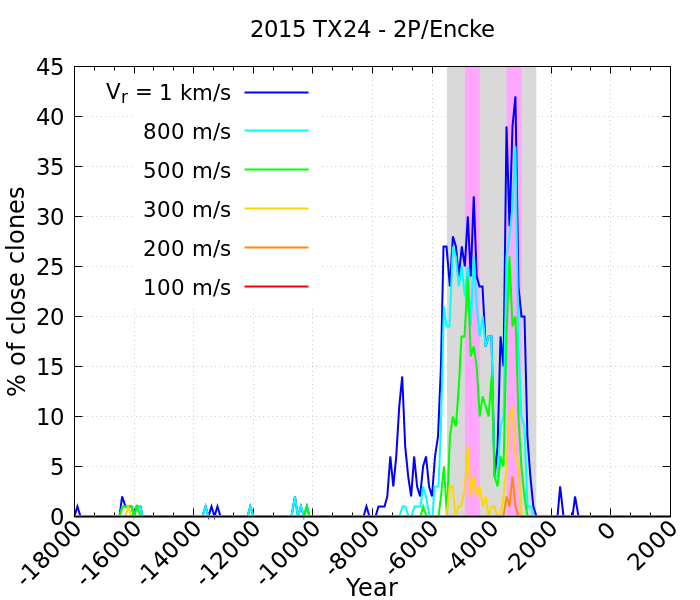}
  \caption{Percentage of clones of NEAs 2004 TG10, 2005 TF50, 2005 UR and 2015 TX24 approaching at least one clone of comet 2P/Encke with a MOID below 0.05 AU and different relative velocities limits.}
  \label{fig:different_relVs}
\end{figure*}

\newpage
\textbf{D3  Sensitivity to 2015 TX24's initial orbit}\\

\begin{figure*}[!ht]
  \centering
  \includegraphics[width=.8\textwidth]{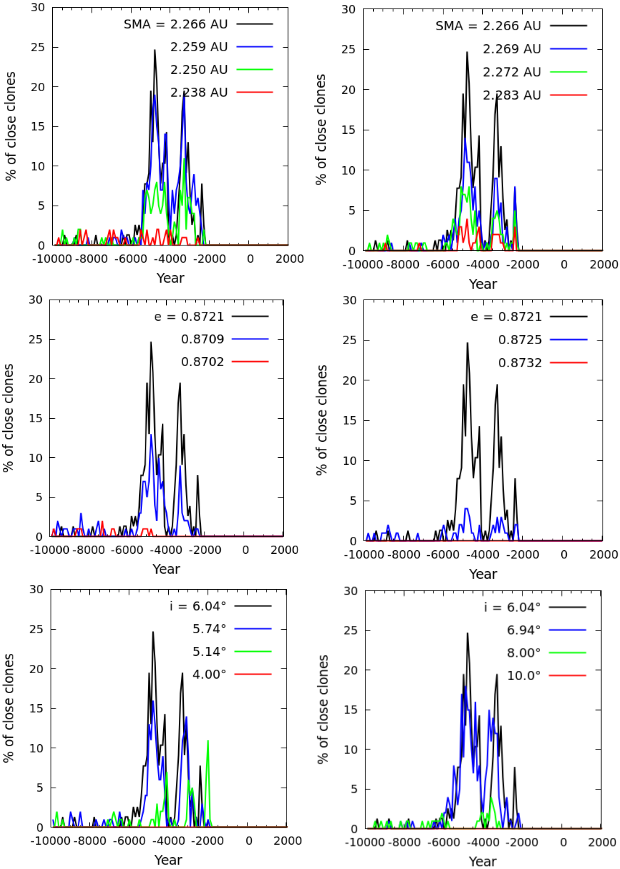}
  \caption{Percentage of clones of NEA 2015 TX24 (with slightly modified initial orbital elements as shown in the subplot legends) approaching at least one clone of comet 2P/Encke with a MOID below 0.05 AU.}
  \label{fig:2015TX24_variations}
\end{figure*}

\newpage
\textbf{APPENDIX E: COMPLEMENTARY RAPPROCHEMENTS ANALYSIS}\\ \label{appendix:additional_rapprochements} 

\textbf{E1  G5: Extension to $\pm$ 100 000 years} \label{sec:100Kyr}\\

\setfignumprefix{E}
\begin{figure*}[!ht]
  \centering
  \includegraphics[width=.9\textwidth]{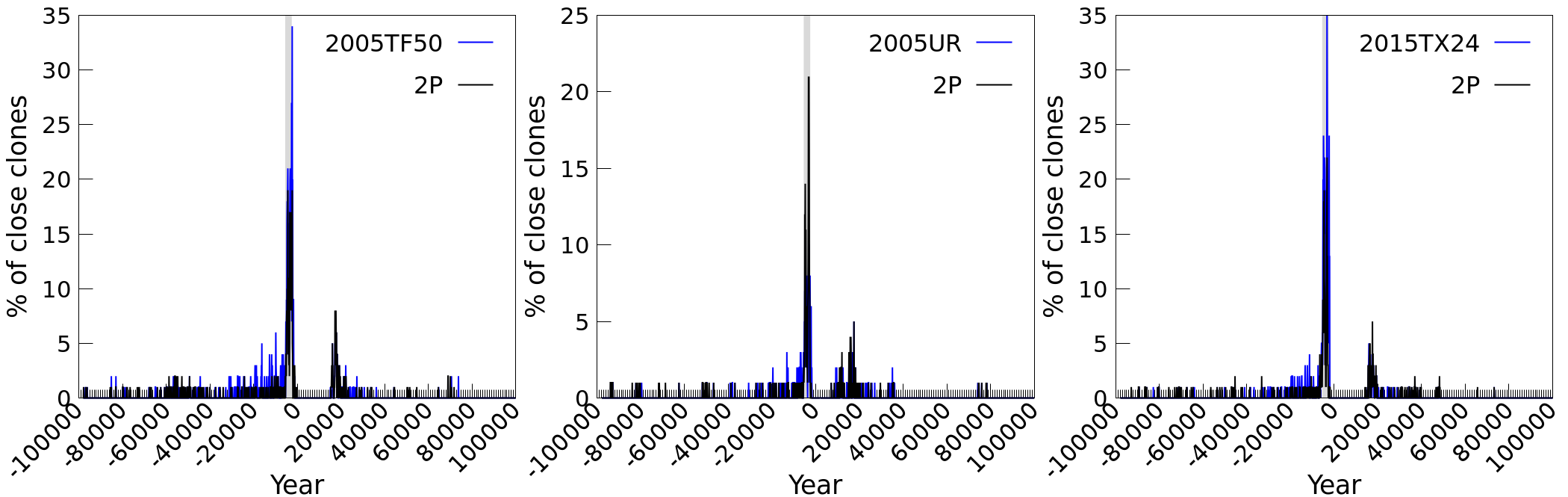}\\
  \includegraphics[width=.9\textwidth]{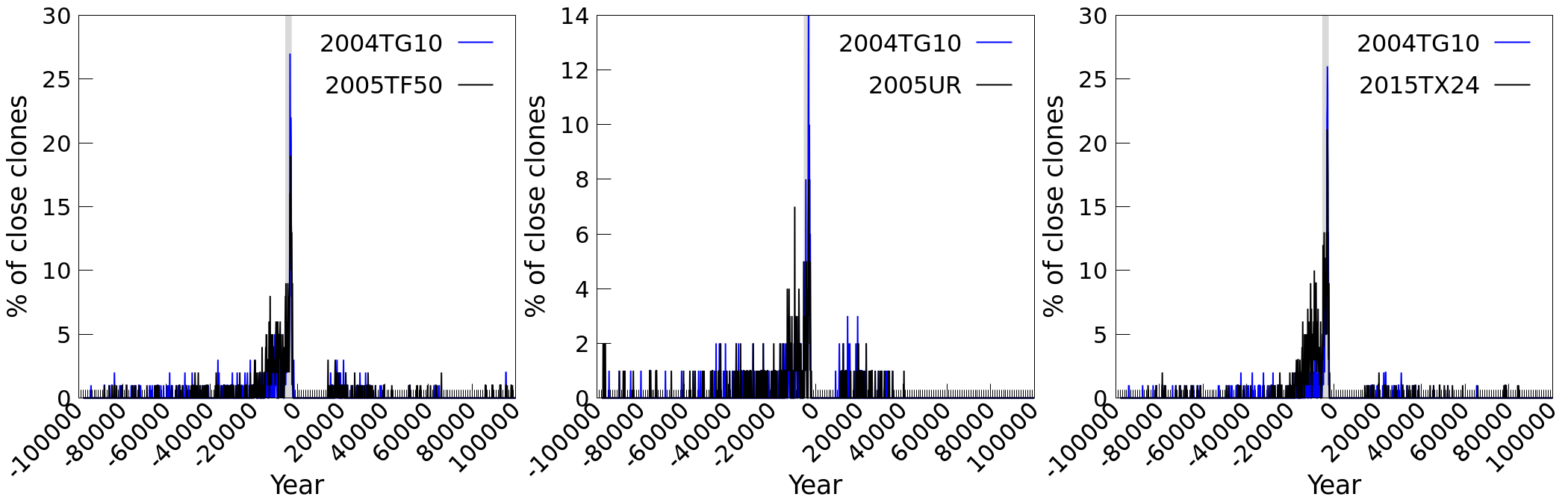}\\
  \includegraphics[width=.9\textwidth]{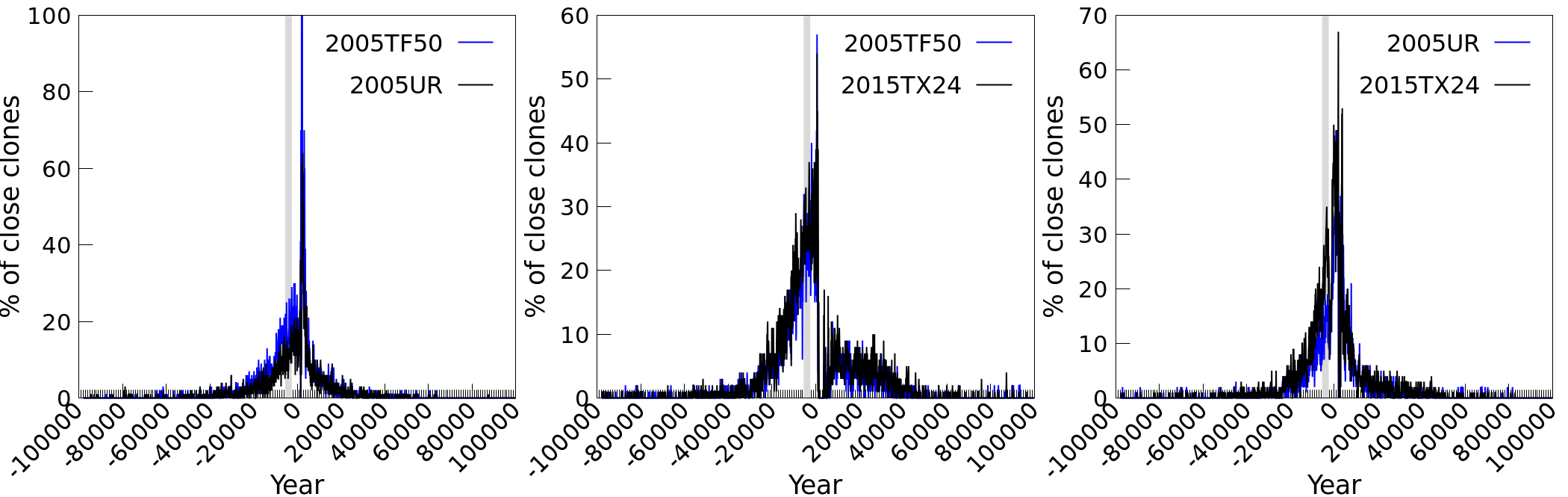}\\
  \includegraphics[width=.9\textwidth]{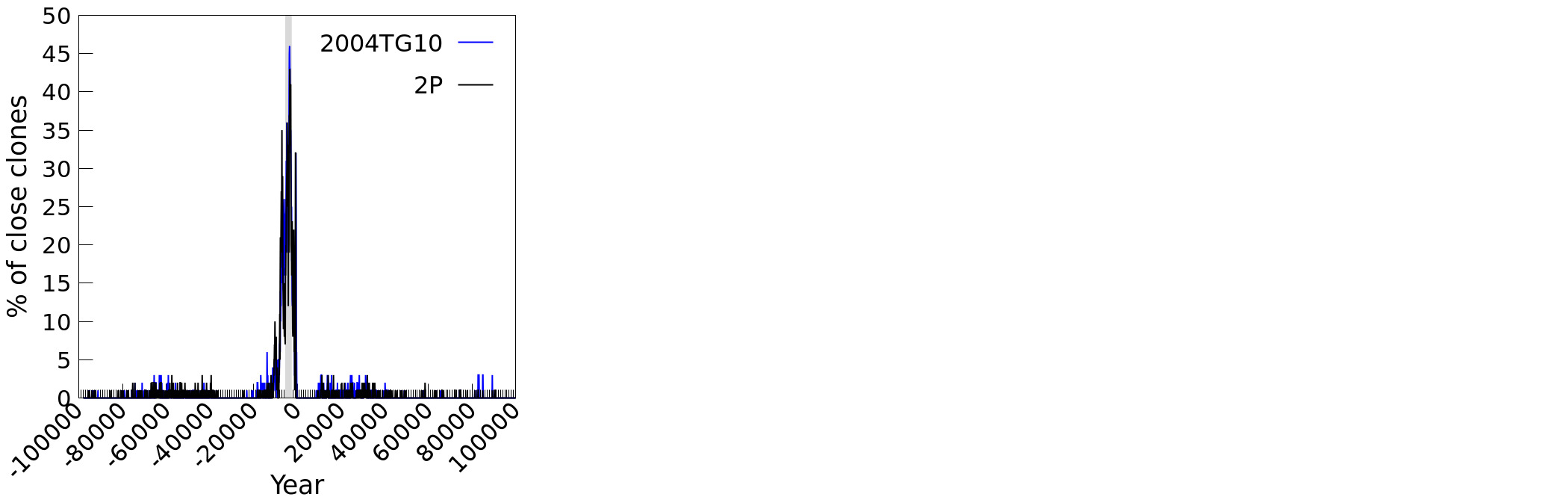}\\
  \caption{Percentage of clones of body \#1 (in blue) and body \#2 (in black) approaching at least a clone of the other object with a MOID smaller than 0.05 AU and a relative velocity below 500 m/s. }
  \label{fig:Main5_100Kyr}
\end{figure*}

\newpage
\textbf{E2  NEA 2005 UY6}\\

\begin{figure*}[!ht]
  \centering
  \includegraphics[width=\textwidth]{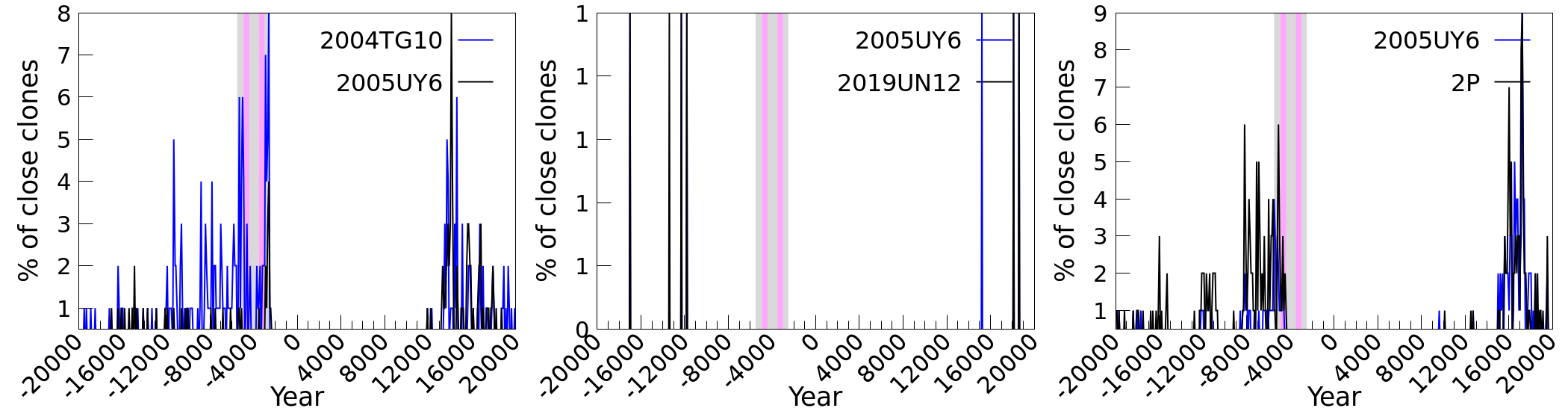}\\
  \caption{Percentage of clones of body \#1 (in blue) and body \#2 (in black) approaching at least a clone of the other object with a MOID smaller than 0.05 AU and a relative velocity below 500 m/s. }
  \label{fig:2005UY6}
\end{figure*}

\textbf{E3  Fireballs' close approaches with 2P/Encke, $V_r$ threshold of 500 m/s}\\

\begin{figure*}[!ht]
  \centering
  \includegraphics[width=\textwidth]{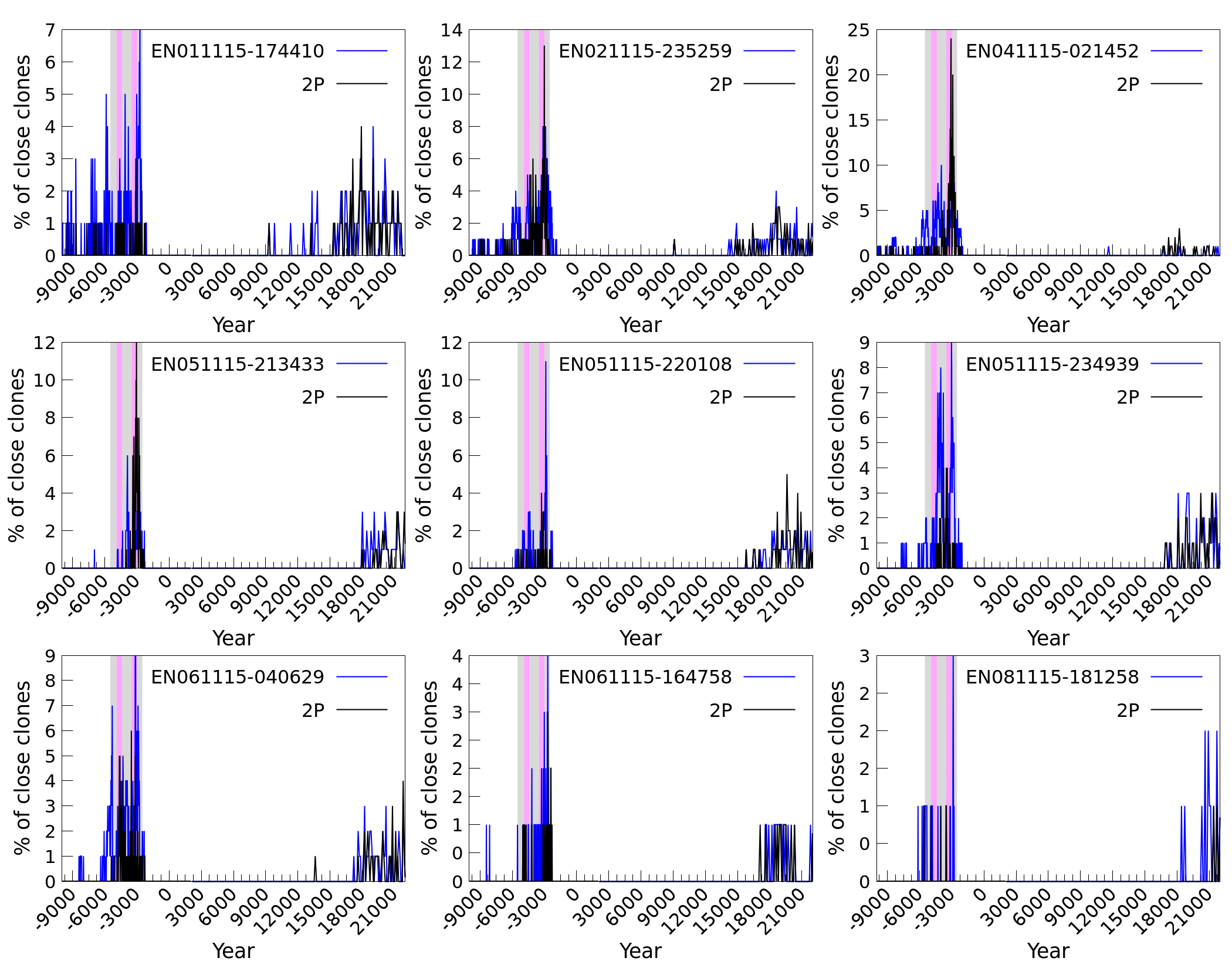}
  \caption{Percentage of clones of the fireballs (in blue) and comet 2P/Encke (in black) approaching at least a clone of the other object with a MOID smaller than 0.05 AU and a relative velocity below 500 m/s. The grey shaded areas encompasses the period 2500 BCE to 5500 BCE, and the two magenta areas the years 3000 BCE to 3500 BCE and 4400 BCE to 4900 BCE respectively.}
  \label{fig:fireballs_with_peak_500ms}
\end{figure*}

\begin{figure*}[!ht]
  \centering
  \includegraphics[width=\textwidth]{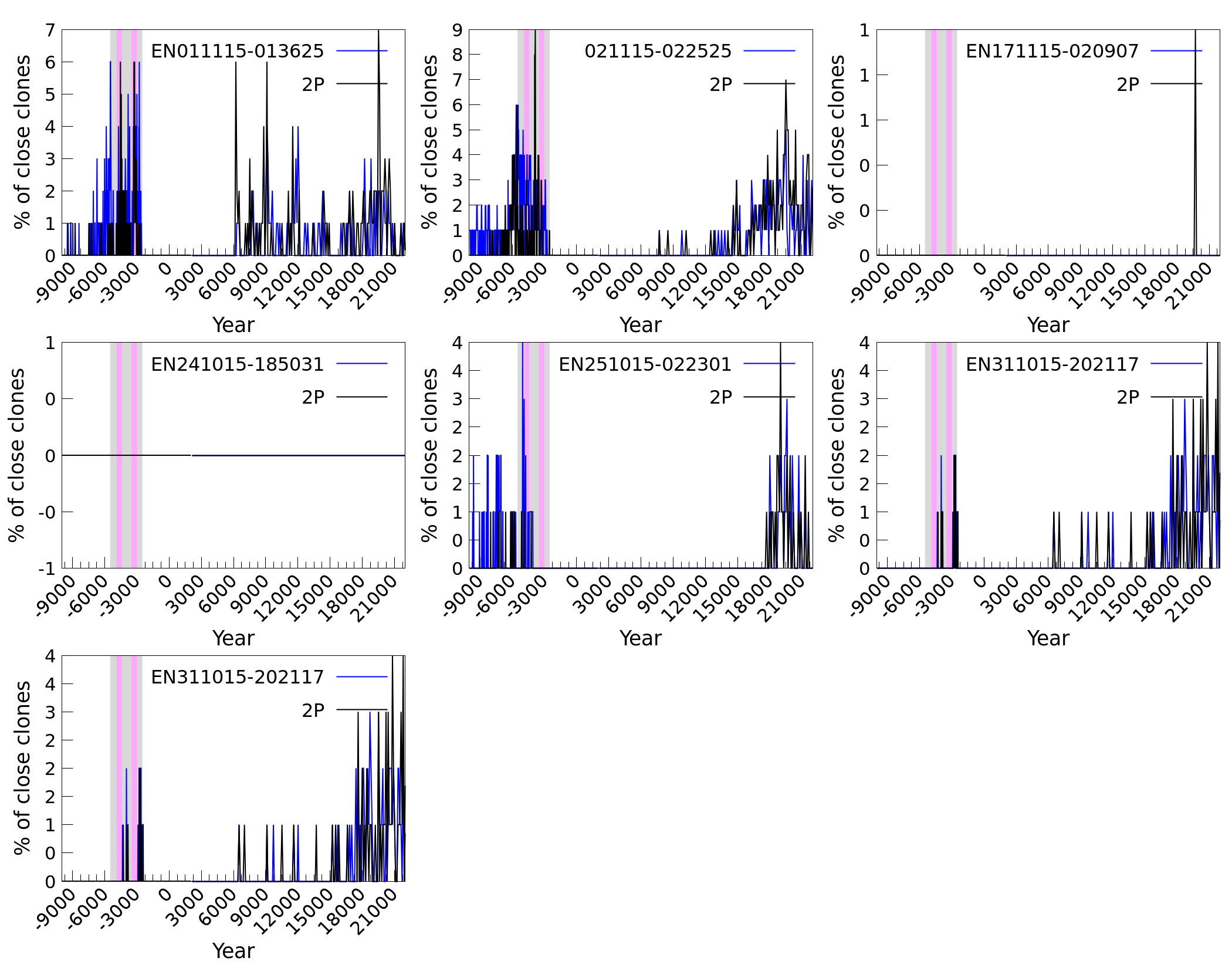}
  \caption{Percentage of clones of the fireballs (in blue) and comet 2P/Encke (in black) approaching at least a clone of the other object with a MOID smaller than 0.05 AU and a relative velocity below 500 m/s. The grey shaded areas encompasses the period 2500 BCE to 5500 BCE, and the two magenta areas the years 3000 BCE to 3500 BCE and 4400 BCE to 4900 BCE respectively.}
  \label{fig:fireballs_without_peak_500ms}
\end{figure*}

\clearpage
\newpage
\textbf{E4  Fireballs' close approaches with 2004 TG10}\\

\begin{figure*}[!ht]
  \centering
  \includegraphics[width=\textwidth]{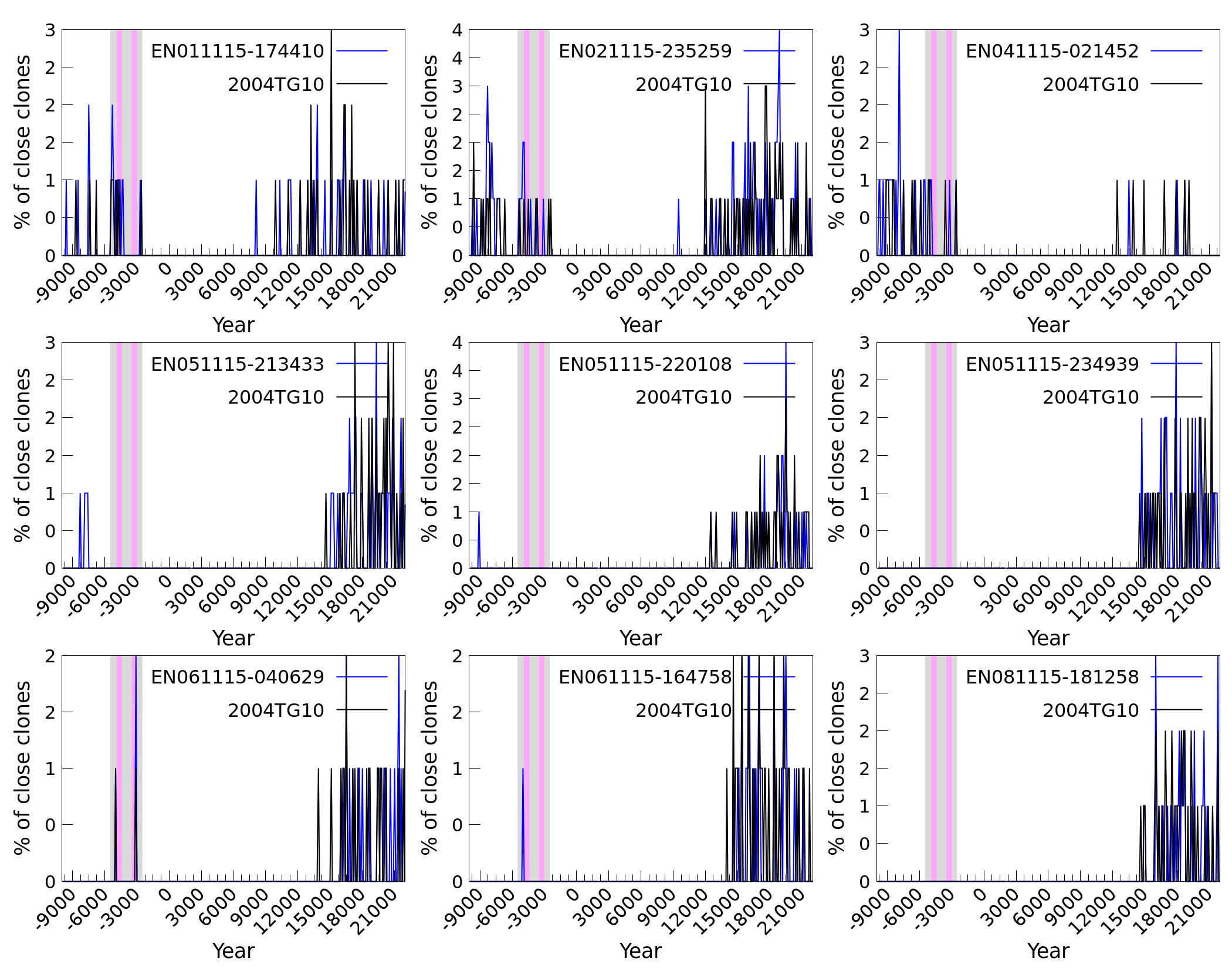}
  \caption{Percentage of clones of the fireballs (in blue) and 2004 TG10 (in black) approaching at least a clone of the other object with a MOID smaller than 0.05 AU and a relative velocity below 500 m/s. The grey shaded areas encompasses the period 2500 BCE to 5500 BCE, and the two magenta areas the years 3000 BCE to 3500 BCE and 4400 BCE to 4900 BCE respectively.}
  \label{fig:fireballs1_2004TG10}
\end{figure*}

\begin{figure*}[!ht]
  \centering
  \includegraphics[width=\textwidth]{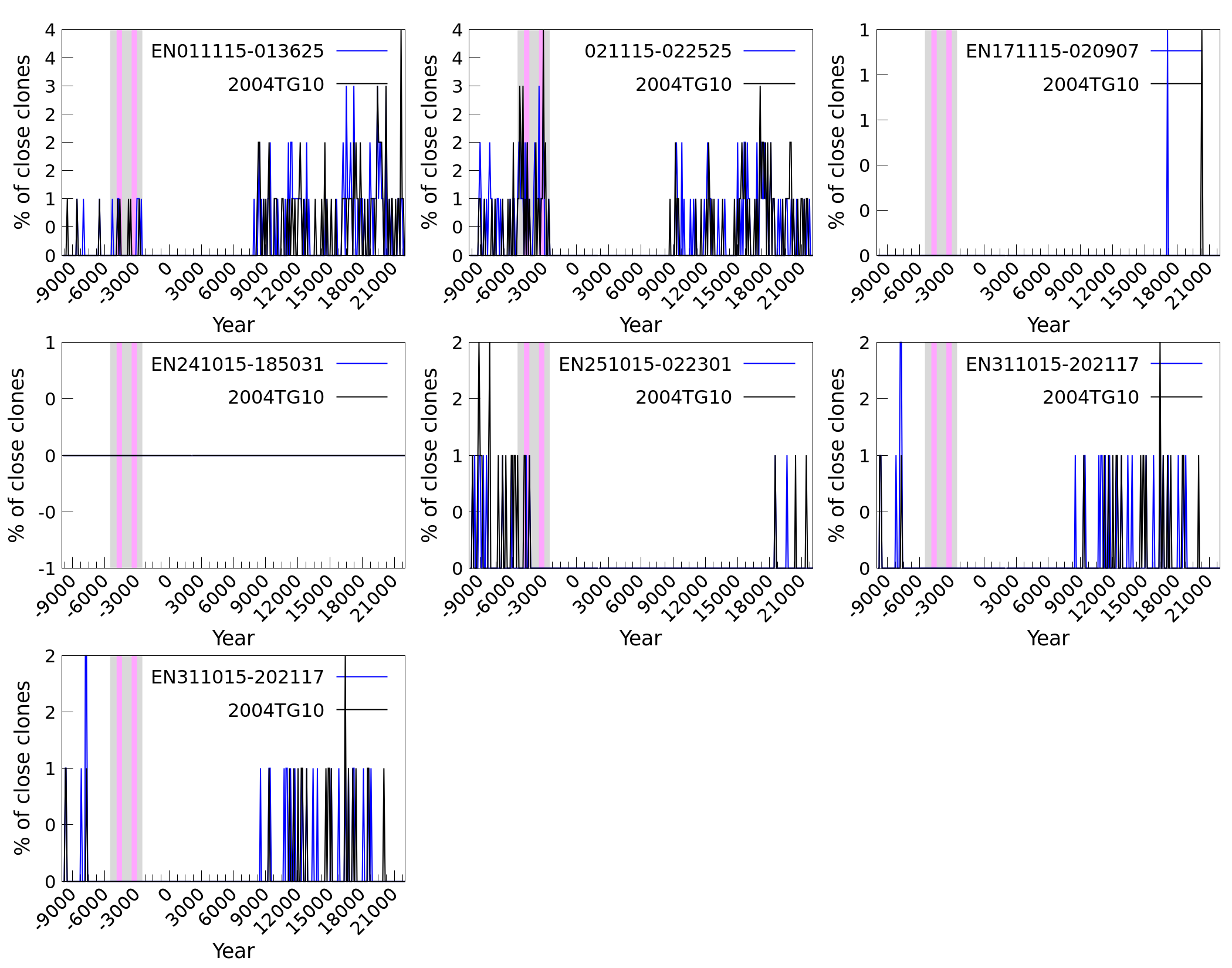}
  \caption{Percentage of clones of the fireballs (in blue) and 2004 TG10 (in black) approaching at least a clone of the other object with a MOID smaller than 0.05 AU and a relative velocity below 500 m/s. The grey shaded areas encompasses the period 2500 BCE to 5500 BCE, and the two magenta areas the years 3000 BCE to 3500 BCE and 4400 BCE to 4900 BCE respectively.}
  \label{fig:fireballs2_2004TG10}
\end{figure*}

\clearpage
\textbf{APPENDIX F: CLONES ORBITAL ELEMENTS AROUND 3200 BCE}\\ \label{appendix:clones_selected} 

\textbf{F1  G5}\\

\setfignumprefix{F}
\begin{figure*}[!ht]
  \centering
  2004 TG10 \hspace{7cm} 2005 TF50\\ 
  \includegraphics[width=.49\textwidth]{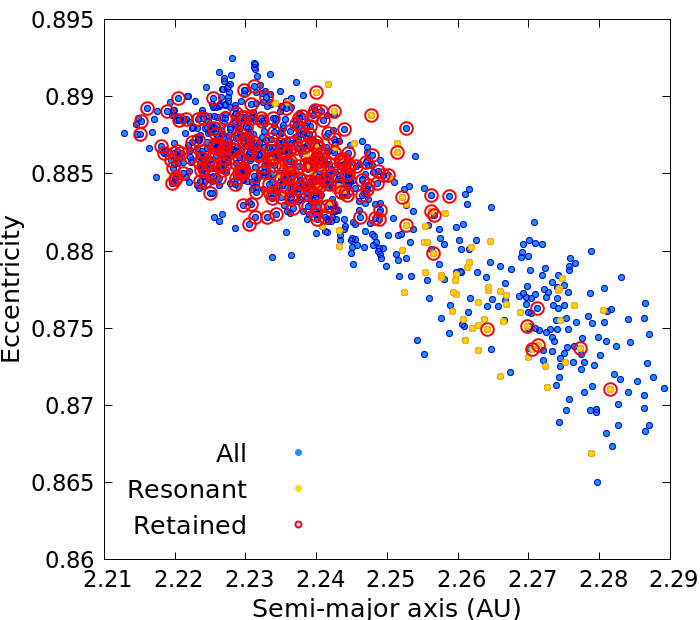}
  \includegraphics[width=.49\textwidth]{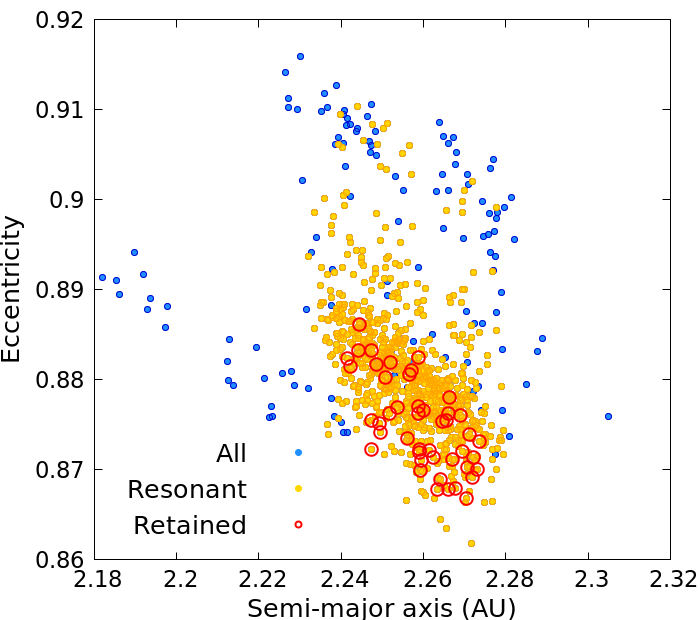}\\
  2005 UR \hspace{7cm} 2015 TX24\\
  \includegraphics[width=.49\textwidth]{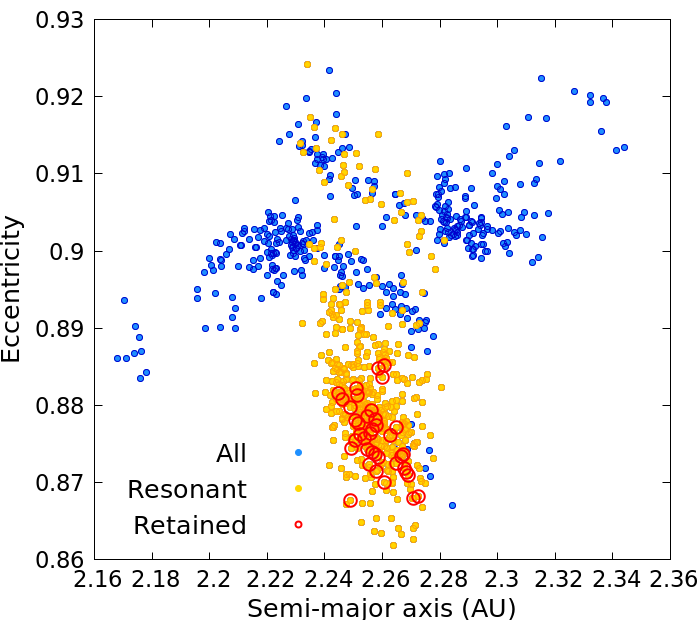}
  \includegraphics[width=.49\textwidth]{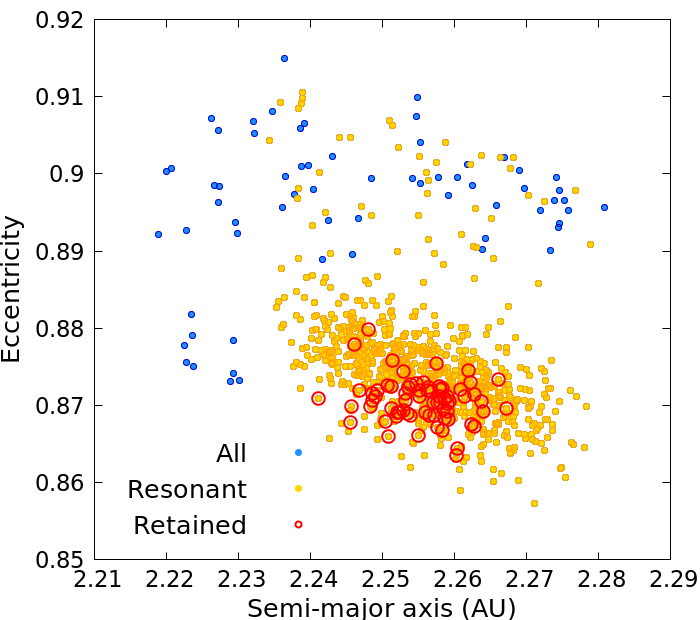}\\
  \caption{Semi-major axis and eccentricity of the clones created from 2004 TG10, 2005 TF50, 2005 UR or 2015TX24's nominal solution and integrated until 10 000 BCE. The plot presents a compilation of the clones' a and e between 3000 BCE and 3500 BCE (in blue). The clones belonging to the 7:2 MMR are presented in yellow, while those approaching at least one clone of comet 2P/Encke with a low MOID and relative velocity are circled in red. \label{fig:clones_selected}}
\end{figure*}

\newpage
\textbf{F2  Fireballs}\\

\begin{figure*}[!ht]
  \centering
  \includegraphics[width=0.32\textwidth]{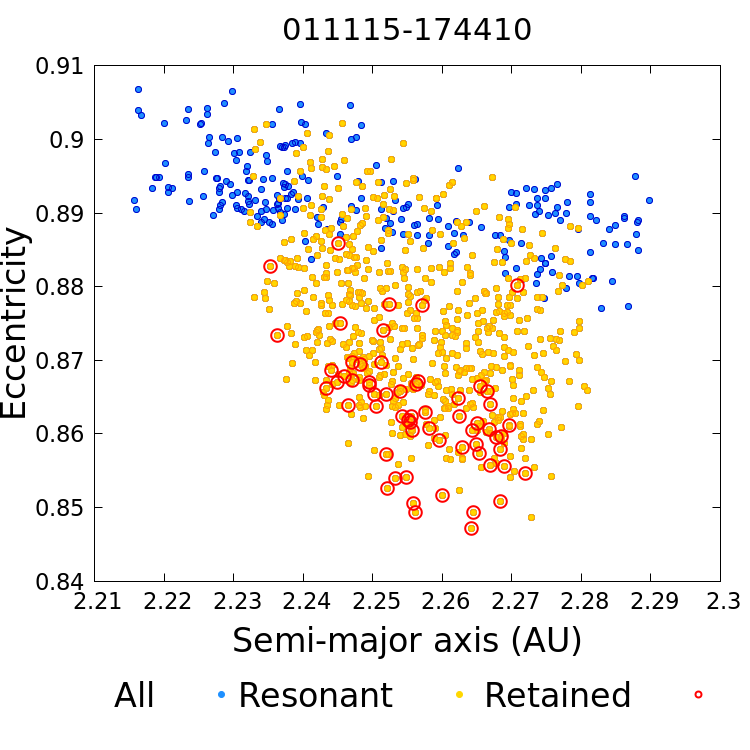}
  \includegraphics[width=0.32\textwidth]{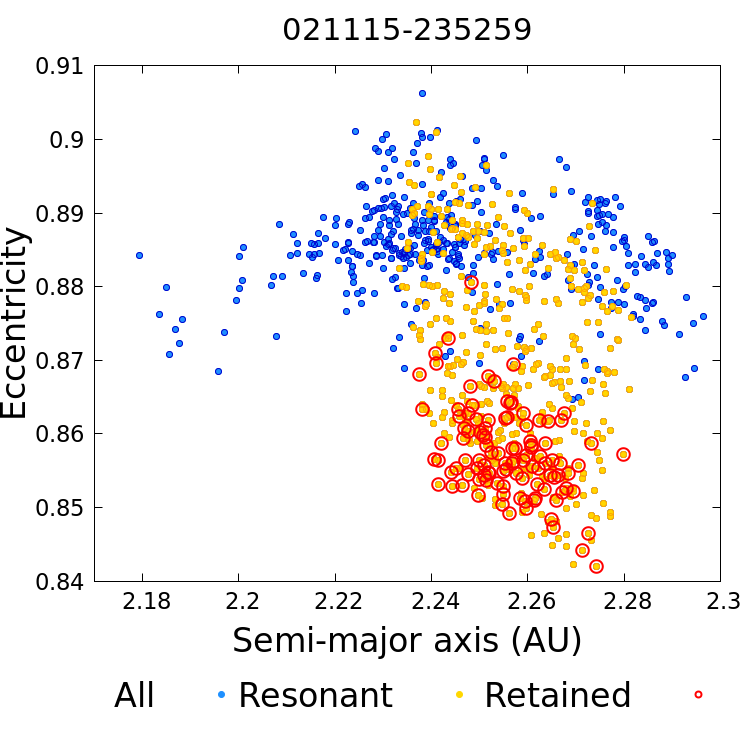}
  \includegraphics[width=0.32\textwidth]{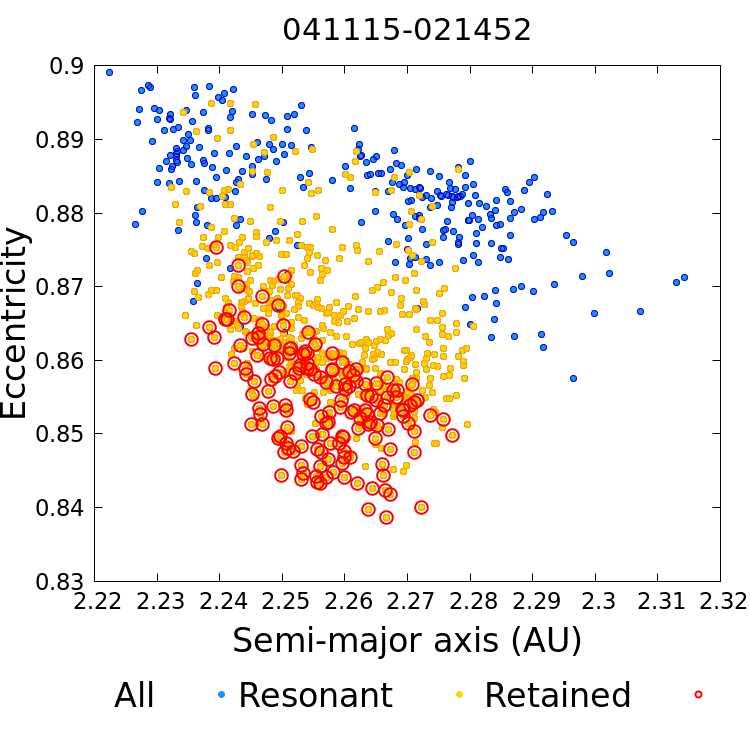}\\
  \includegraphics[width=0.32\textwidth]{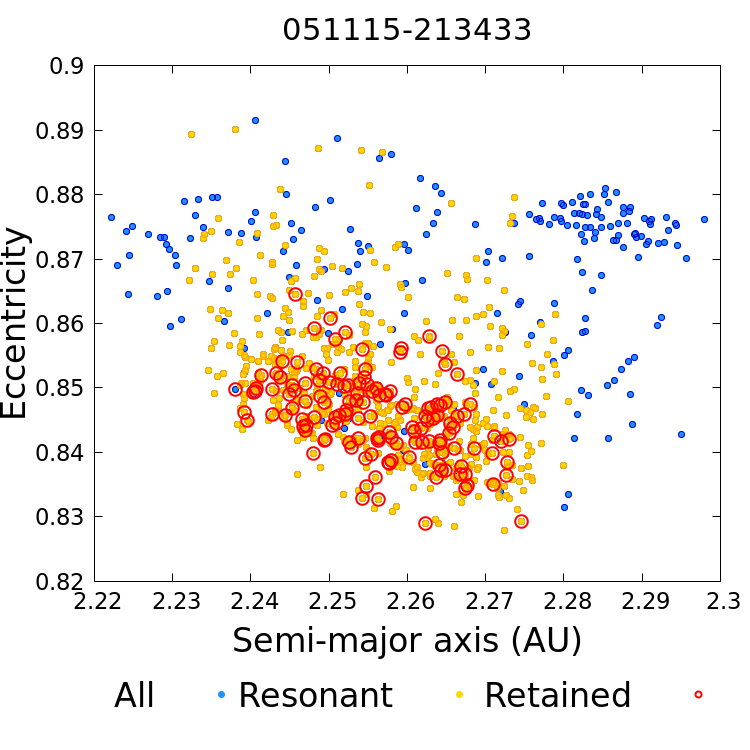}
  \includegraphics[width=0.32\textwidth]{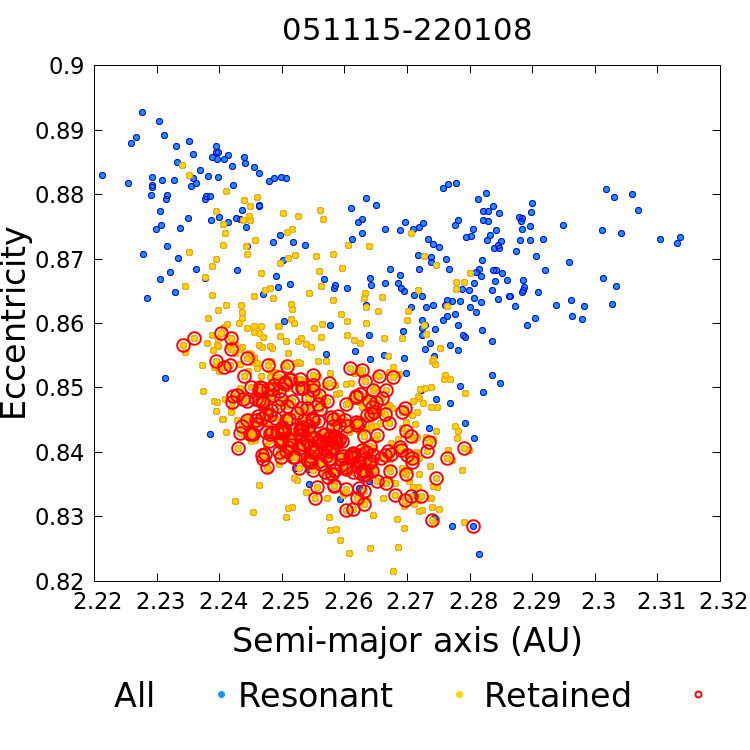}
  \includegraphics[width=0.32\textwidth]{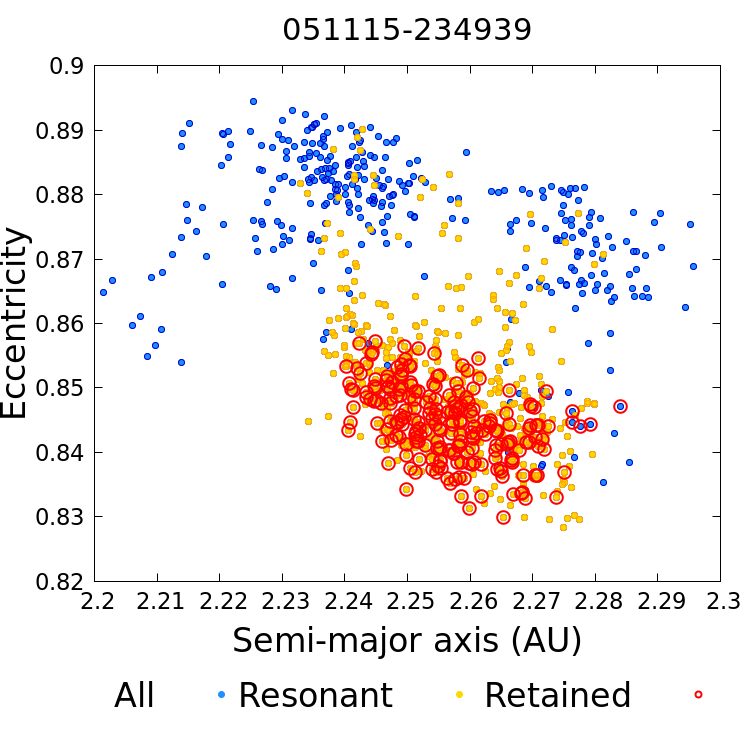}\\
  \includegraphics[width=0.32\textwidth]{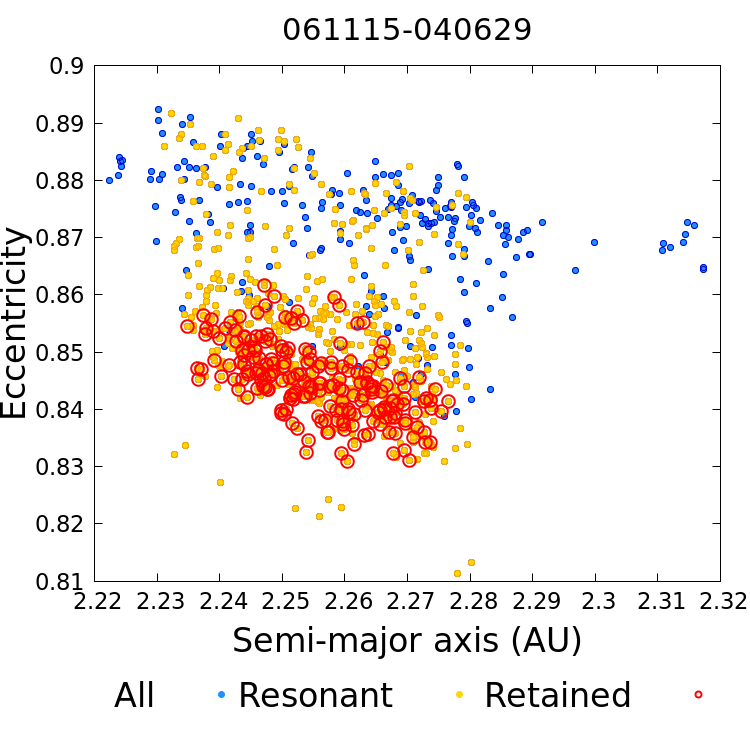}
  \includegraphics[width=0.32\textwidth]{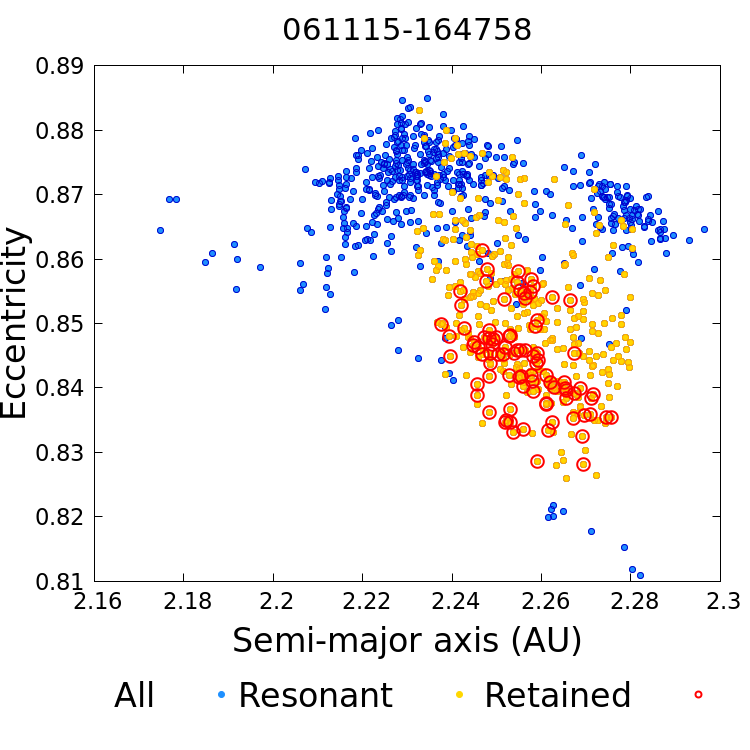}
  \includegraphics[width=0.32\textwidth]{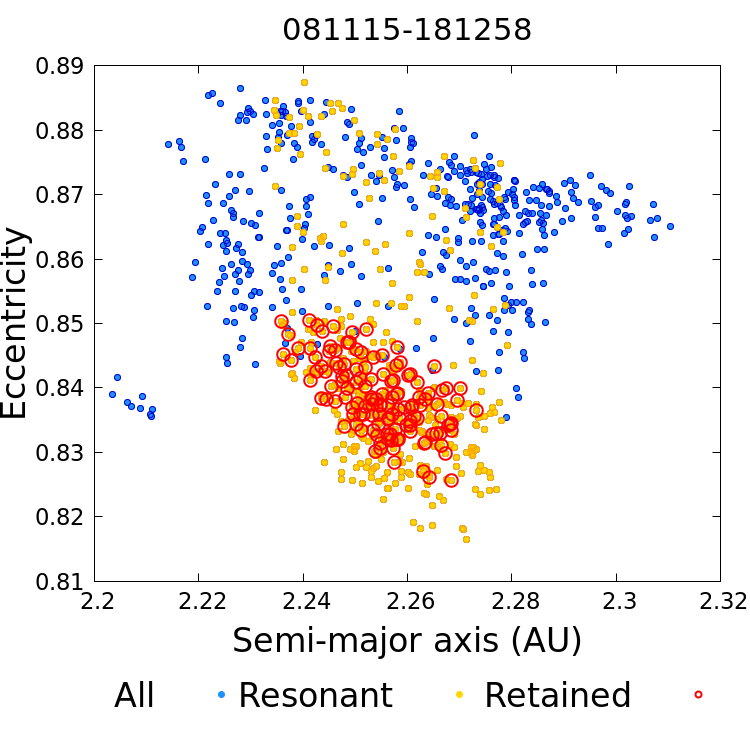}
  \caption{Semi-major axis and eccentricity of the clones created for the fireballs of Figure 12 and integrated until 10 000 BCE. The plot presents a compilation of the clones' a and e between 3000 BCE and 3500 BCE (in blue). The clones belonging to the 7:2 MMR are presented in yellow, while those approaching at least one clone of comet 2P/Encke with a low MOID and relative velocity are circled in red.}
  \label{fig:clones_selected_fireballs1}
\end{figure*}

\begin{figure*}[!ht]
  \centering
  \includegraphics[width=0.32\textwidth]{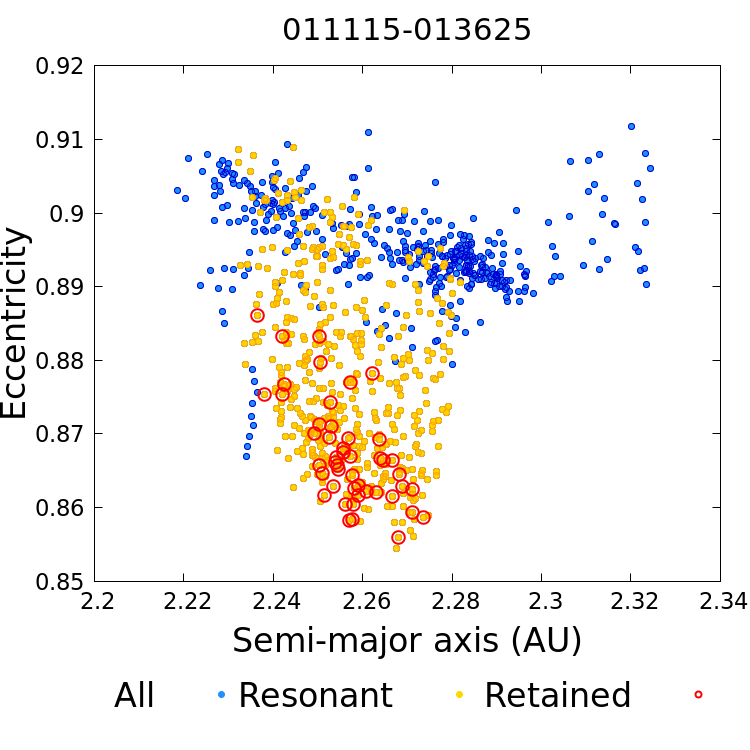}
  \includegraphics[width=0.32\textwidth]{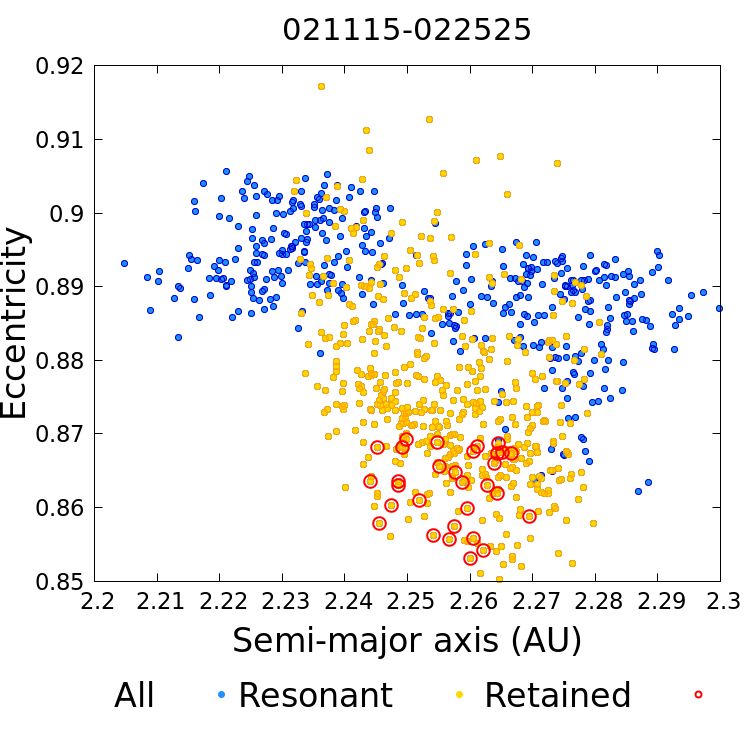}
  \includegraphics[width=0.32\textwidth]{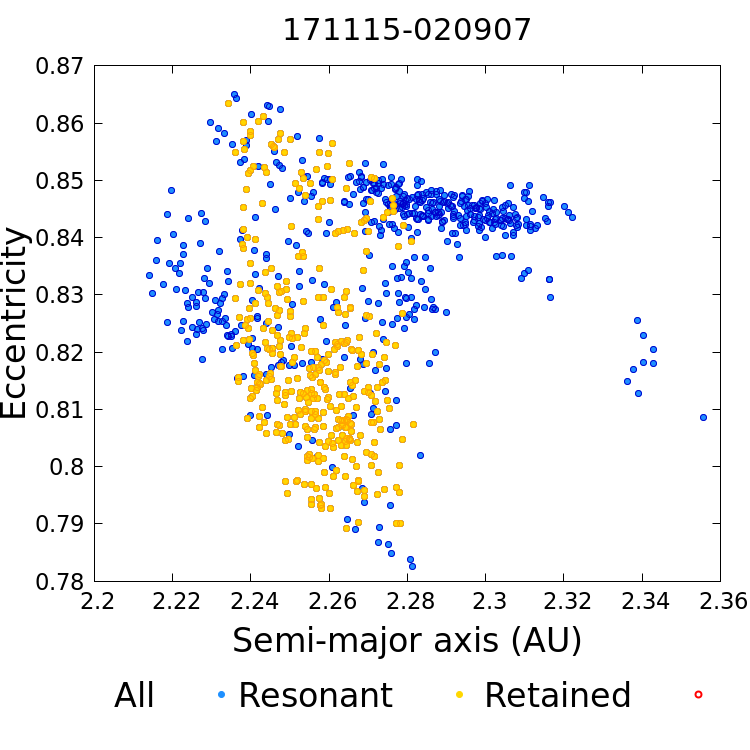}\\
  \includegraphics[width=0.32\textwidth]{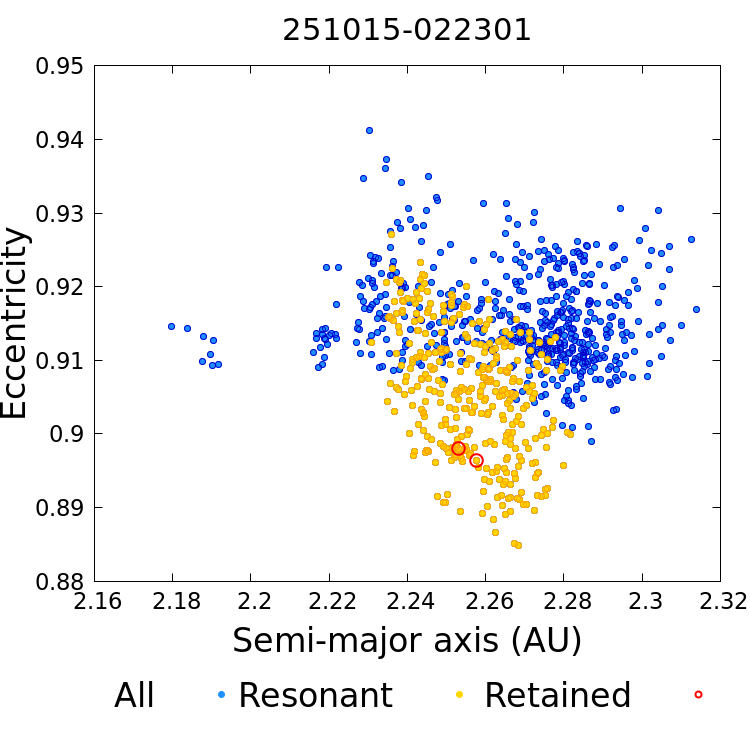}
  \includegraphics[width=0.32\textwidth]{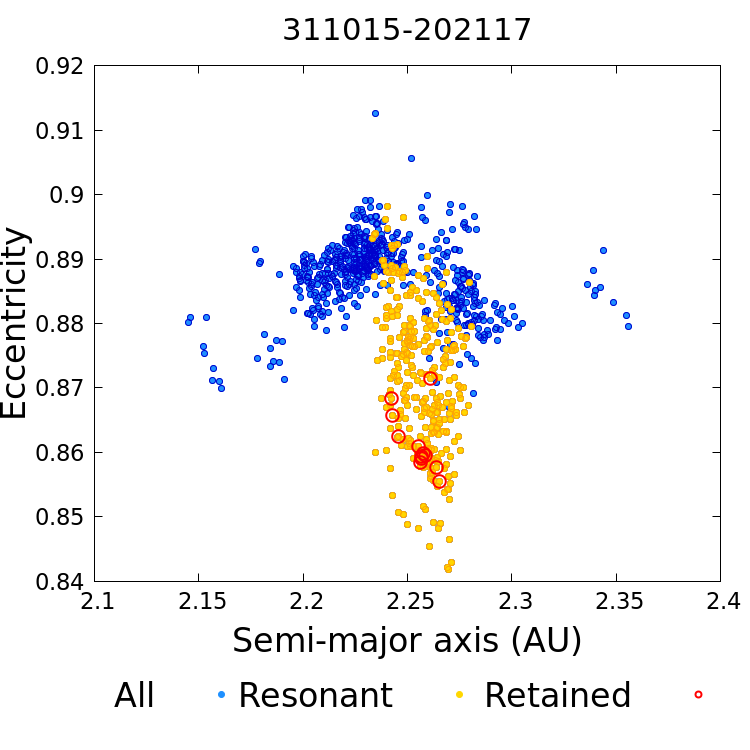}
  \includegraphics[width=0.32\textwidth]{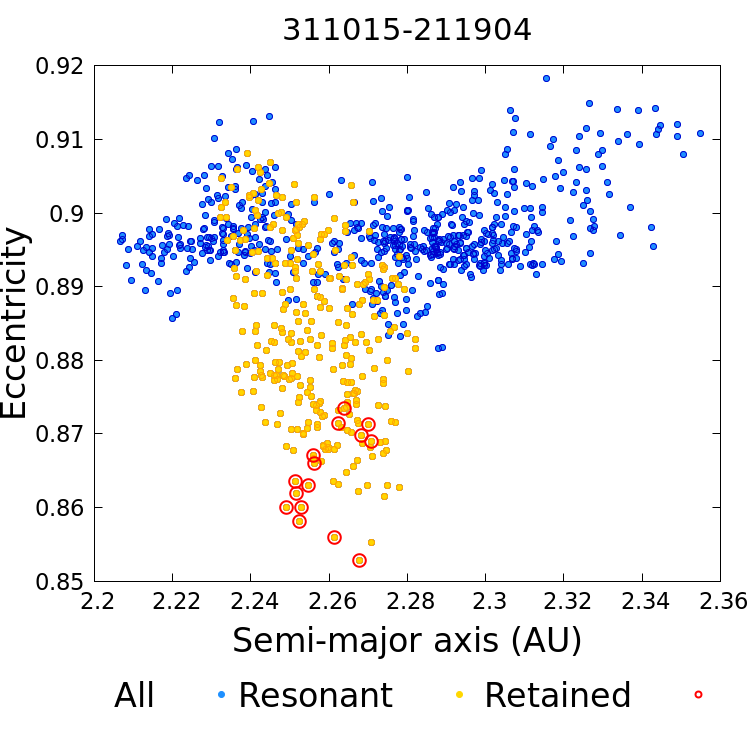}
  \includegraphics[width=0.32\textwidth]{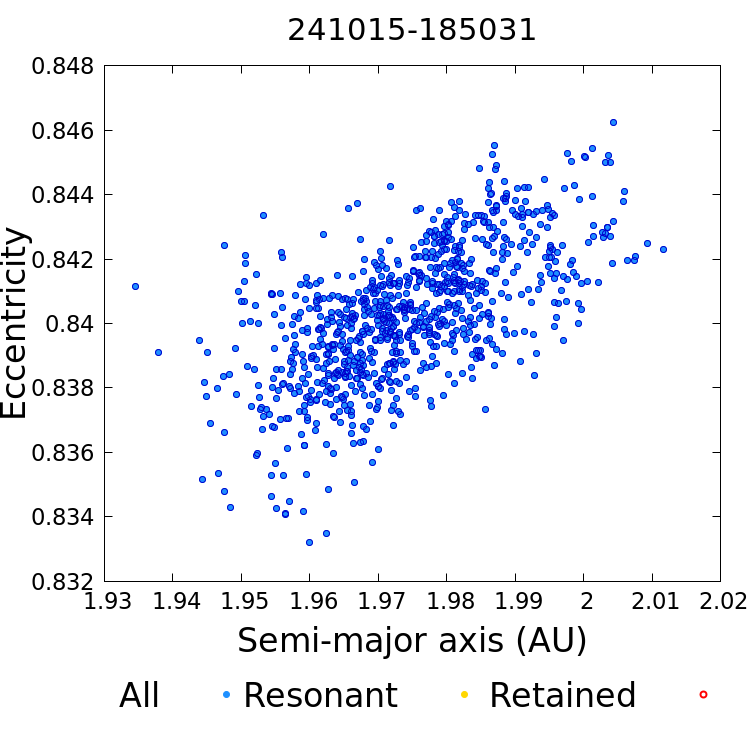}
  \caption{Semi-major axis and eccentricity of the clones created for the fireballs of Figure 13 and integrated until 10 000 BCE. The plot presents a compilation of the clones' a and e between 3000 BCE and 3500 BCE (in blue). The clones belonging to the 7:2 MMR are presented in yellow, while those approaching at least one clone of comet 2P/Encke with a low MOID and relative velocity are circled in red.}
  \label{fig:clones_selected_fireballs2}
\end{figure*}

\clearpage
\newpage
\textbf{APPENDIX G: SIMULATED METEOROIDS EJECTION FROM THE G5 BODIES}\\ \label{appendix:ejection} 

\setfignumprefix{G}
\begin{figure*}[!ht]
  \centering
  \includegraphics[width=.32\textwidth]{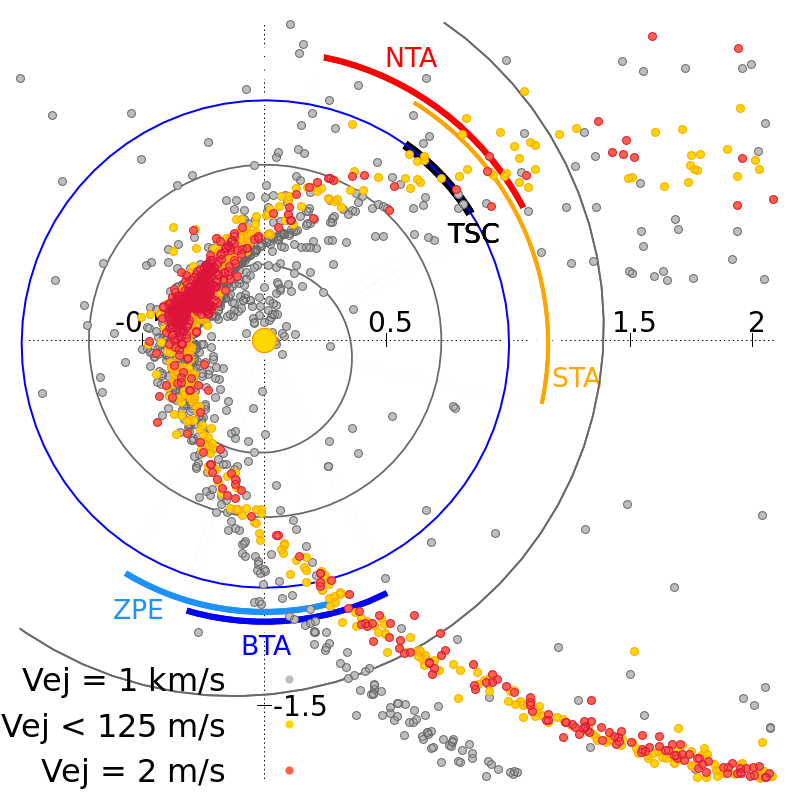}
  \includegraphics[width=.32\textwidth]{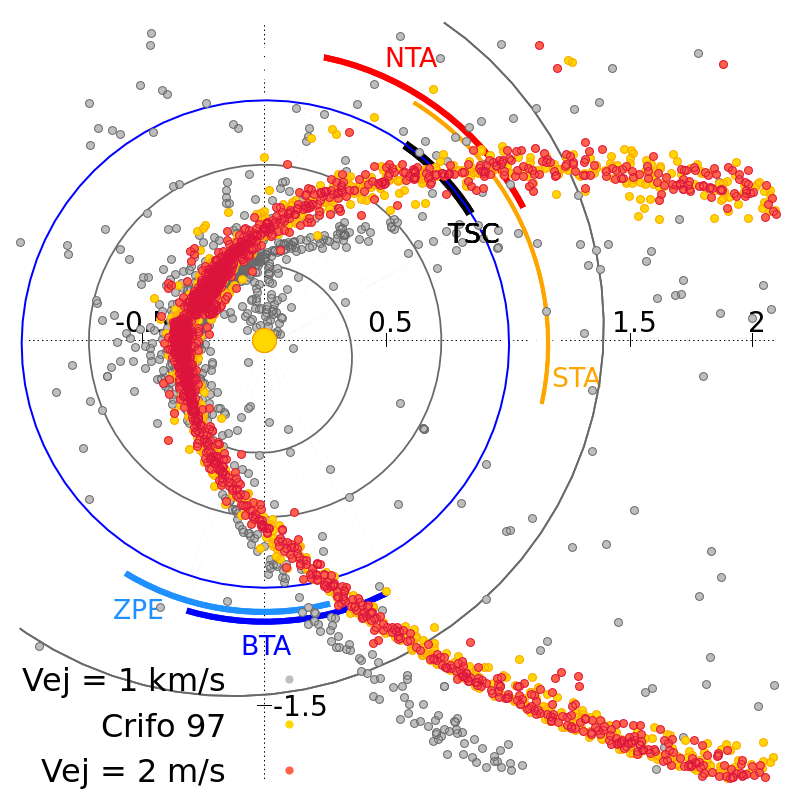}
  \includegraphics[width=.32\textwidth]{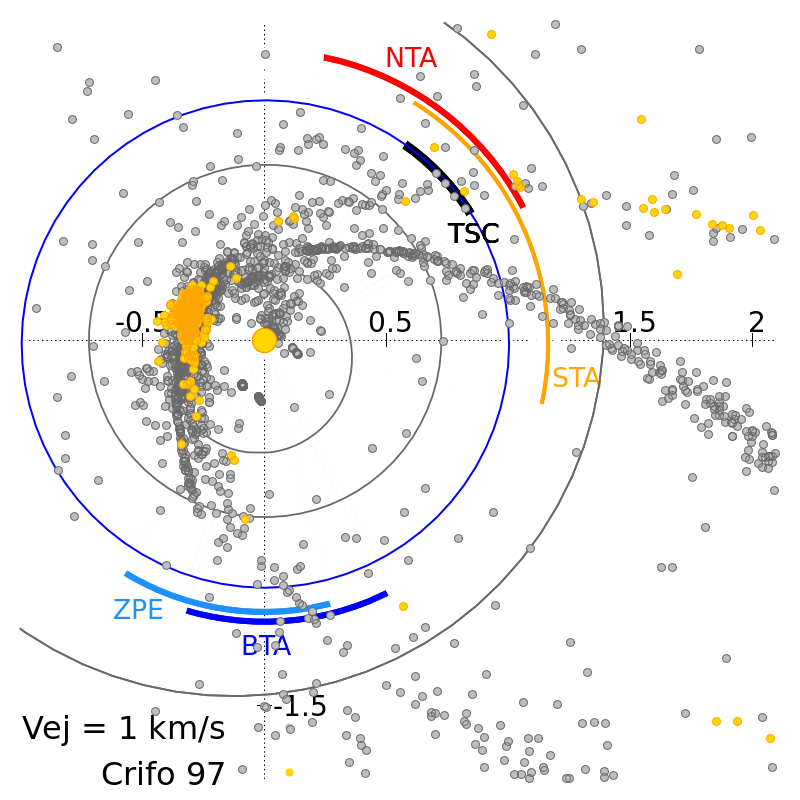}\\
  \large 2P/Encke \hspace{3.5cm} 2004 TG10 \hspace{3cm} 2005 TF50\\[0.2cm]
  \includegraphics[width=.32\textwidth]{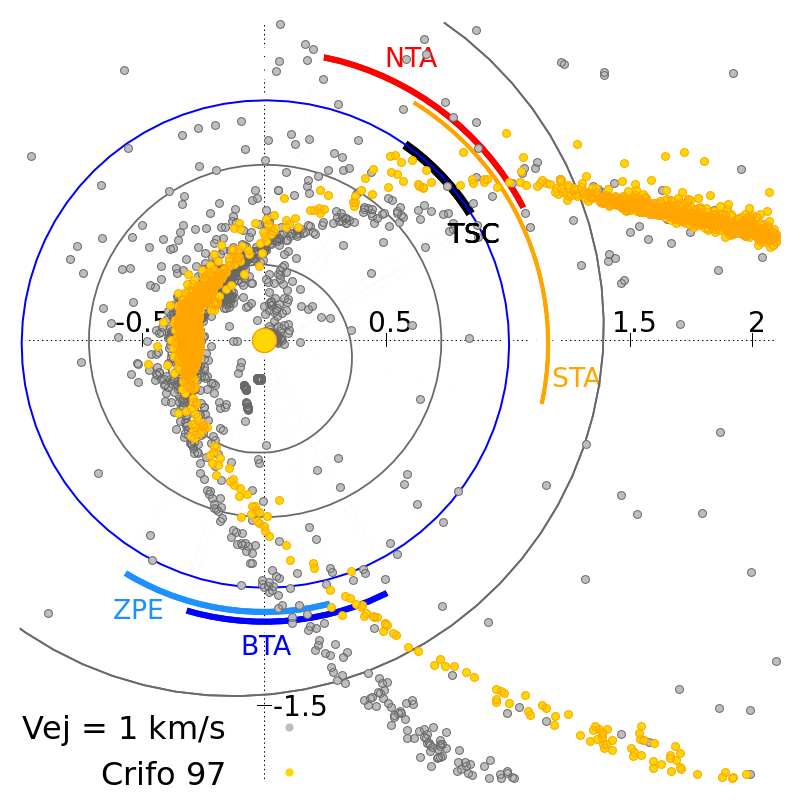}
  \includegraphics[width=.32\textwidth]{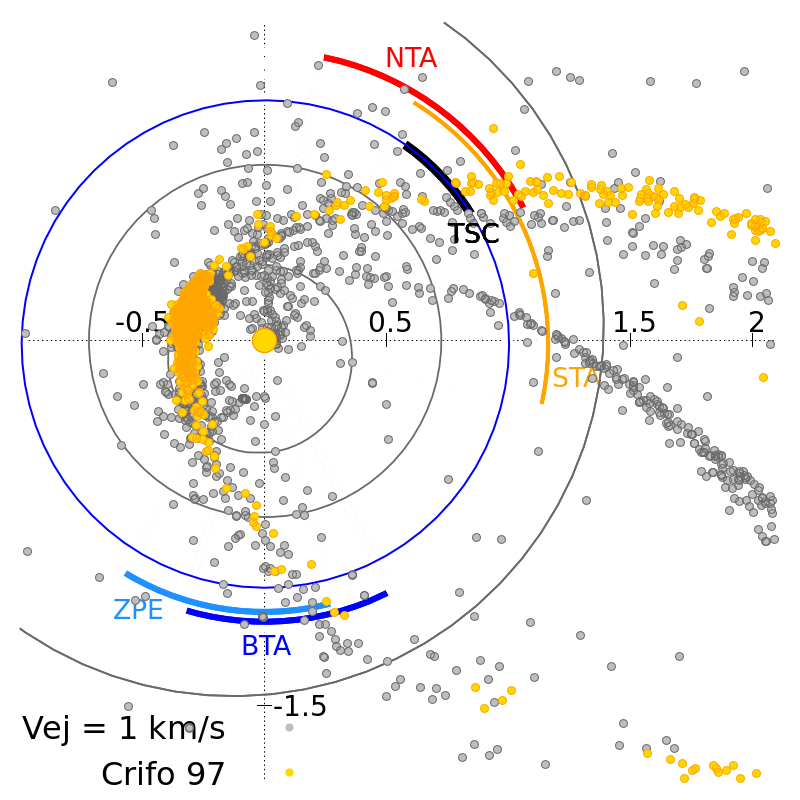}\\
  \large 2005 UR \hspace{4cm} 2015 TX24 \hspace{2cm}\\
  \caption{Distribution of nodal locations for model meteoroids crossing the ecliptic plane in 2015 based on the three ejection scenarios for the G5 bodies' orbits circa 3200 BCE (see Section 6 for details). The nodes are color coded by the meteoroid ejection velocity appropriate to the possible three scenarios discussed in the text, namely with values of 2 m/s (in red), 1 to 125 m/s (in orange) and 1 km/s (in grey). The time periods where the annual Taurid meteor showers (NTA, STA, BTA and ZPE) are active are indicated by the colored arcs of circle close to the Earth's orbit (in blue). The solar longitude range where the Taurid resonant Swarm was observed in 2015 by \protect\cite{Spurny2017} is illustrated by the black circle arc.}
  \label{fig:Ejections_3200BCE}
\end{figure*}

\begin{figure*}[!ht]
  \centering
  \includegraphics[width=.95\textwidth]{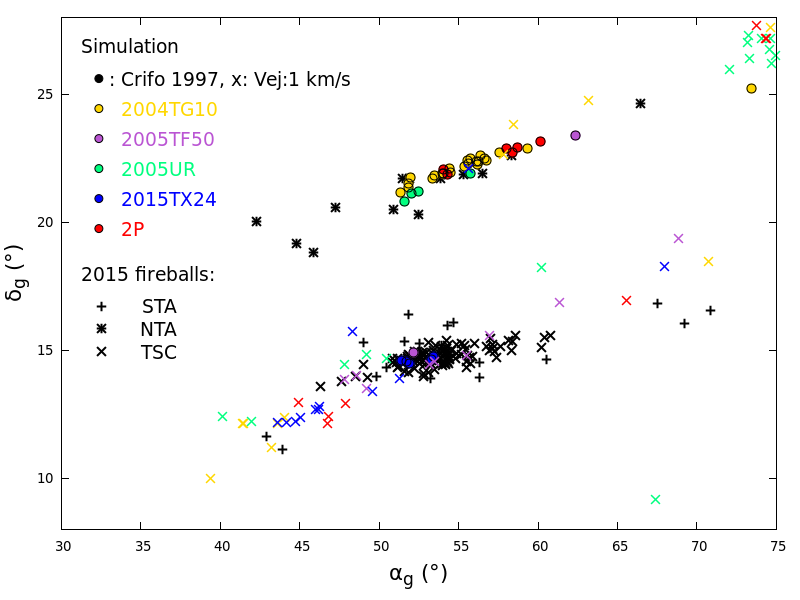}
  \caption{Right ascension and declination of the meteoroids' geocentric radiant as observed and simulated in 2015. The observed radiants of the 2015 fireballs reported by \protect\cite{Spurny2017} are represented by black symbols. The simulated radiants in 2015 corresponding to ejection from 2004 TG10, 2005 TF50, 2005 UR, 2015 TX24 and 2P around 3200 BCE are represented in orange, purple, green, blue and red respectively. Meteoroids ejected with low velocities (corresponding to scenario 2) are represented by circles, while ejecta with velocities of 1 km/s are represented by crosses. Only meteoroids that cross the ecliptic plane at a maximum distance of 0.05 AU from Earth's orbit were selected for radiant computation. See Section 6 for details.}
  \label{fig:radiants}
\end{figure*}

\clearpage
\newpage
\textbf{APPENDIX H: FUTURE POSSIBLE ASSOCIATIONS} \label{sec:future}\\

\setfignumprefix{H}
\begin{figure}[!ht]
  \centering
  \includegraphics[width=\textwidth]{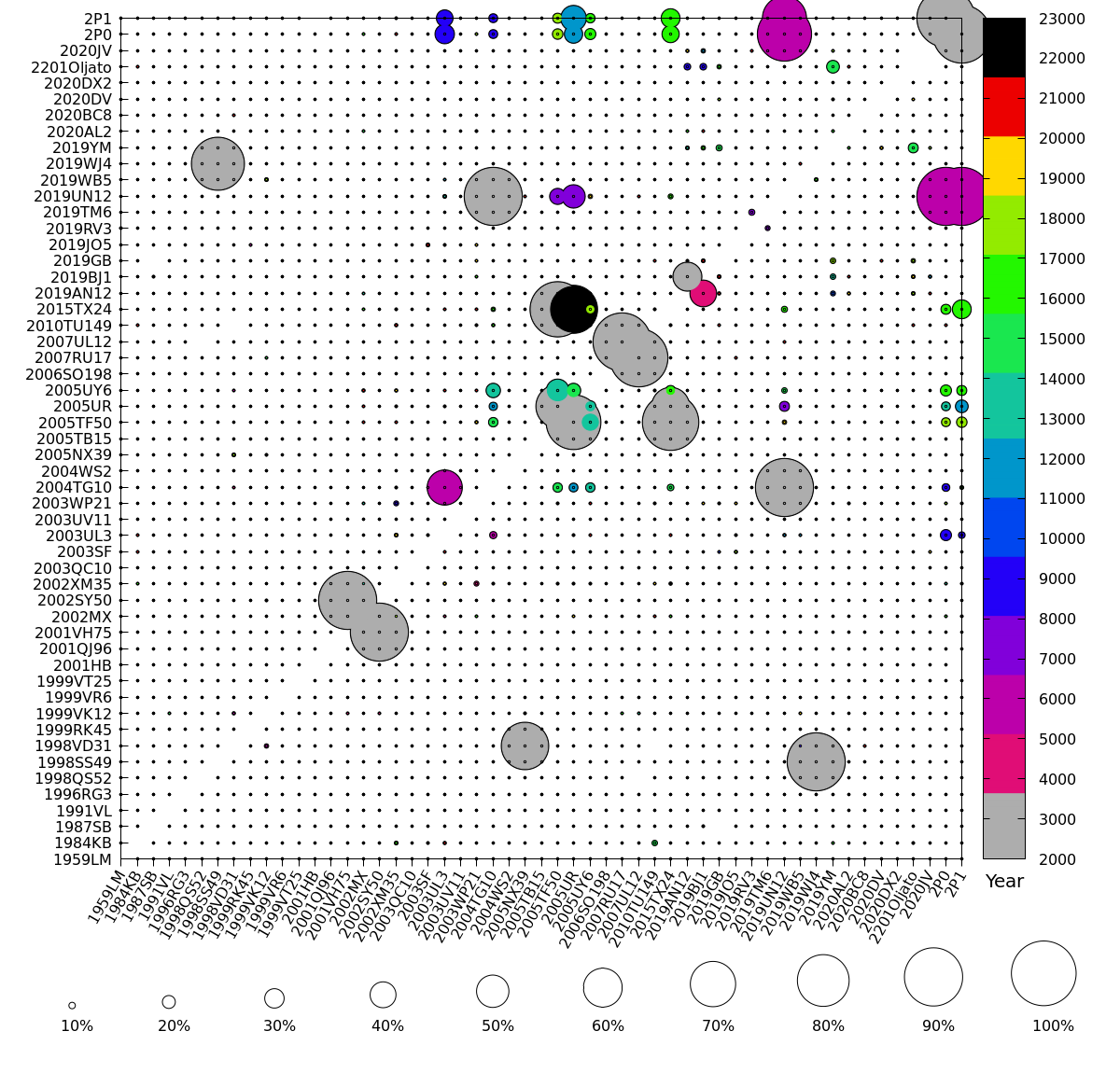}
  \caption{Possible rapprochement between TC candidates over the period 2000 CE to 23 000 CE. Abscissas and ordinates indicate the bodies integrated. Circles at position (body \#1, body \#2) indicate the percentage of clones of body \#1 approaching at least one clone of body \#2 with a MOID smaller than 0.05 AU and a relative velocity below 1 km/s. The size of the circle is proportional to the maximum percentage of clones retained for body \#1 (in abscissa) compared to body \#2 (in ordinate) over the integration period. The color of the circle indicates the epoch when this highest percentage of clones retained is attained.}
  \label{fig:map_future}
\end{figure}

\newpage
\bibliographystyle{mnras}
\bibliography{references2} 